\documentclass[a4paper,fleqn,usenatbib]{mnras}

\usepackage{multicol}
\usepackage[T1]{fontenc}
\usepackage{ae,aecompl}

\usepackage{graphicx}	
\usepackage{amsmath}	
\usepackage{amssymb}	

\title[TESS light curves of CVs -- II]
{TESS light curves of cataclysmic variables -- II -- Superhumps in old 
novae and novalike variables}

\author[A. Bruch]{Albert Bruch
\\
Laborat\'orio Nacional de Astrof\'{\i}sica, Rua Estados Unidos, 154, 
CEP 37500-364, Itajub\'a, MG, Brazil
}

\date{Accepted XXX. Received YYY; in original form ZZZ}

\pubyear{2019}

\begin{document}
\label{firstpage}
\pagerange{\pageref{firstpage}--\pageref{lastpage}}
\maketitle

\begin{abstract}
Superhumps are among the abundant variable phenomena observed in the light
curves of cataclysmic variables (CVs). They come in two flavours as positive
and negative superhumps, distinguished by periods slightly longer or
shorter, respectively, than the orbital periods of these interacting
binary systems. Positive superhumps are ubiquitous in superoutbursting
short period dwarf novae of the SU~UMa type but are less common in longer
period systems with accretion disks in a permanent bright state such as
novalike variables and most old novae. Negative superhumps do not 
seem not to have a preference for a particular type of CV. Here, I take 
advantage of the long high cadence light curves provided by TESS for
huge number of stars, selecting all old novae and novalike
variables with past reported superhumps for which TESS light curves are
available and have not yet been analysed in previous publications in order
to study their superhump behaviour. 
In combination with information taken from the literature the
results enable to compile the most complete census of superhumps in these
stars so far. As a corollary, for the eclipsing systems in the present 
sample of objects eclipse epochs derived from the TESS light curves and
in some cases from archival light curves are listed and used to update
orbital ephemeris and to discuss period changes.
\end{abstract}




\section{Introduction}
\label{Introduction}

Variability in cataclysmic variables (CVs) occurs in a multitude of different
forms and on a wide range of time scales. Most of it is associated to mass
transfer or aspect variations in these close binary systems composed of a 
white dwarf primary star and a Roche-lobe filling late type secondary
component transferring matter to the primary which -- in the absence of a 
strong magnetic field of the white dwarf -- forms an accretion disk around
the compact star before it is accreted onto its surface.

A good characterization of many of the variable phenomena requires extensive
observations of the respective stars with a suitable time resolution  over as
long a time base as possible. In this respect the long continuous high cadence
light curves provided by the Kepler mission have been extremely beneficial
\citep[e.g.,][to cite only a few examples]{Osaki13, Ramsay16, Bruch22a}.
However, Kepler observed but a few CVs. This changed considerably with
the launch of the {\it Transit Exoplanet Survey Satellite} 
\citep[TESS,][]{Ricker14} which -- although equipped with much smaller
telescopes and pointing at a given sector of the sky for less time -- provided
light curves of many more CVs with a time base and a temporal resolution well
suited to address many issues concerning the variability in these stars.

A particular type of consistent modulations in numerous CVs are the so-called
superhumps (SHs); i.e., variations with a period a few percent different from
the orbital period of the binary. SHs come in two flavours: positive
superhumps (pSHs) with periods slightly longer than the orbital period, 
and negative superhumps (nSHs) the periods of which are a bit shorter than 
the orbit.

pSHs were first observed in dwarf nova type CVs of the SU~UMa subclass
during superoutburst \citep{Vogt74} and have since become the hallmark of
this particular outburst stage of SU~UMa stars \citep[][and other publications
of this series]{Kato09}. pSHs are thought to arise when the accretion disk
expands such that the revolution period of matter at its outer rim reaches
the 3:1 resonance radius with the binary orbit. This condition is most easily 
attained
during large scale (super-) outbursts in short period dwarf novae, i.e., 
SU~UMa stars. The tidal interaction of the disk with the secondary star then
induces an elliptical deformation in the former \citep{Whitehurst88,
Whitehurst91}. Whenever the secondary star
passes close to the elongated part of the disk tidal, stresses cause an
increase of the disk luminosity. This occurs on a period slightly longer 
than the orbital period because of a prograde precession of the deformed
disk.

pSHs are not restricted to SU~UMa stars in superoutburst but are also
observed in increasing number in non-outbursting CVs such as novalike
variables (NLs) and old novae (which have an accretion disk in a similar
state as the NLs). Most of these have periods longer than the SU~UMa stars.
Since the secondary star mass in CVs increases systematically with the
orbital period their mass ratio $q = M_{\rm prim}/M_{\rm sec}$ is on average
also higher and may reach (or even surpass, see Sect.~\ref{Discussion}) the
theoretical limit for the condition required to generate SHs. This limit
is contested but appears to lie somewhere in the range $q = 0.22 \ldots 0.39$
\citep{Whitehurst91, Pearson06, Smak20}.

While the basic physics of pSHs are thus thought to be reasonably well
understood, this is not the case for negative superhumps. Phenomenologically,
they are explained to arise in a warped accretion disk or a disk inclined
with respect to the orbital plane. In such a configuration, depending on the
variable aspect between the disk and the stream of infalling matter from the
secondary star, the latter penetrates more or less deeply into the 
gravitational well of the white dwarf before its hits the accretion disk and
thus liberates a variable amount of energy, leading to a modulation of the
disk luminosity. The inclined disk precesses retrogradely such that the same
aspect between the disk and the infalling stream of matter repeats on a
period slightly less than the orbital period. While this scenario is widely
accepted to explain nSHs there is no consensus about the mechanism which
causes a warp or an inclination of the accretion disk in the first place
\citep{Montgomery09, Thomas15}. Thus, there are no theoretical constraints
for the appearance of nSHs. Observationally, they are found in short period
dwarf novae \citep{Wood11, Osaki14} as well as in long period NLs
\citep{Kimura20}.

The periods of both, pSHs and nSHs, are not strictly constant but exhibit 
small variations depending on details of the distribution of mass within the
the accretion disk.

Recently, \citet{Bruch22b} (hereafter referred to as Paper~I) investigated 
the TESS light curves of a sample of NLs and old novae and identified SHs in
several systems which were hitherto not known to be superhumpers. In extension
of that study I investigate here the TESS light curves of all NLs and old
novae with SHs, either positive or negative, reported in the literature
which have been observed by TESS, and the TESS data of which have not already
been the subject of other publications. The data used and methods applied
are briefly outlined in Sect.~\ref{Data and data handling}. Thereafter, the
individual systems are discussed in Sect.~\ref{Results}. For some eclipsing
CVs additional eclipse epochs, an update of the orbital period, and an 
assessment of period variations are also included as a corollary. In 
Sect.~\ref{Discussion}, I present a census of all superhumping NLs and old 
novae. A summary of the results concludes this study in Sect.~\ref{Summary}.

\section{Data and data handling}
\label{Data and data handling}

The details of the data used in this study and their handling are largely
the same as in Paper~I and were described there. Therefore, I will only 
give a summary here. TESS SAP data with a time resolution of 2~min were 
downloaded from the Barbara A.\ Misulski Archive for Space Telescopes 
(MAST)\footnote{https://archive.stsci.edu}. For one object (KIC~8751494)
data from the Kepler mission, retrieved from the same source, are also
used. Whenever observations from
different TESS sectors were obtained in immediate succession they were
combined into a single light curve. Different light curves of the same
object are referred to as LC\#1, LC\#2, etc. The start and end epochs of
the light curves are listed in 
Table 1.

\begin{table}
\label{Table: obs-log}	\centering
	\caption{Journal of observations.}

\begin{tabular}{lccc}
\hline

Name & LC     & Start time & End time \\
     & number & \multicolumn{2}{c}{BJD 2450000+} \\
\hline
PX And    & 1 & 8764.69 & 8788.92 \\ [1ex]
UU Aqr    & 1 & 9447.70 & 9473.16 \\ [1ex]
KR Aur    & 1 & 9474.17 & 9550.63 \\ [1ex]
BZ Cam    & 1 & 8816.88 & 8868.83 \\ 
          & 2 & 9010.26 & 9035.13 \\
          & 3 & 9390.65 & 9418.85 \\ [1ex]
V592 Cas  & 1 & 8764.69 & 8789.68 \\
          & 2 & 8955.79 & 8982.27 \\ [1ex]
RR Cha    & 1 & 9333.86 & 9389.72 \\ [1ex]
V751 Cyg  & 1 & 8711.37 & 5737.41 \\ [1ex]
V1974 Cyg & 1 & 8738.65 & 8763.32 \\
          & 2 & 9418.99 & 9446.58 \\ [1ex]
BB Dor    & 1 & 8325.29 & 8682.36 \\
          & 2 & 9036.28 & 9389.72 \\ [1ex]
BH Lyn    & 1 & 8842.51 & 8868.83 \\
          & 2 & 9579.82 & 9606.94 \\ [1ex]
BK Lyn    & 1 & 8870.44 & 8897.79 \\ [1ex]
AH Men    & 1 & 8325.30 & 8353.17 \\
          & 2 & 8410.90 & 8436.83 \\
          & 3 & 8596.78 & 8682.36 \\
          & 4 & 9036.28 & 9060.64 \\
          & 5 & 9333.86 & 9389.72 \\ [1ex]
RR Pic    & 1 & 8354.11 & 8595.68 \\ 
          & 2 & 8624.97 & 8682.36 \\
          & 3 & 9036.28 & 9060.64 \\
          & 4 & 9088.24 & 9332.58 \\
          & 5 & 9361.29 & 9389.72 \\ [1ex]
AO Psc    & 1 & 9447.70 & 9473.16 \\ [1ex]
AY Psc    & 1 & 9447.69 & 9498.81 \\ [1ex]
V348 Pup  & 1 & 9201.74 & 9254.07 \\ [1ex]
RW Tri    & 1 & 8790.66 & 8814.27 \\ [1ex]
UX UMa    & 1 & 8711.36 & 8763.32 \\
          & 2 & 8899.32 & 8954.88 \\ [1ex]
DW UMa    & 1 & 8870.46 & 8897.78 \\
          & 2 & 9607.94 & 9635.97 \\ [1ex]
HS 1813+6122&1& 8683.35 & 8841.14 \\
          & 2 & 8870.17 & 9037.40 \\
          & 3 & 9419.99 & 9456.58 \\
          & 4 & 9579.81 & 9664.31 \\ [1ex]
RX J2133.7+5107 & 1 & 8711.36 & 8763.32 \\ [1ex]
KIC 8751494&1 & 8711.37 & 8737.41 \\
          & 2 & 9390.66 & 9446.58 \\ [1ex]
KIC 9406652&1 & 8683.36 & 8710.21 \\
          & 2 & 8711.37 & 8737.41 \\
          & 3 & 8085.66 & 9446.58 \\ [1ex]
NSV 1907  & 1 & 9174.23 & 9200,23 \\
\hline
\end{tabular}
\end{table}

Frequency analysis of the data were performed with Fourier techniques
applying the Lomb-Scargle algorithm \citep{Lomb76, Scargle82} or following
\citet{Deeming75}. Unless variations on longer time scales were targeted
these were removed by subtraction of a \citet{Savitzky64} filtered version 
of the light curve, using a cut-off time scale of 2~d and a 4$^{\rm th}$ order
smoothing polynomial for the filter. 
The frequency errors of power spectrum signals were
estimated using the prescription of \citet{Schwarzenberg-Czerny91} which due
to flickering and window patterns of real variations may overestimate
the true errors. 

Comparing TESS light curves with terrestial data it should be kept in mind
that the TESS passband encompasses a wide range between 6\,000 and 
10\,000~\AA, centred on the Cousins $I$-band.

For the deeply eclipsing CVs in the present sample eclipse epochs were
measured to enable an update of the orbital ephemeris and for future
reference. Instead of measuring individual eclipse timings the light 
curves were folded on the orbital period, choosing the epoch such  that the 
centre of the primary eclipse coincides with phase 0. This was done 
individually for the data of each TESS sector, yielding more than one
eclipse epoch for the light curves combined from several sectors. The
corresponding results are listed in 
Table~2.
For four CVs (UU~Aqr, V348~Pup, RW~Tri and UX~UMa) additional eclipse
epochs were measured in light curves downloaded from the American Association
of Variable Star Observers (AAVSO) archives \citep{Kafka21} and the 
data bank of the Observat\'orio do Pico dos 
Dias\footnote{http://databank.lna.br} (LNA Data Bank)
as the minima of polynomials of suitable degree fitted to the eclipse profiles.
They are listed in Appendix~A. The transformation from JD to BJD was 
performed using the on-line tool of \citet{Eastman10}.

\section{Results}
\label{Results}

\begin{table}
\label{Table: eclipse epochs}
	\centering
	\caption{Representative eclipse epochs.}

\begin{tabular}{llll}
\hline
Star   & Light curve & Epoch (BJD) & Cycle number$^1$ \\
\hline
PX And & LC\#1 & 2458779.1330 & 65187 \\ [1ex]
UU Aqr & LC\#1 & 2459462.1633 & 47111 \\ [1ex]
RR Cha & LC\#1 & 2459363.0160 & 0     \\ [1ex]
AH Men & LC\#1 & 2458340.0867 & 0     \\
       & LC\#2 & 2458425.0527 & 668   \\
       & LC\#3 (part 1) & 2458611.0146 & 2130 \\
       & LC\#3 (part 2) & 2458646.1193 & 2406 \\
       & LC\#3 (part 3) & 2458671.0497 & 2602 \\
       & LC\#4 & 2459051.1099 & 5590 \\
       & LC\#5 (part 1) & 2459348.1110 & 7925 \\
       & LC\#5 (part 2) & 2459373.0414 & 8121 \\ [1ex]
BH Lyn & LC\#1 & 2458857.1359 & 74911 \\
       & LC\#2 & 2459591.1530 & 79620 \\ [1ex]
AY Psc & LC\#1 (part 1) & 2459462.1062 & 54476 \\
       & LC\#1 (part 2) & 2459488.1850 & 54596 \\ [1ex]
V348 Pup& LC\#1 (part 1)& 2459211.0279 & 104276 \\
       & LC\#1 (part 2) & 2459238.0154 & 104541 \\ [1ex]
RW Tri & LC\#1 & 245880.0263 & 22112 \\ [1ex]
UX UMa & LC\#1 (part 1) & 2458726.0252 & 37658 \\
       & LC\#1 (part 2) & 2458746.0858 & 37760 \\
       & LC\#2 (part 1) & 2458914.0431 & 38614 \\
       & LC\#2 (part 2) & 2458944.1338 & 38767 \\ [1ex]
DW UMa & LC\#1 & 2458885.0575 & 56178 \\
       & LC\#2 & 2459622.0503 & 61573 \\ [1ex]
HS 1813-6122 & LC\#1 (part 1) & 2458693.0878 & 0 \\
       & LC\#1 (part 2) & 2458723.0351 & 203 \\
       & LC\#1 (part 3) & 2458753.1285 & 407 \\
       & LC\#1 (part 4) & 2458773.0430 & 542 \\
       & LC\#1 (part 5) & 2458803.1372 & 746 \\
       & LC\#1 (part 6) & 2458833.0826 & 949 \\
       & LC\#2 (part 1) & 2458880.1437 & 1268 \\
       & LC\#2 (part 2) & 2458910.0900 & 1471 \\
       & LC\#2 (part 3) & 2458940.0367 & 1674 \\
       & LC\#2 (part 4) & 2458970.1298 & 1878 \\
       & LC\#2 (part 5) & 2459000.0758 & 2081 \\
       & LC\#2 (part 6) & 2459020.1399 & 2217 \\
       & LC\#3          & 2459429.0664 & 4989 \\
       & LC\#4 (part 1) & 2459589.1243 & 6074 \\
       & LC\#4 (part 3) & 2459649.0186 & 6480 \\ [1ex]
NSV 1907&LC\#1 & 2459189.1635 & 7710  \\
\hline
\end{tabular}
$^1$ cycle count convention according to:
this work (UU~Aqr, RR~Cha, AH~Men, RW~Tri, UX~UMa, HS~1813-6122);
\citet{Hellier94} (PX~And);
\citet{Andronov89} (BH~Lyn);
\citet{Diaz90} (AY~Psc);
\citet{Dai10} (V348~Pup);
\citet{Boyd17} (DW~UMa);
\citet{Hummerich17} (NSV~1907)
\end{table}

\subsection{PX And: no superhumps in the TESS light curve}

\begin{figure}
	\includegraphics[width=\columnwidth]{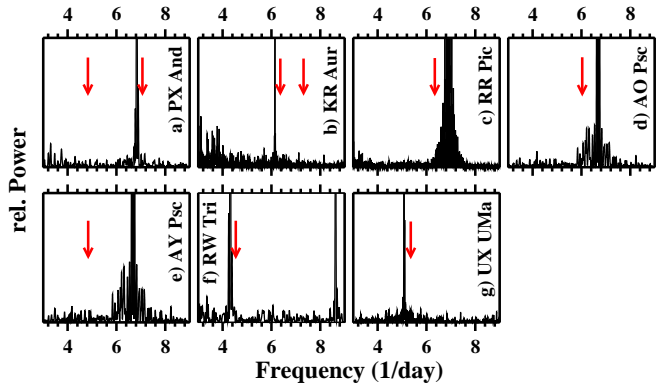}
      \caption[]{Power spectra in the range around the orbital frequency
                 of six CVs with superhumps reported in the past but with
                 no indications for superhumps in the TESS light curves. 
                 The orbital signals are heavily truncated in order to 
                 better visualize any faint signals in their vicinity.
                 The frequencies of previously identified periodic signals
                 are marked with red arrows. Note that the broad base around
                 the orbital frequency (in particular in the RR~Pic, AO~Psc and
                 AY~Psc power spectra) are not independent signals but the 
                 unavoidable sidelobes of the main peak.}
\label{negatives}
\end{figure}

PX~And is an eclipsing novalike variable. \citet{Stanishev02} derived the 
most accurate value for the orbital period: 0.146352739(11)~d. The same
authors also found a periodic variation at 0.1415~d which they attribute
to a nSH, and another one at 0.207~d the origin of which
remained unexplained. The only significant signals in the power spectrum of 
the single TESS light curve correspond to the orbital period and its 
overones. Fig.~\ref{negatives}a shows the frequency range around the
orbital signal (which is truncated in order to better visualize any faint
signals in its vicinity).  No trace of a SH, either positive or negative, 
or of the 0.207~d period appears. Their frequencies are marked by red 
arrows in the figure. A representative eclipse epoch derived from the
TESS data is listed in 
Table 2.

\subsection{UU Aqr: The superhump vanished}

UU~Aqr is an eclipsing novalike variable. Although known as a variable
star for almost a century \citep{Beljawsky26} it was identified as a CV 
only in 1986 by \citet{Volkov86}. Superhumps were observed by 
\citet{Patterson05} but were absent in extensive photometry of 
\citet{Bruch19a}. \citet{Lima21} make no mention of SHs but
claim to see a photometric period of 54.4~min, and of 25.7~min in circular
polarization.

\begin{figure}
	\includegraphics[width=\columnwidth]{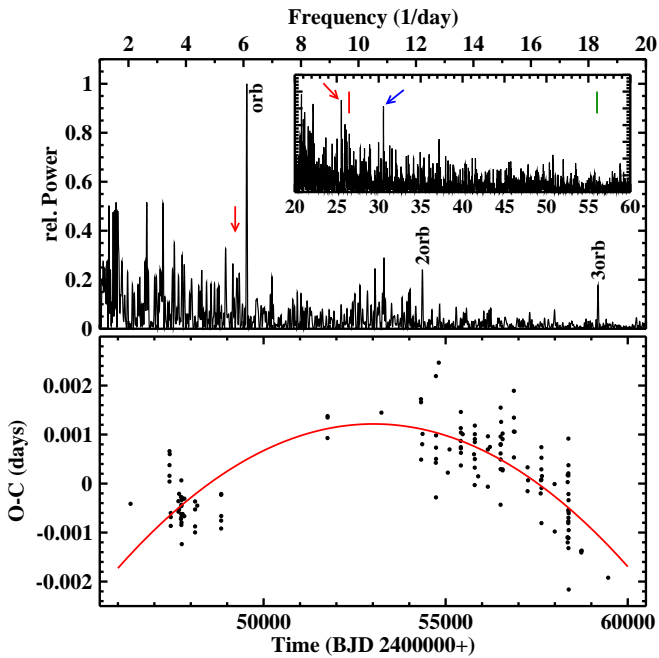}
      \caption[]{{\it Top:} Low frequency part of the power spectrum of the 
                UU~Aqr light curve. The insert shows the adjacent higher
                frequency part at on expanded power scale. The marks drawn
                into the figure are explained in the text.                
                {\it Bottom:} $O-C$ diagram of eclipse timings of UU~Aqr with
                respect to linear ephemeris according to 
                Eq.~\ref{UU Aqr: linear ephemeris}. The red graph represents
                the best fit second order polynomial.
                }
\label{uuaqr}
\end{figure}

TESS observed UU~Aqr in a single sector. The power spectrum of the light 
curve, after masking eclipses and removing variations on time scale above
2~d, is shown in the upper frame of Fig.~\ref{uuaqr}. In the low 
($<$20~d$^{-1}$) frequency range the orbital signal and the first two
overtones stand out moderately strong above a multitude of peaks with
decreasing power towards higher frequencies which
can be attributed to non-coherent fluctuations in the brightness of
UU~Aqr on time scales of hours. An increase of power between 10 and
11~d$^{-1}$ may be significant. No outstanding signal is present near 
5.711~d$^{-1}$ (marked by a red arrow in the figure), i.e., the frequency 
of the SH which is so prominent in the observations of \citet{Patterson05}
(see their figure 3). Thus, the superhump was not active during the epoch of
the TESS observations. 

At higher frequencies (insert in Fig.~\ref{uuaqr}) the fourth overtone 
of $F_{\rm orb}$ is the only signal which can be attributed to orbital variations
(blue arrow). A stronger signal at 25.55~d$^{-1}$ ($P = 56.36$~min; red arrow) 
is distinct
from the third orbital overtone, but is possibly related to the 54.4~min
photometric periodicity mentioned by \citet{Lima21}. Although the
corresponding frequency (red bar) is somewhat higher, the power spectrum 
in figure~7 of \citet{Lima21} contains numerous alias peaks reaching out
until well beyond 56.36~min. I note, however, that the light curves of 
\citet{Bruch19a} do not contain an indication for variations in this 
period range. On the other hand, the power spectrum of those data has a 
marginally significant peak compatible with the 25.7~min polarimetric 
period which \citet{Lima21} take as an indication
for an intermediate polar nature of UU~Aqr. But no such signal can be 
discerned in the TESS data (green bar in Fig.~\ref{uuaqr}). 

The orbital period of UU~Aqr was last refined by \citet{Baptista95} almost
30~yr ago. It is based on eclipse epoch measurements over a time base of
about 7~yr. I am not aware of any published eclipse timings since then
which would enable to enlarge the time base for period determination.
A representative eclipse epoch derived from the TESS light curve is 
listed in 
Table~2.
The AAVSO archives and the LNA Data Bank contain many more time 
resolved light curves of UU~Aqr observed between 2000 and 2019 which can 
be used to measure additional eclipse timings. After
rejecting a couple of eclipses because their timings led to excessive $O-C$ 
values, I am left with 99 additional eclipse
epochs which are listed in 
Table~A1. 
These new data extend the time base for period determination by more than a 
factor of 5.

Combining the new eclipse timings with those listed by \citet{Baptista94},
(assigning weight 10 to the TESS eclipse epoch because it is based on many
individual eclipses, and 1 to all others),
neglecting the slight difference between BJD and HJD used in the earlier
publications, and choosing an epoch close to the centre of all available 
eclipse epochs as zero point for cycle counts, 
the revised linear orbital ephemeris for UU~Aqr are:
\begin{equation}
\label{UU Aqr: linear ephemeris}
T_{\rm min} = BJD\, 2451755.72717(8) + 0.163580440(2) \times E
\end{equation}

The resulting $O-C$ curve for the eclipse epochs are plotted in the lower
frame of Fig.~\ref{uuaqr}. There is a clear trend over time which can very
well be described by a parabola, indicating that the orbital period of
UU~Aqr changes gradually over time. Thus, the eclipse epochs are better
described by quadratic ephemeris:
\begin{eqnarray}
\label{UU Aqr: quadratic ephemeris}
T_{\rm min} & = & BJD\, 2451755.72834(4) \\ \nonumber
          &   & + 0.163580465(1) \times E \\ \nonumber
          &   & - 1.84(4)\, 10^{-12} \times E^2
\end{eqnarray}
The period decreases currently at a rate of
${\rm d}P/{\rm d}t = -2.240(4) \times 10^{-11}$
and the relative period decrease is 
$\dot{P}/P = 5.010(9) \times 10^{-8}$~yr$^{-1}$.

\subsection{KR Aur: The superhumps subsided}

KR~Aur is a well known novalike variable of the VY~Scl subtype. The long
term behaviour has been extensively monitored in the literature 
\citep[see, e.g.,][]{Honeycutt04}. For the early history of the system, see 
\citet{Kato02a}. The orbital period was measured 
spectroscopically by \citet{Hutchings83} and \citet{Shafter83} and
more recently photometrically by \citet{Rodriguez-Gil20} who provide the
most accurate value of 0.162771641(49)~d. 
Apart from the frequent low states which characterize KR~Aur as a VY~Scl
star the system exhibits variability also on short time scales,
i.e., the usual flickering seen in all CVs, but stronger than in most NLs
\citep{Bruch21}. Significant
signals with unstable periods on the time scale of several hundred seconds
have been seen by \citet{Singh93} and \citet{Kato02a}.
\citet{Biryukov90} claim the presence of 25~min variations, but these 
are quite unstable and can at most be classified as quasi-periodic 
oscillations (QPOs). In contrast,
\citet{Kozhevnikov07} reports the presence in 2004, January and February,
of a nSH at a period of 0.15713(2)~d.
Similar signals were, however, not detected in observations of
\citet{Kato02a}. In contrast, more recently,
in 2021 January, \citet{Boeva21} observed a nSH at a period
of 0.1367(2)~d, significantly shorter than that seen by 
\citet{Kozhevnikov07}. All these observations were performed in high states.

Some months later, between September and November of the same year, 
again in a high state, TESS 
observed KR~Aur in three sectors in subsequent time intervals. Apart from
multiple signals at frequencies $<$4~d$^{-1}$ due to random variations on
longer time scales, only a strong signal at the orbital frequency is
outstanding. No indications for superhumps can be detected 
(Fig.~\ref{negatives}b). Thus, the
variations seen by \citet{Boeva21} had subsided. The appearance of 
superhumps in KR~Aur is consequently not permanent but an intermittent 
phenomenon. Moreover, the high frequency part of the power spectrum does 
not contain power in excess of the usual red flickering noise on the times 
scales indicated by \citet{Singh93}, \citet{Kato02a} or \citet{Biryukov90}.

\subsection{BZ Cam: Lots of unstable signals}
\label{BZ Cam}

The long term photometric behaviour of BZ~Cam, classified as a novalike
variable, is somewhat unusual for its class. For many years it remained
at a seemingly stable magnitude of $\sim$13 after a low state at $\sim$14~mag 
in 1928 \citep{Garnavich88}. Another low state occurred in 1999
\citep{Greiner01, Kato01}. This would make BZ~Cam a typical VY~Scl star. But 
the AAVSO long term light curve, starting in late 2000, contains several
excursions to a brighter state around 12~mag (apart from a short glitch
to the low state level). 

Combining spectroscopic and photometric data
\citet{Patterson96} derived an orbital period of 0.153693(7)~d. They also
saw very complicated structures in the power spectra of their light curves
with a concentration of multiple signals in the frequency range below
$\sim$20~d$^{-1}$. They tried to isolate specific signals and to discuss
them in terms of positive and negative SHs, but admitted that their
interpretation is not unique. \cite{Kato01}, in contrast, claim the
presence of a pSH at 0.15634(1) during the 1999 low state
of BZ~Cam.

\begin{figure}
	\includegraphics[width=\columnwidth]{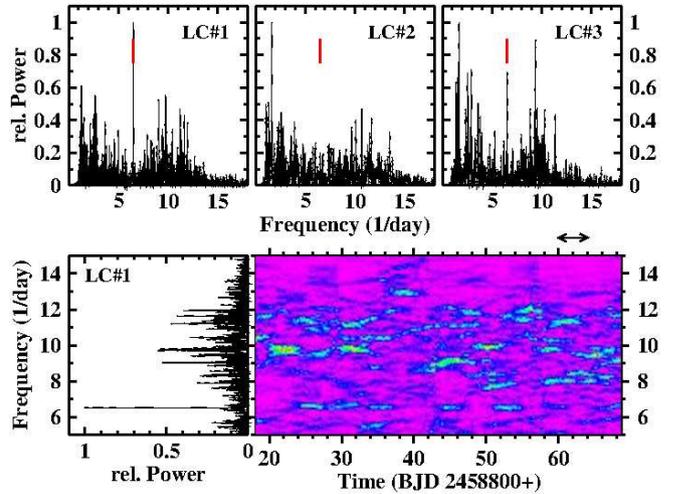}
      \caption[]{{\it Top:} Power spectra of the light curves of BZ~Cam.
                the red vertical lines indicate the orbital frequency.
                {\it Bottom:} Part of the power spectrum of LC\#1 in
                a conventional representation (left) together with 
                a three dimensional plot of the power (colour coded) as a 
                function of frequency and time (right). The double arrow above
                the plot indicates the width of the time intervals used to
                calculate the power spectrum and thus its time resolution.}
\label{bzcam-ps}
\end{figure}

The TESS observations of BZ~Cam can be combined into three light curves.
Their power spectra (upper frames of Fig.~\ref{bzcam-ps}) are similar to
those shown by \citet{Patterson96} with a concentration of peaks in the
range between 8 and 14~d$^{-1}$ (periods between 1.7 and 3~h). 
The orbital frequency is marked by red vertical
bars in the figure. Orbital variations clearly manifest themselves in LC\#1
and LC\#3, but are absent in LC\#2. No trace of the SH seen by
\citet{Kato01} is present in the power spectra. 

In order to further investigate the occurrence of multiple apparently quite
unstable periods of a few hours, time resolved power spectra were calculated
using a sliding window with a widths of 4~d. The results for LC\#1 are shown
in the lower right frame of Fig.~\ref{bzcam-ps} (those for the other light 
curves are qualitatively similar). For comparison, the conventional power
spectrum of the entire light curve is reproduced in the left frame. The
time resolved spectrum is dominated by a profusion of signals which appear at
random, vanish after a couple of days and can change their frequency during
their life time. This is the typical behaviour of QPOs 
which are not uncommon in CVs but normally have shorter periods in the
range of minutes to some tens of minutes. Their behaviour in BZ~Cam is,
however, somewhat reminiscent of CP~Pup (see Paper I). These QPOs having been
seen by \citet{Patterson96} in 1994-95 and by TESS between 2019 and 2021
suggest that they are a permanent property of BZ~Cam.

Another interesting feature in the time resolved power spectrum is the
coming and going of the orbital signal which semi-periodically appears
and vanishes on the time scale of several days.

\subsection{V592 Cas: No nSH and drastically changed pSH waveform}
\label{V592 Cas}

V592~Cas was discovered by \citet{Greenstein70} as LSI~55$^o$-8. The
orbital period was measured spectroscopically to be 0.115063(1)~d
by \citet{Taylor98}. The latter authors also found strong pSHs
at a period of 0.12228(1)~d. Additionally, in 1997-1998 they saw a weak 
signal at 0.11193(5)~d which they interpreted as a nSH.
This was not detected in the 1993 observing season.

\begin{figure}
	\includegraphics[width=\columnwidth]{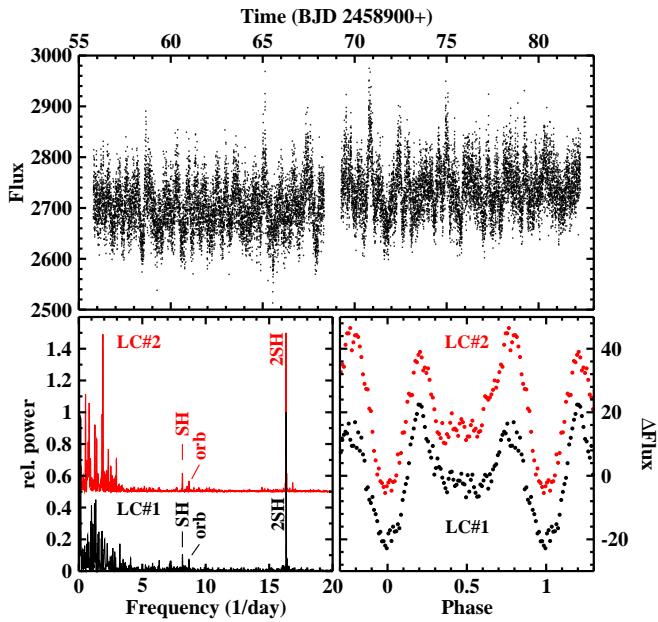}
      \caption[]{{\it Top:} Light curve LC\#2 of V592~Cas.
                {\it Bottom:} Power spectra (left) and superhump
                waveforms (right) derived from light curves LC\#1 (black)
                and LC\#2 (red, shifted vertically for clarity).}
\label{v592cas}
\end{figure}

The overall properties of the two available TESS light curves taken about 
6 months apart are similar. The upper frame of Fig.~\ref{v592cas} shows LC\#2.
It is characterized by regular but non periodic variations on the time scale
of a day. This is reflected in the power spectra in the lower left frame
of the figure which contains numerous peaks at frequencies below 4~d$^{-1}$.
A faint peak corresponding to a period of 
0.1151(1)~d
in both power spectra can be identified with the orbital period of V592~Cas. 
I consider another signal at a slightly lower frequency,
corresponding to a period of
0.1225(1)~d,
very similar to that reported by \citet{Taylor98}, as being caused
by a positive superhump. However, its first overtone is vastly stronger.
This is explained by the SH waveform shown in the lower right
frame of Fig.~\ref{v592cas} which consists of two maxima separated by
minima of quite different depth. Except for a reversal of the slightly
different heights of the two maxima, the waveform is the same in both
lightcurves. It is drastically different from the simple saw-tooth shape 
observed by \citet{Taylor98}. 

With one notable exception,
apart from overtones and simple arithmetic combinations of the orbital and 
SH frequencies the power spectra contain no indications of other
significant periodicities. In particular, even after applying the same
technique as \citet{Taylor98}, i.e., subtracting the pSH
variation from the light curve, the present data reveal no trace
of a nSH. The exception is a peak at low frequencies (period: 0.53~d)
seen in LC\#2 which is about as strong as the dominant first overtone of the 
superhump. It is not present in LC\#1. One might therefore suspect it to be 
due to an accidental alignment of the random low frequency variations
of V592~Cas. But this seems not to be the case because it is 
equally present at very nearly the same frequency in 
the first and in the second half of the light curve and thus persists at least 
over its total time base. This periodicity has no obvious relationship to
the orbital or the SH period. Its nature remains unclear.

Finally, there is a broad enhancement of power between
~65 and 170~d$^{-1}$ (8.5 -- 22~min) which may explain the 22~min oscillation
observed by \citet{Kato02b}.

\subsection{RR Cha: Superhumps and a revision of the WD spin period}

Few detailed studies of the quiescent phase of Nova Chamaeleontis 1953
(RR~Cha) have been published. Most relevant in the present context is
the paper of \citet{Woudt02} who discovered eclipses in RR~Cha, recurring
at a period of 0.1401~d. Moreover, they detected positive as well
as negative SHs at periods of 0.14442~d and 0.13529~d, respectively. 
They also identify
a signal in the power spectra of their data corresponding to a period
of 32.5~min and suspect RR~Cha to be an intermediate polar.
Further evidence for this is provided by \citet{Rodriguez-Gil03}
who observed circular polarization in RR~Cha which ``appears to be
modulated on the spin period of the primary and harmonics of the positive
superhump period''.

\begin{figure}
	\includegraphics[width=\columnwidth]{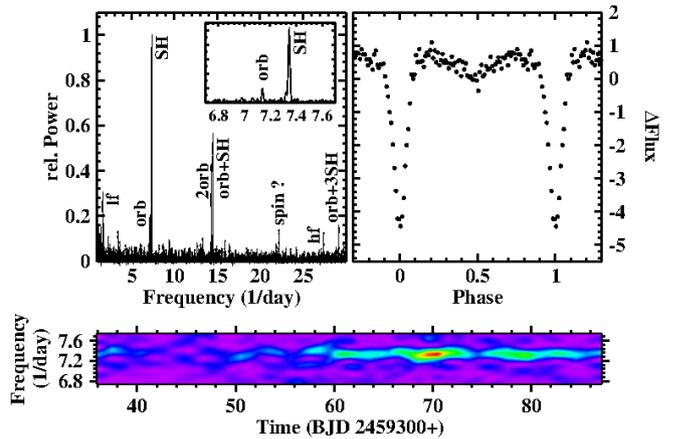}
      \caption[]{{\it Top Left:} Power spectrum
                of RR~Cha. On the scale of the plot the orbital and SH
                signals are only marginally resolved. Therefore, the insert
                shows the corresponding frequency range on an expanded scale.
                {\it To Right:} Average waveform of the orbital variations 
                of RR Cha.
                {\it Bottom:} Time resolved power spectrum of the range 
                around the superhump frequency.}
\label{rrcha}
\end{figure}

Due to the faintness of RR~Cha, the TESS light curve presents itself to the
eye as almost featureless. However, a closer look reveals some interesting
properties. The power spectrum of the original data is dominated by a signal
at the orbital frequency. The period measured by \citet{Woudt02} is based
on observations over a time interval of just above 2~d. The TESS light
curve with a time base of almost 56~d should therefore permit to increase
the accuracy of the orbital period by more than an order of magnitude.
Folding the data on the inverse of the orbital frequency derived from the
power spectrum yields a representative epoch for the eclipse
minimum (Table 2) and the ephemeris
\begin{equation}
T_{\rm ecl} = BJD\, 2459363.0160(14) + 0.14006(1) \times E 
\end{equation} 
where the error of the epoch is arbitrarily taken to be 1\% of the period.
The orbital waveform is shown in the upper right frame of 
Fig.~\ref{rrcha}. Out of eclipse it is characterized by a symmetrical 
double hump.

The power spectrum of RR~Cha, after masking the eclipses, is reproduced in
the upper left frame of Fig.~\ref{rrcha}. It is dominated by a signal at
$F_{\rm SH} = 7.346(1)$~d$^{-1}$ ($P_{\rm SH} = 0.13613(2)$~d),
just above the orbital frequency. The period is very close to that
of the nSH seen by \citet{Woudt02}. A time resolved power
spectrum reveals a significant evolution of the strength of this signal.
It starts rather weakly, gains strength in the third quarter of the light
curve, and then looses power again, as can be seen in the time resolved
power spectrum shown in the lower frame of Fig.~\ref{rrcha}. 
The TESS light curve does not contain 
the pSH detected by \citet{Woudt02}.

Other signals in the power spectrum can be identified as the first 
overtone of the orbital frequency and arithmetic combinations of the orbital
and the superhump frequencies. However, a peak at
22.215(3)~d$^{-1}$
cannot be explained in this way. The corresponding period of
64.821(9)~min is almost exactly twice the 32.5~min period seen by
\citet{Woudt02} (with no error margin attached to it) and which they
interpret as either the white dwarf spin period or its beat with the
orbital period. This can hardly be a coincidence. Depending on the detailed
conditions in an individual system the interplay between orbital motion and
the white dwarf spin can lead to many different periodic modulations in the
light curves \citep{Warner86, Norton96}. Signals at twice the spin frequency
are, for instance, seen in the intermedidate polar AO~Psc 
(see Sects.~\ref{AO Psc}). It is therefore conceivable that
the period seen by \citet{Woudt02} is the first overtone of the spin period 
(which is not seen in the present TESS data) and that a change in the system 
configuration leads to a signal at the fundamental period seen now. 

Two more apparently significant signals can be detected 
in the power spectrum, one identified in Fig.~\ref{rrcha} as 
$ F_{\rm lf} = 1.622(2)$~d$^{-1}$ ($P_{\rm lf} = 0.6164(7)$~d),
the other as
$ F_{\rm hf} = 27.381(3)$~d$^{-1}$ ($P_{\rm hf} = 0.036522(4)$~d).
Neither $P_{\rm lf} = 14.8\, {\rm h}$ nor $P_{\rm hf} = 52.6\, {\rm min}$ 
has an obvious relationship to other periods in RR~Cha. Thus, their
origin remains unexplained. 

\subsection{V751 Cyg: Nothing new}

\begin{figure}
	\includegraphics[width=\columnwidth]{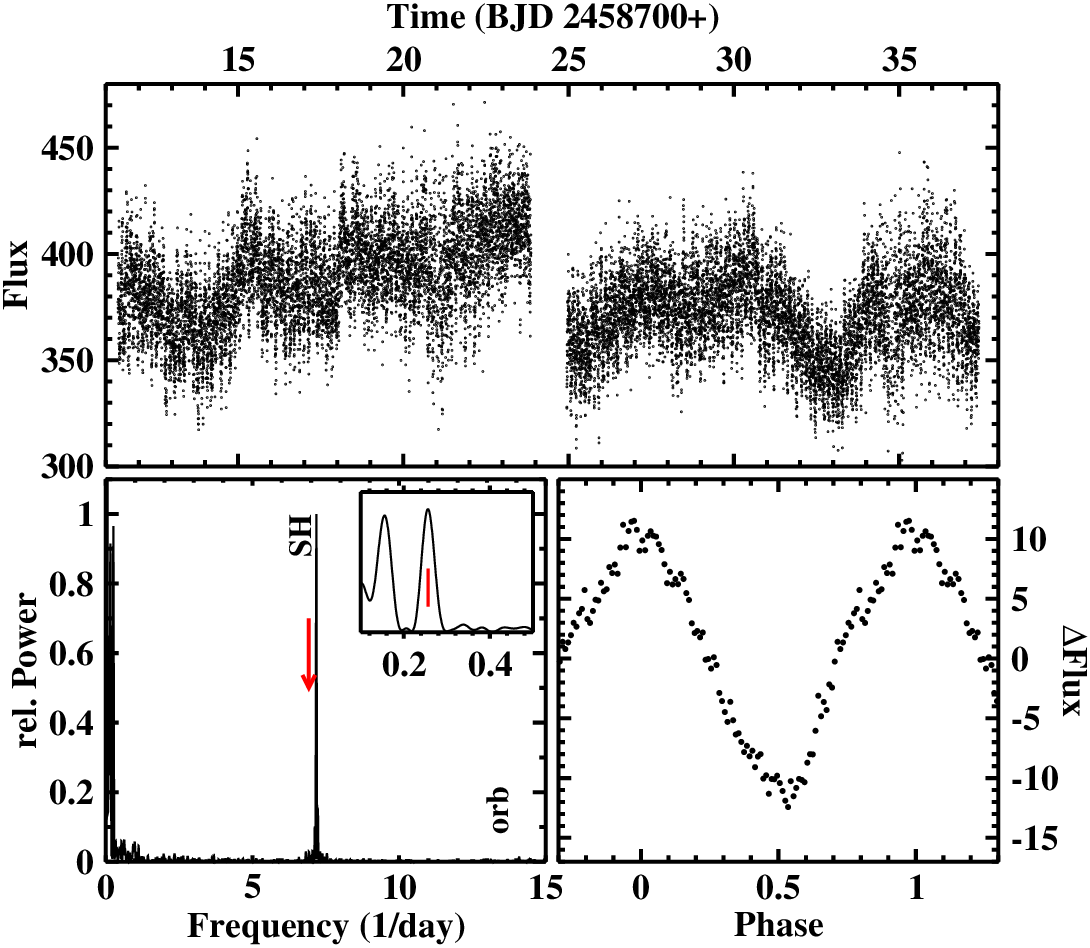}
      \caption[]{{\it Top:} Light curve of V751~Cyg. {\it Bottom left:}
                Power spectrum of V751~Cyg. The red arrow indicates the
                frequency of the orbital period which is not detectable in
                the light curve. The insert shows the low frequency part
                the power spectrum where the beat frequency between orbital
                and the superhump signals is marked by a red bar.
                {\it Bottom right:} Waveform of the superhump modulation.}
\label{v751cyg}
\end{figure}

\citet{Patterson01b} measured a spectroscopic orbital period of 0.1445(1)~d
for V751~Cyg and found a nSH at 0.13948(7)~d. The latter
was also seen by \citet{Papadaki09}. The single
TESS light curve confirms the continued presence of the superhump at a
period of
0.13930(2)~d with a very nearly sinusoidal waveform. Fig.~\ref{v751cyg}
shows the light curve, the power spectrum and the superhump waveform.
The orbital frequency is marked by a red arrow in the power spectrum,
indicating that an orbital signal is notably absent in the light curve.
Instead, a clear signal at 1.87~d, i.e., the beat between orbit and superhump,
is clearly seen (marked by a red bar in the insert of the lower left frame of
Fig.~\ref{v751cyg}). It also appears in the power spectrum of 
\citet{Patterson01b}
The power spectrum does not contain other coherent signals but an enhancement
of power between ~60 and 75~d$^{-1}$ (20 -- 24~min), possibly due to QPOs. 

\subsection{V1974 Cyg: Superhumps and a 1.3~d variation}
\label{V1974 Cyg}

V1974~Cyg (Nova Cygni 1992) exhibits two distinct photometric periodicities.
The first one, at 0.0812585(5)~d \citep{DeYoung93, Retter97}, is considered
to be orbital. A second slightly variable period close to 0.0850~d 
\citep{Semeniuk94, Semeniuk95, Retter97} can be interpreted as a positive 
superhump. A third periodicity of 0.08304~d was seen in 1994 by 
\citet{Retter97} but did not show up in 1995.

\begin{figure}
	\includegraphics[width=\columnwidth]{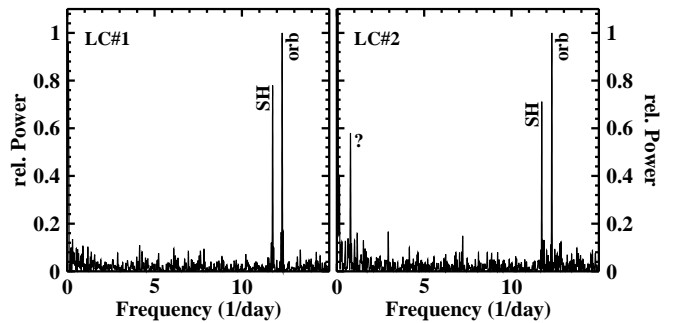}
      \caption[]{Power spectra of the two TESS light curves of V1974~Cyg.}
\label{v1974cyg}
\end{figure}

The two available TESS light curves, separated by two years, confirm the
presence of the orbital signal at
0.08127(2)~d
while the superhump signal yields slightly different periods of
0.08504(2) and 0.08525(2)~d in LC\#1 and LC\#2 (Fig.~\ref{v1974cyg}). Both 
signals are of comparable strength in the power spectra. Additionally, the 
power spectrum of LC\#2 contains another peak, marked with a question
mark in the figure, which is almost as strong as the superhump signal. It 
corresponds to a period of
1.281(7)~d. 
Note that this is {\it not} the beat period between the orbital and
superhump modulations. Its origin remains unclear but is reminiscent
of the 0.53~d period seen in LC\#2 of V592~Cas (Sect.~\ref{V592 Cas}). 
The 0.083~d period reported by \citet{Retter97} cannot be deteced in the TESS 
light curves.

It is noteworthy, however, that in addition to the superhump 
\citet{Semeniuk94} found a period of 3.75~d in their 
data. Although they do not quote error limits, their fig.~7 suggests that
this period is compatible with twice the beat between the orbital and
the superhump periods (1.84~d) \citep[see also][]{Semeniuk95}, similar to 
what has been seen by \citet{Bruch18} in V603~Aql. 

\subsection{BB Dor: Four periods, but not the orbital one}

BB~Dor (= EC 05287-5847) was identified as a cataclysmic variable by
\citet{Chen01}. Their tentative classification as a VY~Scl type star was
confirmed by \citet{Rodriguez-Gil12}. The latter authors observed long
term variations with a period of 36.43~d as well as a spectroscopic
orbital period of 0.154095(30)~d. A period of 0.14923(7)~d --
observed and thought to be orbital by \citet{Patterson05} -- would then
be due to a nSH, while a weaker signal at 0.1633~d indicates
a pSH. Another still much weaker signal in their power
spectrum has a frequency of 12.833~d$^{-1}$, very nearly the sum of the
two superhump frequencies.

\begin{figure}
	\includegraphics[width=\columnwidth]{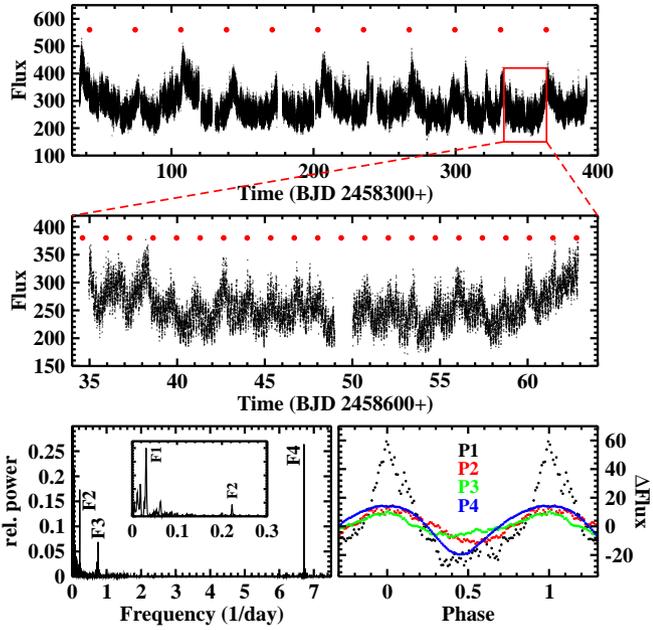}
      \caption[]{{\it Top:} Light curve LC\#1 of BB Dor.
                {\it Middle:} Light curve of TESS sector 12.
                The red dots above the light curves mark predicted times
                of maxima of variations with a period/quasi-period of
                32.15~d (top) and 1.35~d (middle).
                {\it Bottom left:} Power spectrum of LC\#1. The insert contains
                an expanded view of the low frequency range.
                {\it Bottom right:} LC\#1 folded on the four independent
                periods identified in the light curves (after subracting
                variations on longer time scales).}
\label{bbdor-lc}
\end{figure}

TESS observed BB~Dor in no less than 20 sectors. The star remained in a 
stable high state all the time. The individual light
curves can be combined into two long ones, both with baselines of almost a
year (there is a gap of 27~d in the second light curve). LC\#1 is shown in
the upper frame of Fig.~\ref{bbdor-lc}. The middle frame contains the light
curve of sector 12. Periodic or semi-periodic variations are
obvious on three different time scales: some tens of days, just over one day,
and a fraction of a day. Predicted times of maxima for the first two of
these, based on a formal period measurement (see below) are marked by red
dots on top of the light curves in the figure.

The power spectrum LC\#1 is reproduced in the lower left frame of 
Fig.~\ref{bbdor-lc}. That of LC\#2 is practically identical. Four
independent signals can be identified. The corresponding frequencies
$F_1$ - $F_4$ are marked in the figures and listed together with the
corresponding periods in 
Table~3. On a low
power level (not resolved in the figure) several more significant peaks
appear, but they all occur at frequencies equal to simple arithmetic
combinations of the main signals and are therefore not independent.
$P_1$ varies significantly, $P_2$ -- $P_4$ only
slightly between LC\#1 and LC\#2. In fact, in the power spectrum of the
combined data the corresponding peaks split up into two. Therefore, values 
derived from both light curves are listed in the table.

The bottom right frame of Fig.~\ref{bbdor-lc}, finally, shows LC\#1
folded on the four periods, after variations on appropriate longer time
scales have been subtracted in the case of $P_2 - P_4$. The waveforms
derived from LC\#2 are not significantly different. 

\begin{table}
\label{Table: BB Dor frequencies}	
\centering
	\caption{Independent frequencies and their corresponding periods
                 identified in the light curves of BB Dor.}

\begin{tabular}{llll}
\hline

     & LC     & Frequency (d$^{-1}$) & Period (d) \\
\hline

$F_1$ & LC\#1 & 0.0311 (2) & 32.1 (2) \\
      & LC\#2 & 0.0267 (2) & 37.5 (3) \\ [1ex]
$F_2$ & LC\#1 & 0.2225 (2) & \phantom{0}4.495 (3) \\
      & LC\#2 & 0.2231 (1) & \phantom{0}4.483 (3) \\ [1ex] 
$F_3$ & LC\#1 & 0.7549 (2) & 1.3247 (3) \\
      & LC\#2 & 0.7533 (2) & 1.3227 (4) \\ [1ex]
$F_4$ & LC\#1 & 6.71402 (2) & \phantom{0}0.1489422 (4) \\
      & LC\#2 & 6.71454 (3) & \phantom{0}0.1489305 (6) \\
\hline
\end{tabular}
\end{table}

The slight differences in time
between the brightness peaks and the red dots in the upper frame of
Fig.~\ref{bbdor-lc} and the significantly different periods found in
LC\#1 and LC\#2 indicate that the $P_1$ variations are not strictly
periodic. Nevertheless, they can clearly be identified with the 
quasi-periodic brightenings observed by \citet{Rodriguez-Gil12}.
The different periods in the two TESS light curves [embracing the 
period of \citet{Rodriguez-Gil12}], separated by a year, indicates that 
they are not caused by a stable clock in BB~Dor. However, their persistence 
over more than 14~yr tells us that this is not just a transient phenomenon.
\citet{Rodriguez-Gil12} speculate that these variations are due to
mass transfer variation caused by migrating star spots on the secondary star
or stunted outbursts. But it is then not obvious why the brightenings
occur with a reasonable well defined periodicity. The convex shape of 
the light between maxima suggests a gradual build-up and subsequent 
decay of the (so far unknown) process responsible for the modulations.

$P_4$ is very close to the main photometric period seen by \citet{Patterson05}
and which is interpreted by \citet{Rodriguez-Gil12} as due to a 
nSH. One of the fainter power spectrum peaks mentioned above 
(better defined in LC\#2 than in LC\#1) has a frequency of 
6.4918~(5)~d$^{-1}$, which on the one hand is within the error margin of the
spectroscopic orbital period and on the other hand is almost identical to
$F_4 - F_2$. Thus, $P_4$ can indeed be identified with a nSH period,
and $P_2$ is correspondingly the nodal precession period
of a warped accretion disk. 

What about $P_3$? Could it be the apsidal precession period of an excentric
disk? In that case the expected frequency of a pSH would be
$F_3 + F_{\rm orb} = 7.245\ (1)$~d$^{-1}$. The closest marginally significant
peak in the power spectrum of LC\#1 is at
7.2558(6)~d$^{-1}$. The period difference between the orbital and the
superhump periods would then be three times as large for the positive
than for the negative superhump. While not impossible, this is significantly
more than the canonical difference of a factor of two. It may also be 
questioned why the superhump signal is then so much fainter than that due
to the apsidal disk motion. An alternative, but equally unsatisfactory
hypothesis is an interplay between $P_1$ and $P_2$. Is it a coincidence
that in both light curves $3 \left( F_1 + F_2 \right) = F_3$ within the 
formal $1\sigma$ error margin? As long as the origin of $P_1$ remains 
unknown it is difficult even to speculate about a reason for such an
interplay.

\begin{figure}
	\includegraphics[width=\columnwidth]{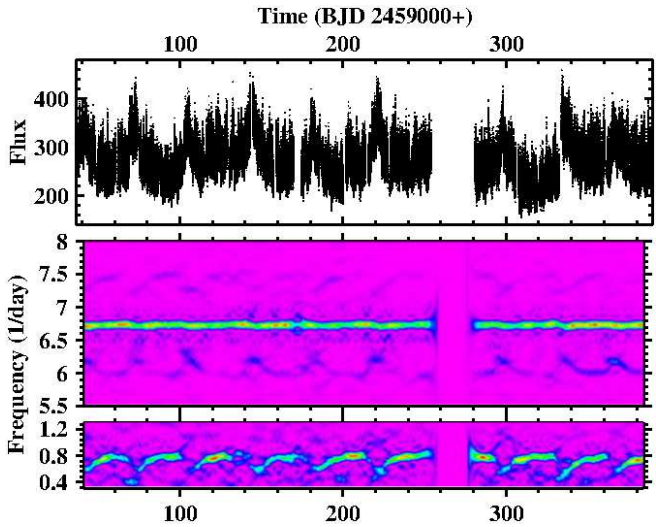}
      \caption[]{{\it Top:} Light curve LC\#2 of BB Dor.
                {\it Middle:} Time resolved power spectrum of the light curve
                in a small range around the nSH frequency.
                {\it Bottom:} The same for a frequency range around $F_3$.}
\label{bbdor-stacked-ps}
\end{figure}

The long light curves permit a closer look at the temporal development
of the periodic signals. The time resolved power spectrum of LC\#2, in the
frequency range of the nSH and of $F_3$, constructed using a sliding window 
with a width of 10~d, is shown in the middle and lower frames, respectively,
of Fig.~\ref{bbdor-stacked-ps}, with the light curve shown in the 
upper frame. The corresponding power spectrum of LC\#1 is very similar.  
The superhump signal itself does not vary significantly with
time. But it is flanked symmetrically on both sides (stronger at frequencies
lower than the SH frequency) by structures modulated with the long period
($P_1$) variations. Their frequency difference with respect to the SH
frequency is equal to the frequency of the $F_3$ signal in the lower frame
of the figure. $F_3$ is approximate constant during ``quiescent'' phases,
subsides at the onset of the brightenings and reappears at a lower
frequency during their maxima.

As a final remark on BB~Dor, I note that the power spectra of the TESS light 
curves contain an excess of power between 30 and 100~d$^{-1}$,
encompassing the range in which \citet{Chen01} observed QPOs.
 
The complex variability of BB~Dra disclosed by the long TESS light
curves certainly deserves a more detailed investigation and interpretation.
But this is beyond the scope of the present paper and must await a specific
study.

\subsection{BH Lyn: Positive Superhumps and QPOs}

BH~Lyn was discovered as PG~0818+513 in the Palomar-Green survey
\citep{Green86}. The system is eclipsing and thus makes it easy to determine
a reliable orbital period which has been derived many times in the past. 
The most precise value of 0.155875577(14)~d was measured by 
\citet{Stanishev06}. They also noted the presence of variations at a
slightly smaller period of 0.1450(65)~d which they interpret as a
nSH. A similar variation at 0.1490(011)~d was also seen by 
\citet{Patterson99}. Additionally, \citet{Stanishev06} observe the presence of
a signal close to 32~d$^{-1}$ in the power spectra of most of their light 
curves which they attribute to QPOs.

\begin{figure}
	\includegraphics[width=\columnwidth]{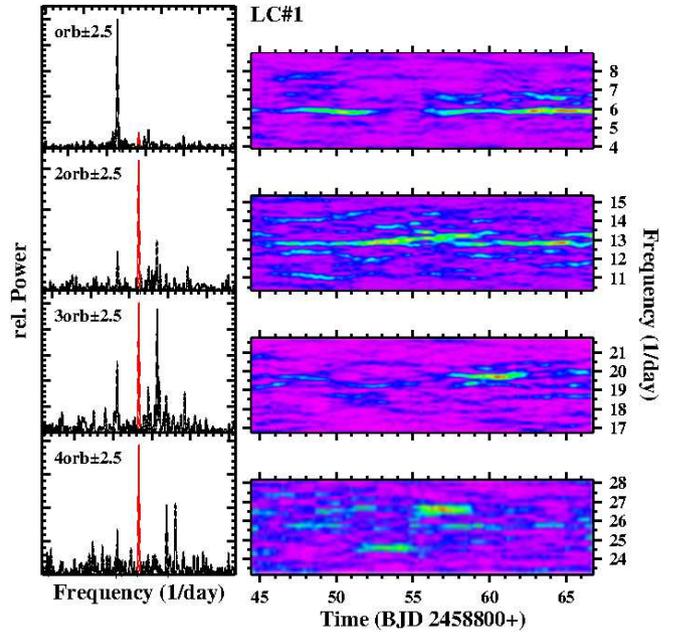}
      \caption[]{{\it Left:} Power spectra of LC\#1 of BH~Lyn in four narrow
                frequency ranges of $\pm$2.5~d$^{-1}$ around the orbital
                frequency and its first, second and third overtone 
                (highlighted  in red).
                {\it Right:} Time resolved power spectra of the same
                frequency range, using a sliding window with a width of
                four days.}
\label{bhlyn-lc2}
\end{figure}

The two available TESS light curves of BH~Lyn exhibit irregular variations on 
time scales of a few days. Apart from the primary eclipse the orbital waveform 
exhibits a clear secondary eclipse. More interesting, however, are the
power spectra (after masking the primary eclipses). On the left side of
Fig.~\ref{bhlyn-lc2} the power spectrum of LC\#1 is reproduced, concentrating
on a frequency range of $\pm 2.5$~d$^{-1}$ around the orbital frequency and its
first, second and third overtone. The peak caused by the orbital variations
is highlighted in red. On the right side of the figure the
time resolved power spectra in the same frequency range are shown, based
on a sliding window with a widths of four days (thus, structures separated
by less than four days are not independent). The general apparence of the
power spectra of LC\#2 is very similar.

While in none of the power spectra a significant peak is detected at a
frequency close to that corresponding to the nSHs seen by
\citet{Patterson99} and \citet{Stanishev06}, the dominant signal in the
upper frames of the figure has a 
frequency just below the orbital frequency. The time resolved power 
spectrum shows that, albeit exhibiting some modulation in its strength, this 
signal is persistent over the whole extend of the light curve. It indicates
thus the presence of a pSH with a period of 
0.17059(5)~d in LC\#1. A similar persistent pSH is also seen
in LC\#2, but at a significantly shorter period of 
0.16484(4)~d. The period excess $\epsilon = \left( P_{\rm SH} - P_{\rm orb}
\right) / P_{\rm orb}$ thus drops from 0.094 in LC\#1 to 0.058 in LC\#2.
While it is known that superhump periods can change over time, such a large
difference of $\epsilon$ at different epochs is unusual.

Apart from the orbital and SH signals the power spectra contain
a multitude of peaks in narrow frequency ranges around the orbital
frequency and its overtones. They appear less clearly also at higher
overtones than shown in Fig.~\ref{bhlyn-lc2} [note that the QPOs claimed
by \citet{Stanishev06} at 32~d$^{-1}$ are very close to the forth overtone
of the orbital frequency]. No simple relation between 
them is apparent, meaning that they are independent from each other. 
The time resolved
power spectra reveal that they can appear for considerable time intervals
with varying strength and frequency changes. These are characteristics of
QPOs. Their concentration around the orbital frequency and its overtones
is noteworthy. While not as extreme, this behaviour is reminiscent of 
similar properties observed in the old nova CP~Pup (see Paper~I) and BZ~Cam
(Sect.~\ref{BZ Cam}).

Finally, representative eclipse epochs of BH~Lyn are listed in 
Table~2 for future reference.
  
\subsection{BK Lyn: Positive superhump, yes, but no negative one}
\label{BK Lyn}

BK~Lyn (=PG 0917+342) is a novalike variable with some curious peculiarities.
At an orbital period of 0.07498(4)~d \citep{Ringwald96} it is one of very few 
(non-magnetic) novalike variables below the CV period gap. In a never before
seen transition, in 2005 BK~Lyn morphed into a ER~UMa star \citep{Patterson13},
i.e., an SU~UMa star with many normal outbursts in quick succession and a very
short supercycle. However, in 2014 the system returned 
to the more stable high brightness state of a NL, as is evident from the
AAVSO long term light curve. Superhumps were first seen in BK~Lyn by
\citet{Howell91} but misinterpreted as orbital variations. \citet{Skillman93}
then correctly identified a 113.1~min modulation with a slightly varying
period as a pSH which was later confirmed by \citet{Misselt95}.
In a more extensive photometric study \citet{Patterson13}, in addition
to pSHs, also detected their negative counterparts in some
observing seasons. Finally, \citet{Yang17} claim the presence of a long
term period of 42.05(1)~d in BK Lyn.

\begin{figure}
	\includegraphics[width=\columnwidth]{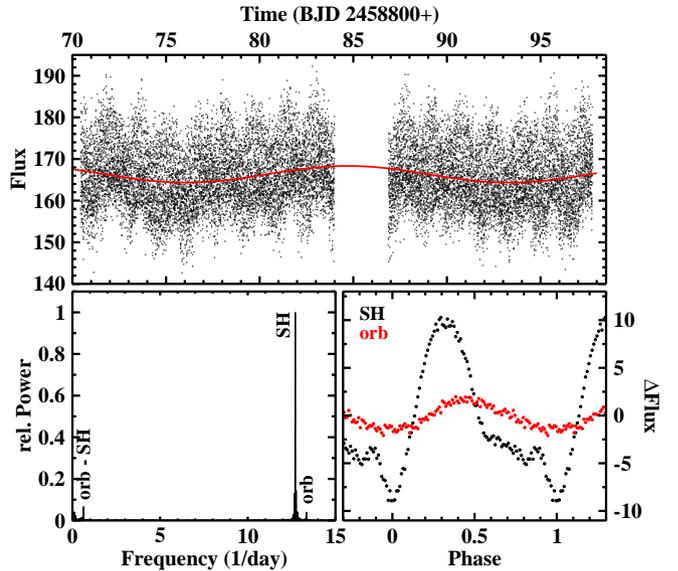}
      \caption[]{{\it Top:} Light curve of BK Lyn. The red graph represents
                a least squares sine fit with a period of 17.29~d.
                {\it Bottom:} Power spectrum of BK~Lyn (left) and 
                waveforms of the superhump and orbital modulations (right).}
\label{bklyn}
\end{figure}

The only TESS light curve available (upper frame of Fig.~\ref{bklyn})
shows a clear modulation on the time scale of about 1.5~d superposed on
a longer period variation. The latter is well fit by a sine wave with
a period of 
17.29(3)~d (the red line in the figure). But since the light curve covers
only just about 1.5 cycles it is by no means clear that this modulation
is, in fact, periodic and persistent. The origin of the more rapid 
variations becomes immediately clear looking at the power spectrum in
the lower left frame of Fig.~\ref{bklyn} which is dominated by a strong
signal at
$F_{\rm SH} = 12.7096(3)$~d$^{-1}$ ($P_{\rm SH} = 0.078681(2)$~d).
It is evidently due to the pSH observed on
previous occasions and now has a slightly longer period. A much
fainter signal is present at a higher frequency of
$F_{\rm orb} = 13.344(2)$~d$^{-1}$ ($P_{\rm orb} = 0.074942(9)$~d).
Within the error quoted by \citet{Ringwald96} this period is identical
to the spectroscopic orbital period but has a higher precision.
Another signal appears at a low frequency of
$F_{\rm b} = 0.633(1)$~d$^{-1}$ ($P_{\rm b} = 1.579(3)$~d).
Within the error margins this is the difference between $F_{\rm orb}$ and
$F_{\rm SH}$ and thus the period of the apsidal motion of an eccentric disk
in the canonical interpretation of positive superhumps. It explains the
shorter time scale modulations seen in the light curve. At higher frequencies
signals at simple arithmetic combinations of $F_{\rm orb}$ and $F_{\rm SH}$
appear, but the data contain no trace of the nSH seen by
\citet{Patterson13}. 

The waveform of the superhump variations (in black, together with the
unconspicuous quasi-sinusoidal orbital waveform in red) is shown in
the lower right frame of Fig.~\ref{bklyn}. Over the years the superhump
shape varies. In the TESS data it comprises only the first have of the
cycle. The second half is characterized by an only slighly declining
level in the phase range 0.55 -- 0.85 before dropping to the minimum 
at phase 0. In \citet{Patterson13} (their figure 3)
the waveform is clearly double humped, while \citet{Skillman93} observed an
almost sinusoidal superhump (their figure 6).  

\subsection{AH Men: Shallow eclipses and a strong negative superhump}

\citet{Buckley93} published the first encompassing photometric study of
AH~Men (1H05551-819). Apart from QPO-like variations in the range of 
600 - 2400~s they detected the presence of a quasi-sinusoidal modulation at
0.1392202(9)~d. Radial velocities measurements 
confirm that this variation ``occurs
at, or very near to, the orbital period''. This, however, is at odds with
later observations of \citet{Patterson95} who noted the continuous presence
of a signal at 0.1229934~(6)d. While in 1995 AH~Men did not exhibit other
coherent variations, between December 1993 and February 1994 Patterson
saw strong variations at 0.127208~d (which he considers orbital), 0.12300~d,
0.062517~d and 3.7~d, as well as oscillations in the range of 17 -- 22~min. 
In a later paper \citet{Patterson98} mentions a superhump period of
0.1385(2)~d. Should this [and the period seen by \citet{Buckley93}] be
due to a pSH, while the 0.12300~d period points at a nSH?

\begin{figure}
	\includegraphics[width=\columnwidth]{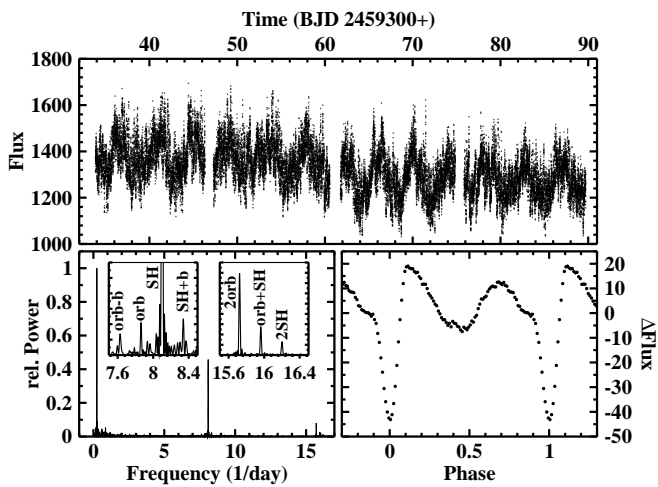}
      \caption[]{{\it Top:} Part of LC\#5 of AH~Men. {\it Bottom left:}
                Power spectrum of the light curve. The inserts contain small
                frequency intervals of this power spectrum on an expanded
                vertical scale. Here, ``b'' stands for the beat frequency
                between the orbital and superhump signals.
                {\it Bottom right:} Orbital waveform of
                AH~Men (average of all light curves).}
\label{ahmen}
\end{figure}

The upper frame of Fig.~\ref{ahmen} shows one of the five available TESS
light curves (LC\#5). It is dominated by regular variations which, as we will 
see, are due to the beat between the orbital and a nSH period.
While this signal shows up in the power spectra of all light curves it is
obvious to the eye only in LC\#3 and later. Apart from this periodicity
the power spectrum of LC\#5 (lower left frame of the figure) just as that of
all other light curves is dominated by a strong peak close to 8.09~d$^{-1}$ and
some much fainter satellite lines, as shown in the lower left frame of the 
figure and on an expanded scale in the left insert. I interpret the strong
$F_{\rm SH} = 8.09$~d$^{-1}$ signal as being due to a negative superhump and 
a smaller
peak at $F_{\rm orb} = 7.86$~d$^{-1}$ as orbital. Other signals are related to
the beat between them. At higher frequencies another group of signals 
appears (right insert), the strongest occurring at 
$2F_{\rm orb} = 15.72361(8)$. 
At still higher frequencies additional peaks are seen which can all be 
interpreted as linear combinations of $F_{\rm orb}$ and $F_{\rm SH}$.

Why do I consider $F_{\rm orb}$ to be orbital in nature? This becomes obvious 
when folding the light curves on the corresponding period. The average of all
folded light curves is shown in the lower right frame of Fig.~\ref{ahmen}. 
It clearly reveals a double-humped structure and a rather shallow eclipse.
This waveform explains why the first overtone of $F_{\rm orb}$ 
in the power spectra is so
much stronger than the fundamental frequency. 
On the other hand, folding
the light curves on $1/F_{\rm SH}$ yield a nearly sinusoidal waveform, albeit
with a much larger amplitude. Based on all light curves (eclipse timings are
listed in Table~2), the following
ephemeris for the eclipse minimum is derived:
\begin{displaymath}
T_{\rm min} = BJD 2458340.0869 (2) + 0.12719550 (5) \times E
\end{displaymath}

Although rather stable, the period of the superhump varies slightly.
The corresponding values measured in the individual light curves are:
0.12364(3) (LC\#1),
0.12383(3) (LC\#2),
0.123485(5) (LC\#3),
0.12337(2) (LC\#4) and
0.123446(6) (LC\#5).

\subsection{RR Pic: Only transient superhumps}

RR~Pic is a well studied old nova. The orbital period reveals itself in
the form of a
clear and persistent hump first seen by \citet{vanHouten66} and ever since.
The most precise value of 0.145025959~d was derived by 
\citet{Fuentes-Morales18}. They also saw pSHs with a
period of 0.1577 in 2007. \citet{Schmidtobreick08} report another superhump 
instance in 2005 with the same period and which went along with a signal at 
the beat between the SH and orbital periods. 

The numerous sectorial data sets of RR~Pic observed by TESS can be combined 
into 5 contiguous light curves, two of which encompass about 8 months. The
orbital signal is outstanding in the respective power spectra, but no
trace of a superhump is visible, even after carefully subtracting the
average orbital waveform from the data (Fig.~\ref{negatives}c). 
It is, however, remarkable that
the waveform is extremely stable [and quite similar to the one shown in
fig.~3 of \citet{Schmidtobreick08}] over the almost 3 years spanned by the
data. Even small details are faithfully repeated in the waveforms derived
from the individual light curves.

\citet{Schmidtobreick08} and \citet{Fuentes-Morales18} analyzed light curves
of RR~Pic from 11 observing seasons and saw superhumps only twice. Adding
to this the TESS data without SHs (over a total time
base of $\approx$3~yr) makes it clear that
superhumps in this system are only rare and transient events.

\subsection{AO Psc: No confirmation of superhumps}
\label{AO Psc}

The optical light curve of the well known intermediate polar AO~Psc is
dominated by the orbital period at 0.1495~d and the orbital side band of
the white dwarf spin period at 14.31~min \citep{Patterson81, Motch81}.
SHs in AO~Psc with a period of 0.149627~d are only mentioned briefly 
by \citet{Patterson01a} who referred details to a publication in 
preparation which, however, never appeared.

The power spectrum of the single TESS light curves contains strong signals at
the orbital frequency $F_{\rm orb}$, the orbital sideband of the white dwarf
spin frequency $F_{\rm spin} - F_{\rm orb}$, and weaker signals at $2F_{\rm orb}$,
$F_{\rm spin}$, $F_{\rm spin} - 2F_{\rm orb}$, $2F_{\rm spin}$ and 
$2F_{\rm spin} - 2F_{\rm orb}$. But there are no indications of superhumps
(Fig.~\ref{negatives}d).

\subsection{AY Psc: No superhumps, but an increasing orbital period}
\label{AY Psc}

In contrast to the other objects in this study, AY~Psc is a dwarf nova. 
It belongs to the Z~Cam stars \citep{Mercado02}, i.e., those dwarf novae 
which occasionally remain in a standstill. The TESS 
light curve, observed in subsequent time intervals in two sectors and
covering 51 days, is rather stable and does not contain the usual 
alternations between outbursts and quiescent phases with a quasi-period 
of 18.3~d \citep{Han17}. It may therefore be concluded that the system
was in a standstill during the entire observing period. During such 
phases the accretion disk is expected to be be in a hot state similar 
to those of NLs and old novae. Therefore, I include AY~Psc in this study.

AY~Psc is an eclipsing system as first noticed by \citet{Szkody89}.
Orbital ephemeris were provided by \citet{Diaz90} and later
refined by \citet{Guelsecen09} and \citet{Han17}. \citet{Guelsecen09}
reported the presence of nSHs with a slightly changing
period between 0.2057 and 0.2073~d during three observing missions in
2003, 2004 and 2005. The total amplitude of the light variations of 
AY~Psc in each of these missions was about 2.25~mag which is similar to
its average outburst amplitude \citep{Han17}. Thus, the system was
observed during a period of normal dwarf nova activity. \citet{Guelsecen09}
did not investigate a dependence of the superhump properties on the phase
of the outburst cycle.  

The power spectrum of the TESS light curve (after masking the eclipses)
does not confirm the presence of SHs Fig~\ref{negatives}e). 
The only significant signals
are at the orbital frequency and its overtones. Thus, if nSHs
occur during the normal activity cycle of AY~Psc, they subsided at least
during the prolonged standstill covered by TESS.

The many eclipses in the TESS light curve permit, together with data taken 
from the literature, an improvement of the orbital period of AY~Psc. 
Representative eclipse timings for the light curves observed in the two 
sectors covered by TESS are listed in 
Table~2. These,
together with the eclipse timings listed by \citet{Diaz90} and \citet{Han17},
including also the eclipse epoch taken from the ephemeris of 
\citet{Guelsecen09} (unfortunately, they do not list individual eclipse
timings), assigning a weight of 1 to the literature timings and 10 to the
representative TESS eclipse epochs, yield the linear ephemeris:
\begin{equation}
\label{AY Psc: linear ephemeris}
T_{\rm min} = BJD\, 2447623.3460(2) + 0.217320654(4) \times E
\end{equation}
Here, the errors are the formal fits error of the linear fit to the
cycle number -- eclipse epoch relation.

\begin{figure}
	\includegraphics[width=\columnwidth]{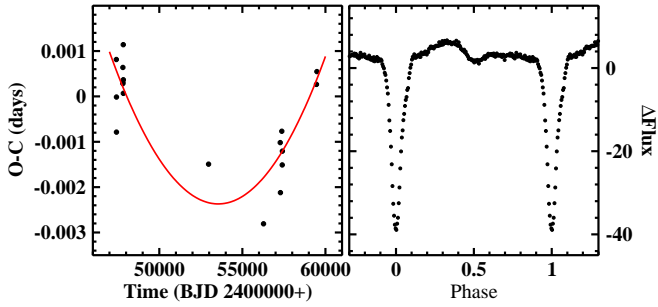}
      \caption[]{{\it Left:} $O-C$ diagram of eclipse timings of AY~Psc with
                respect to linear ephemeris
                (Eq.~\ref{AY Psc: linear ephemeris}). The red graph is a 
                least squares fit of a 2$^{\rm nd}$ order polynomial to the 
                data.
                {\it Right:} Orbital waveform of AY~Psc during standstill.} 
\label{aypsc}
\end{figure}

The $O-C$ diagram with respect to the linear ephemeris is shown in the left
frame of Fig.~\ref{aypsc} where the red graph is a fit of a 2$^{\rm nd}$
order polynomial to the data. It suggests that quadratic ephemeris provide
a better description of the eclipse timings:
\begin{eqnarray}
T_{\rm min} & = & BJD\, 2447623.34640(6) \\ \nonumber
          &   & + 0.217320452(1) \times E \\ \nonumber
          &   & + 3.69(2)\, 10^{-12} \times E^2
\end{eqnarray}
The period increases thus currently at a rate of 
$\dot{P} = 3.40(3) \times 10^{-11}$.
The relative period increase is
$\dot{P}/P = 5.71(4) \times 10^{-8}$~yr$^{-1}$. 

The light curve, folded on the orbital period, yields the waveform shown
in the right frame of Fig.~\ref{aypsc}. Apart from the primay eclipse a
secondary eclipse occurs at phase 0.5. It is preceded by a hump. No hump 
is apparent at the phases before the primary eclipse, in contrast to what
is often seen in other
eclipsing CVs and is attributed to enhanced emission from a bright spot.
The waveform is thus different from that observed by \citet{Guelsecen09}
in white light (see their fig.~3), but this may at least in part be due to
the different passband of TESS.

\subsection{V348 Pup: Superhumps confirmed}

At an orbital period of 0.101838931(14)~d \citep{Rolfe00}, V348~Pup is an
eclipsing novalike variable right at the centre of the CV period gap.
Photometric variations with a period slightly different from the orbital 
one made \citet{Tuohy90} suspect the star to be what would nowadays be
considered an asynchronous polar; a notion which could 
neither be confirmed no rejected in pointed X-ray observations by 
\citet{Rosen94}, while \citet{Froning03} found no evidence for a magnetic 
nature.  Instead, \citet{Rolfe00} reported SHs  with a slightly
variable period in V348~Pup in 1991, 1993 and 1995 which, however, were 
not seen when \citet{Saito16} observed the star at a later epoch. 

\begin{figure}
	\includegraphics[width=\columnwidth]{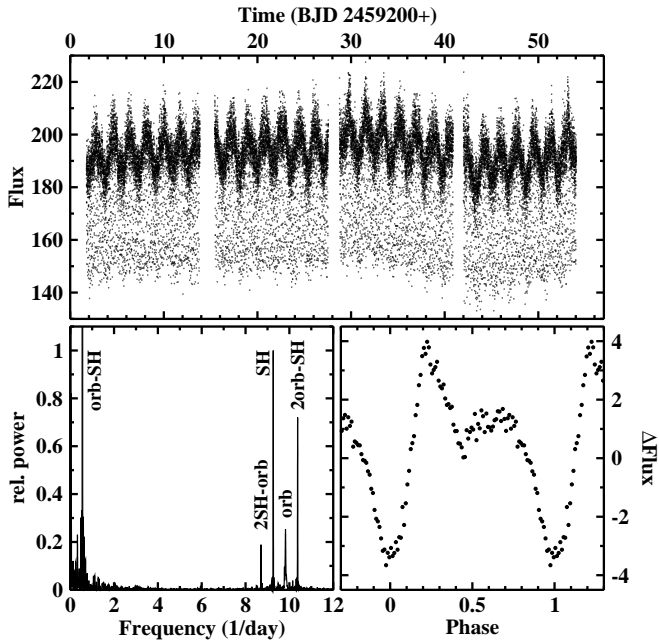}
      \caption[]{{\it Top:} Light curve of V348~Pup.
                {\it Bottom:} Power spectrum (left) and superhump
                waveform (right).}
\label{v348pup-lc}
\end{figure}

The light curve of V348~Pup, combining data from two TESS sectors, is
shown in the upper frame of Fig.~\ref{v348pup-lc}. The strong out-of-eclipse
variations have a period of
1.797(1)~d
and immediately suggest to be due to the beat between the orbital and a
SH period and thus the precession period of an accretion disk.
This is confirmed by the power spectrum (lower left frame of the figure)
which, apart from the dominating signal at the beat frequency, contains a
peak at the orbital frequency $F_{\rm orb}$ and a strong signal at 
$F_{\rm SH} = 9.2626(8)\, {\rm d}^{-1}$ ($P_{\rm SH} = 0.107961(9)$~d).
The period of the latter lies within the range of SH periods observed by
\citet{Rolfe00}, leaving no doubt that it is due to a positive superhump.
Other signals and a multitude of peaks at frequencies beyond the range
shown in the figure can all be expressed as simple arithmetic combinations
of $F_{\rm orb}$ and $F_{\rm SH}$. The waveform of the superhump is significantly
structured as seen in the lower right frame of Fig.~\ref{v348pup-lc}.

Investigating all available eclipse timings available at the time,
\citet{Dai10} concluded that the orbital period of V348~Pup is currently
increasing. Average eclipse epochs derived from the two TESS sector data
(Table~2) and 9 additional epochs measured in
archival light curves retrieved from the LNA Data Bank 
(Table~A2)
permit to extend the total time base. I confirm the period increase and
derive quadratic ephemeris for V348~Pup which are very similar to those
quoted by \citet{Dai10}:
\begin{eqnarray}
T_{\rm min} & = & BJD\, 2448591.668(1) \\ \nonumber
          &   & + 0.1018389(2) \times E \\ \nonumber
          &   & + 2.6(1.7)\, 10^{-13} \times E^2
\end{eqnarray}

\subsection{RW Tri: no superhumps in TESS data}

RW~Tri is a deeply eclipsing novalike variable of the UX~UMa type. 
Low amplitude ($\approx 0.5$~mag) oscillations on time scales of some 
tens of days have been observed by \citet{Honeycutt94}, \citet{Honeycutt01}, 
and even more clearly by \citet{Bruch20}. Apart from this the system is
relatively stable as is corroborated by the long term AAVSO light curve.
\citet{Smak19} reports nSHs with a period of 0.2203~d in his 
observations obtained in 1984 and possibly in 1957. 
They were, however, not present in 2015 -- 2016 
\citep{Bruch20}. The more extensive continuous TESS data permit an additional
verfication of Smak's claim of superhumps in RW~Tri.

\begin{figure}
	\includegraphics[width=\columnwidth]{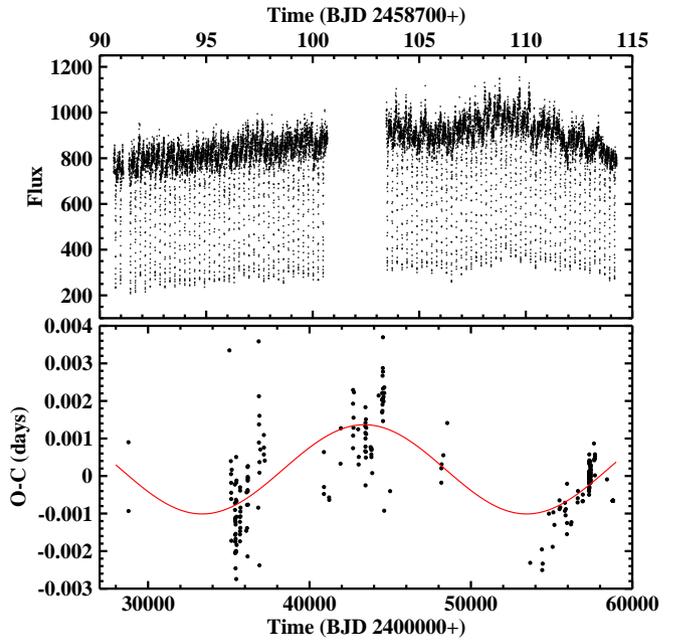}
      \caption[]{{\it Top:} TESS light curve of RW~Tri.
                {\it Bottom:} $O-C$ diagram of eclipse timings of RW~Tri with
                respect to linear ephemeris 
                (Eq.~\ref{RW Tri: linear ephemeris}). The red graph is a 
                least squares sine fit.}
\label{rwtri}
\end{figure}

A single TESS light curve of the system is available (upper frame of
Fig.~\ref{rwtri}). Over its 
24~d time base a gradual rise and subsequent decline in brightness occurs
which is roughly compatible with the oscillations mentioned above. The
flux level at the bottom of the eclipses follows these variations. This 
means that the responsible light source is not
eclipsed. 

The power spectrum does not contain significant signals at frequencies other
than the orbital frequency and its overtones (Fig.~\ref{negatives}f). 
Thus, at least during the epoch
of the TESS observations no superhumps were excited in RW~Tri.

As in the case of UU~Aqr the last update of the orbital period of RW~Tri
occurred decades ago \citep{Robinson91}. I retrieved numerous light 
curves from the AAVSO archives observed between 2005 and 2018 and measured
95 additional eclipse epochs in the same way as was done for UU~Aqr and
V348~Pup. The results are listed in
Table~A3.
They were used together with the representative eclipse epoch derived from
the TESS light curve 
(see Table~2), which got 
10 times the weight of the individual eclipses, and those
listed by \citet{Mandel65}, \citet{Africano78} and \citet{Robinson91}
to recalculate orbital ephemeris for RW~Tri:
\begin{equation}
\label{RW Tri: linear ephemeris}
T_{\rm min} = BJD\, 2453672.6246(4) + 0.231883245(6) \times E
\end{equation}
Not surprisingly, the orbital period is very close to the value quoted by
\citet{Robinson91}. As in many other CVs the $O-C$ curve, now covering over
80~yr (albeit only sparsely covered during the first $\sim$15~yr), exhibits
systematic variation on the time scale of years, indicating small fluctuations
of the orbital period which cannot be attributed to the secular evolution of
the system. These variations are, however, much more gradual in RW~Tri than
the very rapid period changes in UX~UMa (see Sect.~\ref{UX UMa}).
With some good will one might even suspect a cyclic variation with a period
of 55~yr (red curve in the figure), but covering only just about one cycle 
I am reluctant to claim it to be persistent.

\subsection{UX UMa: Superhumps remain to be isolated events}
\label{UX UMa}

UX~UMa is the prototype of novalike variables, in particular of those system 
which, in contrast to VY~Scl stars, have never been observed to go into a low 
state. As the prototype of its class and the
brightest eclipsing novalike variable, UX~UMa has been extensively studied 
in the past \citep[see][for a summary of previous observations]
{Neustroev11}. The orbital period was last refined by
\citet{Baptista95}. In extensive photometric observations during the 2015
observing season \citet{deMiguel16} found a modulation with a period of 
3.680~d in the light curve of UX~UMa which they
interpret as being due to a retrograde precession of the accretion disk.
An associated nSH at the beat period of the precession and
the orbit is also seen. \citet{Bruch20} confirmed this behaviour but also
noted that it was restricted to that particular season and did not repeat
itself in previous or following years.  

\begin{figure}
	\includegraphics[width=\columnwidth]{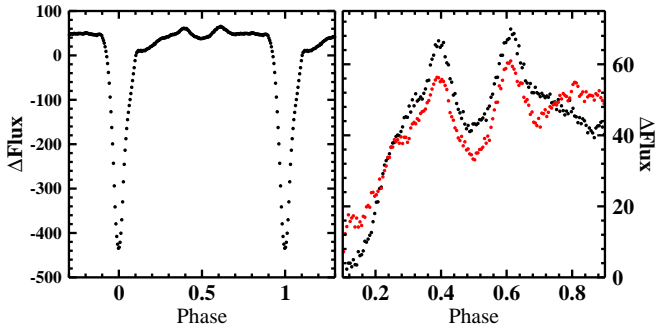}
      \caption[]{{\it Left:} Orbital waveform of UX~UMa (average of LC\#1
                and LC\#2).
                {\it Right:}  The same on an expanded vertical scale, 
                restriced to out-of-eclipse phases and shown separately 
                for LC\#1 (black) and LC\#2 (red).}
\label{uxuma-waveform}
\end{figure}

Is there any trace of the unusual behaviour observed in 2015 to be found
in the TESS observations taken in 2019, Aug-Oct and 2020, Feb-Apr? No,
there is not. The power spectra of the two TESS light curves do not contain
any significant signal other than the orbital one and its overtones
(Fig.~\ref{negatives}g). The orbital waveform (left frame of 
Fig.~\ref{uxuma-waveform}) is characterized -- apart
from the primary eclipse -- by a single hump, interrupted by the secondary
eclipse which gives the hump the structure of two separate ``horns''. The
primary eclipse egress exhibits a clear change of gradient, i.e., the typical
sign of a retarded egress of a hot spot. The flux level just after eclipse
egress is considerably lower than just before ingress. This waveform is 
different from that normally observed at shorter wavelengths 
\citep[see fig.~20 of][]{Bruch20}. A comparison of the
waveforms out of eclipse resulting from LC\#1 (black) and LC\#2 (red) is 
shown in the right frame of the figure. Only slight variations occur in the 
6 months between the light curves.  

Just as for UU~Aqr and RW~Tri, the orbital period of UX~UMa was last 
refined almost 30~yr ago \citep{Baptista95}. Again, I took advantage
of light curves observed between 1999 and 2022 found in the AAVSO archives, 
as well as of some unpublished light curves taken between 1977 and
1992, provided by R.E.~Nather and E.L..~Robinson (private communication).
They yielded no less than 291 useful additional eclipse
epochs 
(Table~A4).
Representative eclipse epochs derived from the TESS light curves are listed
in Table~2.

Combining the new eclipse timings with those listed by \citet{Nather74},
\citet{Africano76}, \citet{Kukarkin77}, \citet{Quigley78}, 
\citet{Rubenstein91}, \citet{Rutten92} and \citet{Baptista95}
(as usual assigning weight 10 to the TESS eclipse epochs and 1 to all others) 
yielded the following revised orbital ephemeris for UX~UMa:
\begin{equation}
\label{UX UMa: linear ephemeris}
T_{\rm min} = BJD\, 2451319.779(3) + 0.19667127(7) \times E
\end{equation}

\begin{figure}
	\includegraphics[width=\columnwidth]{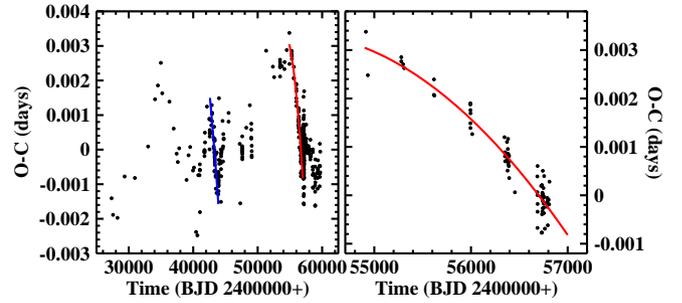}
      \caption[]{{\it Left:} $O-C$ diagram of the eclipses in UX~UMa.
                The blue and red lines mark intervals of particularly
                strong gradients.
                {\it Right:} Detail of the $O-C$ diagram corresponding to
                the interval marked by the red line in the left frame,
                together the least squares parabolic fit to the data.}
\label{uxuma-o-c}
\end{figure}

It turns out that within the $1 \sigma$ error the period is identical to the
one derived by \citet{Baptista95}\footnote{Using only the data available to
\citet{Baptista95} I could reproduce, of course, their period precisely, but 
the error is 10 times larger.}. Indeed, in spite of the much longer time base 
the formal error increased. This is explained by the $O-C$ diagram, reproduced 
in the left frame of Fig.~\ref{uxuma-o-c}, which reveals an increased
non-random scatter of the data points in recent years.

In the early years, a 29~yr cyclic variation of the orbital period of UX~UMa 
was first suspected by \citet{Mandel65} and further discussed by 
\citet{Nather74} and \citet{Africano76}, but then called into question 
by \citet{Kukarkin77} and \citet{Quigley78}. Regarding the $O-C$ diagram in
fig.~5 of \citet{Baptista95} this hypothesis can clearly be rejected. This is
impressively confirmed by the present results (Fig.~\ref{uxuma-o-c}) which
extends the time base by about a factor of 2. The $O-C$ diagram indicates
non-periodic but systematic variations of the the orbital period on widely
varying time scales. The most rapid $O-C$ (and consequently period) changes 
are highlighted by blue and red lines in the figure which have very 
nearly the same gradient. The right frame of
Fig.~\ref{uxuma-o-c} contains an enlarged version of one of these $O-C$ 
diagram sections. The red line represents a least squares 2$^{\rm nd}$ 
order polynomial fit to the data which yields a period change of 
$\dot{P} = -1.97 \times 10^{-10}$.
This corresponds to a time scale for the period decrease as short as 
$P/\dot{P} = 2.73 \times 10^6$~yr.

On secular time scales
the periods of cataclysmic variable are expected to decrease
due to angular momentum loss of the system via magnetic breaking and
gravitational radiation. Such variations are monotonic and
occur on vastly longer time scales than observed here. Period variations
on time scales of years and with changing sign are not uncommon in CVs. 
If they are cyclic they are often explained by the presence of a third
body in the system (a hypothesis more often than not disproved by additional
observations). Alternatively, the Applegate mechanism \citep{Applegate92}
is frequently invoked with mixed success. However, it appears fair to 
say that so far no generally excepted idea to explain the often erratic
period variations has been put forward.

\subsection{DW UMa: A positive, yes, but no negative superhump}

The eclipsing system DW~UMa has been subjected to many photometric studies
which revealed the presence of positive and negative superhumps. The most
extensive investigation was performed by \citet{Boyd17} who also cite
references to other relevant papers. The orbital period is 0.1366065324(7)~d.
SHs were observed with slightly varying periods around 0.133~d (nSH)
and 0.145~d (pSH). The beat period between the orbit and the superhumps
is also clearly seen.

\begin{figure}
	\includegraphics[width=\columnwidth]{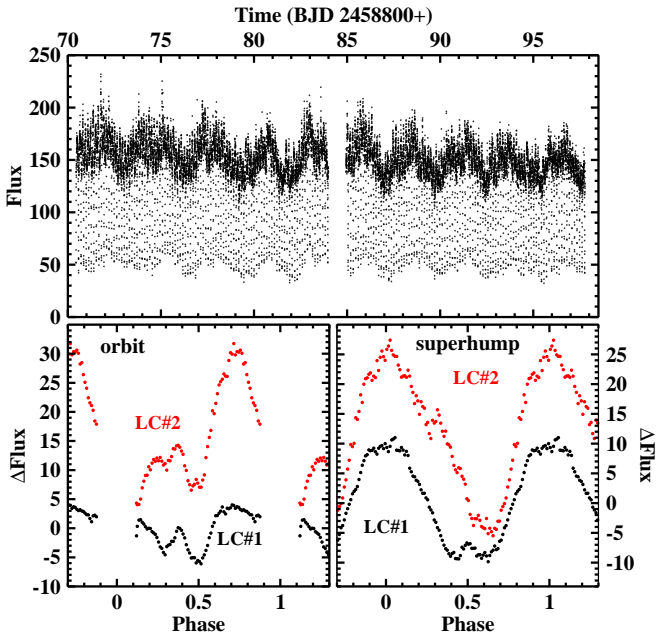}
      \caption[]{{\it Top:} Light curve LC\#1 of DW~UMa.
                {\it Bottom:} Orbital (left) and superhump
                waveforms (right) derived from light curves LC\#1 (black)
                and LC\#2 (red, shifted vertically for clarity).}
\label{dwuma}
\end{figure}

The latter feature is impressively confirmed in the two TESS light curves,
the first of which is reproduced in the upper frame of Fig.~\ref{dwuma}.
The power spectra (after masking the eclipses) basically confirm the earlier 
results with the noticeable exception that no trace of a nSH is present.
 Without assessing their formal significance I identified no less than
40 (LC\#1) and 35 (LC\#2) peaks up to the Nyquist frequency which appear to 
stand out above the surrounding ``continuum''. All except 3 (LC\#1) and 5
(LC\#2) can be explained as $nF_{\rm orb} + mF_{\rm b}$ where 
$F_{\rm b} = F_{\rm orb} - F_{\rm SH}$ and $n$ and $m$ are integer values in the
range $1 \le n \le 46$ and $-4 \le m \le 2$. The pSHs have
significantly different periods of
0.14387(2)~d (LC\#1) and
0.14479(3)~d (LC\#2)
in the two light curves separated by $\approx$2~yr. The orbital (eclipses
masked) and superhump waveforms are shown in the lower frames of 
Fig.~\ref{dwuma}. Large (orbital) and moderate (superhump) differences
between LC\#1 and LC\#2 are evident. For future reference representative 
eclipse epochs for
the time intervals covered by the light curves are listed in 
Table~2.

\subsection{HS 1813+6122: Transient superhumps, transient outbursts,
and partial eclipses}

Almost all our limited knowledge about the configuration and physics
of HS~1813+6122 (HS~1813 hereafter) comes from a single publication:
\citet{Rodriguez-Gil07}. They performed photometric and spectroscopic
observations on various occasions between 2000 and 2004 and derived an
orbital period of 3.55~h. Their photometry also contained a modulation at
3.39~h which they interpreted as a nSH. In contrast to
what is seen in the present TESS data (see below) \citet{Rodriguez-Gil07}
apparently did not observe outburst of HS~1813, nor could they identify
eclipses. Based on the spectroscopic evidence they classify the system as
a SW~Sex type star.

TESS observed HS~1813 in many sectors, permitting to construct two almost
half year long light curves (LC\#1 and LC\#2), separated by just one
month, and two shorter light curves at later epochs 
(see Table~\ref{Table: obs-log}).
The flux level differs strongly and systematically from one sector to the 
next. This cannot be real and must be attributed to difficulties to define
the zero point of flux for the TESS light curves. The discontinuous flux
levels thus required to add or subtract constants when stitching together
data from different sectors to form a continuous light curve. This 
introduces considerable uncertainties in the general trend of the resulting
curves, but should not affect relative variations within a given sector
or high frequency variations.

\begin{figure}
	\includegraphics[width=\columnwidth]{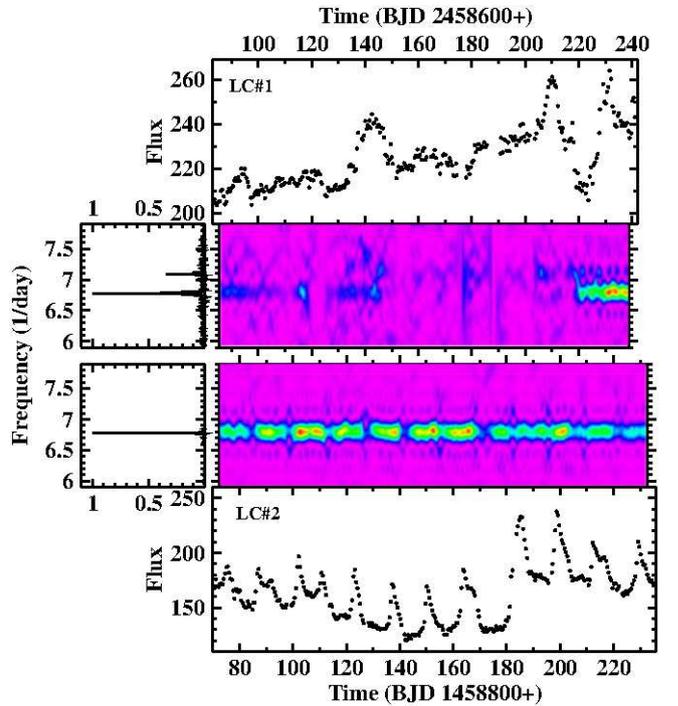}
      \caption[]{Light curves LC\#1 and LC\#2 of HS~1813 and their power 
                 spectra. The right
                 upper and lower frames show light curves (binned in
                 intervals of 0.5~d) of a very
                 different general aspect, but close in time. The left
                 frames contain the respective power spectra in a small
                 interval around the orbital (dominating signal)
                 and superhump frequencies (only present in LC\#1).
                 The remaining frames show the time resolved power spectra
                 of the same frequency range and aligned in time with the
                 light curves.}
\label{hs1813-lc}
\end{figure}

The upper and lower right hand frames of Fig.~\ref{hs1813-lc} show LC\#1 
and LC\#2, respectively, binned in intervals of 0.5~d. They contain quite
unusual and surprising features. The single sector light curve LC\#3 only
contains low level variations, while the longer LC\#4 is not unlike LC\#1.
Disregarding the long term trends which may well be artificial (see above), 
the light curve is characterized by some brightenings above an otherwise 
quiescent background (LC\#1), and then rapidly morphs into a decidedly dwarf 
nova like light curve (LC\#2). Due to the uncertainty of the 
flux zero point it is difficult to determine the amplitude of the outbursts.
Assuming that the flux scale of LC\#2 is at least approximately correct, the 
amplitude reached up to $\approx$0.5~mag and thus remained considerably below 
normal dwarf nova
outburst amplitudes. The rapid change in behaviour between LC\#1 and LC\#2
which is also manifest in other than the outburst characteristics (see below)
is rather unique. I am not aware of another CV which has been observed to
behave similarly with the possible exception of the transient ER~UMa-type
behaviour of BK~Lyn (Sect.~\ref{BK Lyn}). This certainly deserves a deeper 
investigation, but is not the topic of this study.

The power spectra of all light curves are dominated by signals at the
orbital frequency and overtones. Their strengths, however, vary considerably
over time as is evident in the time resolved power spectra in the two
middle right hand frames of Fig.~\ref{hs1813-lc} (the frames on the
left side contain the conventional power spectra of the entire light curves
in the same frequency range). In LC\#1 the orbital signal is present during
the first weeks, gets fainter thereafter, and then suddenly increases 
strongly in power at the end of the light curve. Similar fluctuations in
strength are also present in LC\#3 and LC\#4 (not shown). In constrast, in 
LC\#2, i.e., during the episode of dwarf nova-like outbursts, it remains on 
a constant high level.

The superhump reported by \citet{Rodriguez-Gil07} is only present in the
latter part of LC\#1 and at the same frequency of
$F_{\rm SH} = 7.0890(5)\, {\rm d}^{-1}$ ($P_{\rm SH} = 0.141063(9)$~d) observed
by them. It cannot be detected in any of the other light curves and must 
therefore be considered transient.

Folding the TESS light curves on the orbital period derived from the
peak frequency in the power spectra reveals the presence of shallow
partial eclipses in HS~1813. Average eclipse epochs for the individual
TESS sectors are listed in 
Table~2. They are
used to determined the orbital ephemeris:
\begin{equation}
\label{HS 1813: linear ephemeris}
T_{\rm min} = BJD\, 2458693.087\, (2) + 0.1475202\, (6) \times E
\end{equation}
The period error adds up to a phase uncertainty of $\sim$0.15 at the
epoch of the spectroscopic observations of \citet{Rodriguez-Gil07}. 
This makes it unfortunately impossible to obtain a reliable measure of 
a difference between the spectroscopic phase and the conjunction between 
the stellar components of HS~1813 as expected in SW~Sex stars, and thus to 
verify this classification.

\begin{figure}
	\includegraphics[width=\columnwidth]{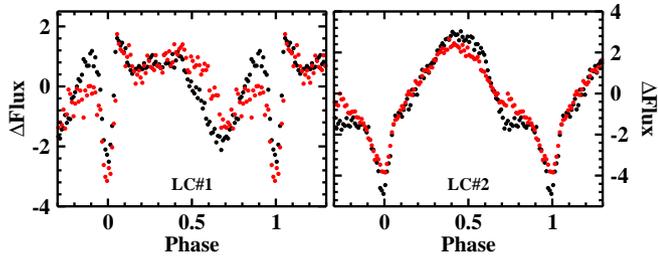}
      \caption[]{Orbital waveform of HS~1813 during epochs of low (LC1\#1, left)
                and high (LC\#2, right) outburst activity. Black symbols refer
                to outburst epochs, red ones to intervals between outbusts.}
\label{hs1813-wf}
\end{figure}

The orbital waveforms, as derived from LC\#1 and LC\#2 are shown in
Fig.~\ref{hs1813-wf}, separately for bright (black) and quiescent (red)
intervals. While differences between bright and quiescent intervals
remain small, significant changes between LC\#1 to LC\#2
are evident. In both light curves the waveform is dominated by a
broad hump, cut in by a shallow V-shaped (and thus partial) eclipse.
In LC\#1 the hump is somewhat more structured (possibly doubles) than in LC\#2. 
Moreover, it shifts by 0.3 units to later phases and the minimum 
broadens in LC\#2. This indicates an obvious change in the structure of the
dominating light sources in HS~1813. It is close at hand to speculate that
this change is somehow related to the onset of the dwarf nova-type 
brightenings. It
is noteworthy, however, that in dwarf novae the waveform normally changes
significantly between quiescence and outbursts while this is not the case
here.

\subsection{RX J2133.7+5107: Superhumps and an unidentified 325~s 
modulation}

This system was detected in the ROSAT Galactic Plane Survey 
\citep{Motch98} and identified by \citet{Bonnet-Bidaud06} as long period 
\citep[$P_{\rm orb} = 0.297431(5)$~d;][]{Thorstensen10}
intermediate polar with a white dwarf spin period of 570.82~s. They also
saw the orbital side band of the spin period in their power spectra. In a 
multi-year campaign \citet{deMiguel17} encountered a nSH
with a slightly varying period in all observing seasons between 2010 and 2016.

\begin{figure}
	\includegraphics[width=\columnwidth]{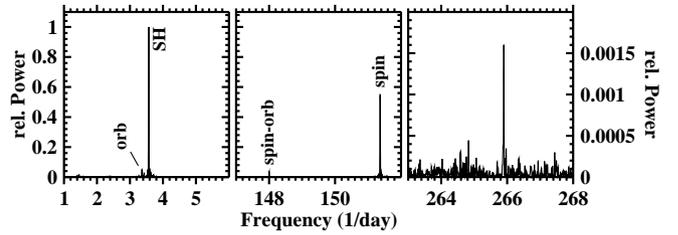}
      \caption[]{Power spectra of the RX~J2133.7+5107 light curve including
                 the superhump and orbital signals (left), the white dwarf
                 spin signal and its orbital sideband (middle) and an 
                 unidentified high frequency signal (right). The
                 vertical scale is the same in the left and middle frames,
                 but greatly expanded in the right frame.}
\label{j2133}
\end{figure}

The TESS light curve confirms these findings. The strongest signal in the
power spectrum (left hand frame of Fig.~\ref{j2133}) corresponds to a 
period of 
0.28916(3)~d 
which is very close to the average nSH period observed by \citet{deMiguel17}.
Just as their observations, the TESS data do not contain a signal at the
beat between orbital and superhump period. But in contrast to the former
the power spectrum of the latter (weakly) reveals orbital variations.
The continued presence of the superhump in all appropriate observations since
2010 allows classifying it as permanent.

Not surprisingly, the power spectrum of the TESS light curves also contains
a strong signal at the white dwarf spin period (measured to be
570.8914(5)~s; middle panel of Fig.~\ref{j2133}), the orbital side band
and the first overtone (not shown). \citet{deMiguel17} also mention signals
at $F_{\rm sp} - F_{\rm orb} + F_{\rm SH}$ and $F_{\rm sp} - F_{\rm orb} - F_{\rm SH}$.
On close inspection, both of these are also weakly present in the TESS light 
curve. \citet{deMiguel17} noted that the
spin period is decreasing over the years. The TESS data permit to add an
additional point which excellently confirms the linear trend. A least
squared fit to all available data yields 
$d{\rm P}/d{\rm t} = -3.8 \pm 0.2$~msec/yr.

While the above results just confirm previous knowledge, the TESS
light curves reveals an additional modulation at a high frequency of
$F = 265.898(2)$~d$^{-1}$ ($P = 324.937(3)$~s;
right hand frame of Fig.~\ref{j2133}). It does not have any obvious relation
to the other periodicities in the system. Time resolved power spectra
reveal that its is persistent -- with some variations in its strength --
throughout the entire light curve. 

The nature of this modulation remains thus unclear. Its coherence over
at least two months makes the accretion disk unlikely as the place of
origin. I am also not aware of any mechanism inherent in the secondary
star which may lead to such short period variations. It may be 
permitted to speculate about white dwarf pulsations as their origin. The
period is within the range observed in well established white dwarf
pulsators in CVs such as GW~Lib 
\citep[][and others]{vanZyl04, Chote16}
and V455~And \citep{Araujo-Betancor05, Szkody13, Bruch20}. But then,
in these stars more than one pulsation mode is excited leading to
multiple power spectrum signals, differently from what is observed
in RX~J2133.7+5107. Moreover, they are short period CVs where the
accretion disk is expected to be weak so as not to outshine the white
dwarf. At the long period of RX~J2133.7+5107 the (precessing) accretion
disk and the accretion regions close to the magnetic poles
of the WD in this intermediate polar will dominate the optical emission
and are thus likely to mask any contribution from the white dwarf itself.  

\subsection{KIC 8751494: Strongly contaminated TESS light curves}

KIC~8751494 was detected as a novalike 
variable, possibly of the SW~Sex subtype, in Kepler data by
\citet{Williams10}. They detected variations with a period of 0.1223(7)~d.
Later \citet{Kato13}, also using Kepler data, found this period to be 
slightly variable and interpreted it as a pSH, the orbital
modulations -- much fainter than the superhump -- having a period of
0.114379(1)~d.

\begin{figure}
	\includegraphics[width=\columnwidth]{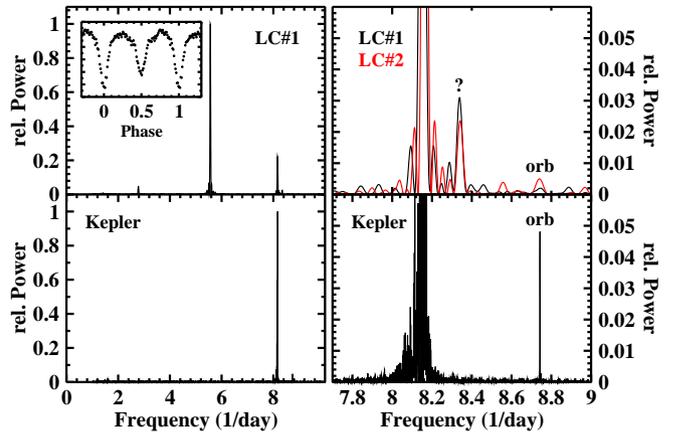}
      \caption[]{{\it Top left:} Power spectrum of LC\#1 of KIC~871494.
                The insert shows the data folded on the period corresponding
                to half of the frequency of the dominating signal. These
                variations can be attributed to contamination from
                the star ATO~J291.0335+44.9915 located close to KIC 871494.
                {\it Top right:} Parts of the Power spectra of LC\#1 (black)
                and LC\#2 (red) on an expanded scale.  
                {\it Bottom:} The same for the power spectrum of the Kepler
                data of KIC~871494.}
\label{2mass-j1924}
\end{figure}

The TESS light curves of KIC~8751494 are heavily contaminated
by light from the variable star ATO~J291.0335+44.9915 \citep{Heinze18}
which is only 42~arcsec (i.e., twice the TESS pixel size) away. 
The power spectra
(upper left frame of Fig.~\ref{2mass-j1924})
are strongly dominated by a signal at 
$F_{\rm ATO} = 5.5635(2)$~d$^{-1}$ ($P_{\rm ATO} = 0.179743(7)$~d)
which is compatible with the period listed by \citet{Heinze18} 
for ATO-J291.335+44.9915.
Signals also appear at multiples of $F_{\rm ATO}$ as well as $F_{\rm ATO}/2$.
The latter must be considered as the fundamental frequency. None of them
is present in the higher spatial resolution Kepler data. Folding the
data on $2P_{\rm ATO}$ (insert in the figure) 
yields a perfect light curve of a short period Algol system
with a secondary eclipse slightly less deep
than the primary eclipse. Again, this is compatible with the classification
as close eclipsing binary of \citet{Heinze18}.
Thus, I attribute these variations exclusively to
ATO~J291.0335+44.9915. For the record, I give ephemeris for the primary
minimum, based on representative minimum epochs measured in the two light
curves:
\begin{displaymath}
T_{\rm min} = BJD\, 2458711.16 (2) + 0.35948 (1) \times E
\end{displaymath}
Based on only two data points, formal errors are not defined. Therefore, I
arbitrarily adopt a period error which would lead to an easily recognizable
phase shift of 0.05 of the minimum over the time base of the observations.

In addition to the contamination from ATO~J291.0335+44.9915 the TESS
power spectra also contain signals coming from KIC~8751494 itself.
The SH is clearly present at slightly different frequencies of
$F_{\rm SH} = 8.152(1)$~d$^{-1}$ ($P_{\rm SH} = 0.12267(2)$~d) in LC\#1 and
$F_{\rm SH} = 8.1617(3)$~d$^{-1}$ ($P_{\rm SH} = 0.12252(5)$~d) in LC\#2.
However, the orbital signal prominently seen in the Kepler data (power spectrum
in the lower frames of Fig.~\ref{2mass-j1924}), is only very weaky present
(see right frames of the figure). Instead, the power spectra of the TESS 
data contain a significant peak between the superhump and the orbital
frequencies, marked with a question mark in the figure. The frequency is 
identical on the $2\sigma$ level in both 
light curves. The average is
$F = 8.335(5)$~d$^{-1}$ ($P = 0.11969(7)$~d).
The nature of this additional periodicity is not immediately obvious and I
leave this question open.
 
\subsection{KIC 9406652: Alternating positive and negative superhumps}
\label{KIC 9406652}

This object was identified as a variable star by \citet{Debosscher11}
in Kepler data. The first detailed investigation, based on low cadence light 
curves taken during Kepler quarters 1 through 15, was performed by
\citet{Gies13} who pointed out its similarity to old novae and novalike
variables. They found the orbital period [later refined by \citet{Kimura20}
to be 0.25451~d], a negative superhump
with a period of 0.2397~d, and a supraorbital period of 4.131~d
which is the beat between the orbital and SH periods. The light curve is
puntuated by semi-regular brightenings which made \citet{Kimura20} 
classify KIC~9406652 as an IW~And-type star, i.e. an unusual type of Z~Cam-type
dwarf novae. KIC~9406652 may thus not be a genuine novalike variable.
Considering that the TESS data were apparently taken during such a
standstill when the accretion disk -- just as in the case of AY~Psc
(Sect.~\ref{AY Psc}) -- is in a similar hot state as those of NLs and
old novae I include KIC~9406652 in this study.

TESS observed the star in 4 sectors. The first and the last two of these,
separated in time by about two years (taken in 2019 and 2021, respectively), 
are consecutive. The data of the latter can be combined into a single 
light curve. However, those of the 
former have drastically different flux levels (which may be an artifact
considering the known uncertainties concerning the absolute flux scale of
TESS light curves) and somewhat distinct properties (see below). Therefore, 
I prefer not to combine them.

\begin{figure}
	\includegraphics[width=\columnwidth]{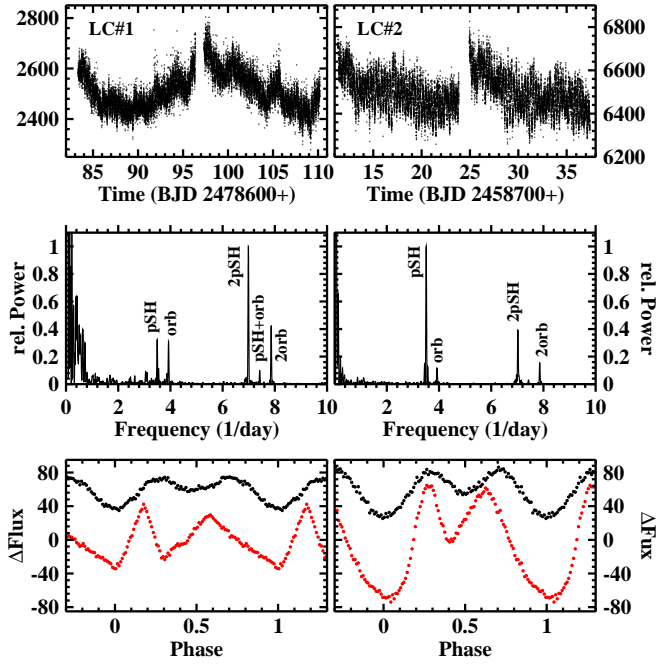}
      \caption[]{Properties of light curves LC\#1 (left column) and
                 LC\#2 (right column) of KIC~9406652: light curves (top),
                 low frequency part of the power spectra (middle), and
                 orbital (black) and superhump (red) waveforms (bottom).}
\label{kic9406652-ep1}
\end{figure}

The behaviour of KIC~9406652 is quite different during the two epochs. I
first concentrate on the 2019 data [LC\#1 and LC\#2; 
Fig.~\ref{kic9406652-ep1}: light curves (top), power spectra (middle), and
orbital and SH waveforms (bottom)]. 
Both light curves exhibit variations on time
scales of several days. The power spectra reveal signals with periods
of several hours. In LC\#1, the strongest signal corresponds to a period
exactly twice that of the second highest peak and must therefore be 
interpreted as the first overtone of a modulation with a period of
$P = 0.2864$(1)~d. This period is 12.5\% longer than the orbital period 
$P_{\rm orb}$ which is also manifest in the power spectrum together with its 
first overtone. It is thus close at hand to identify $P = P_{\rm pSH}$ with
the first appearance of a pSH in KIC~9406652. The power 
spectrum also contains a fainter peak at $F_{\rm pSH} + F_{\rm orb}$ (and still
fainter signals at higher overtones of the principal signals), but the
beat period $1/F_{\rm pSH} - 1/F_{\rm orb}$ and thus the apsidal precession 
period of the implied eccentric accretion disk is notably absent. The
waveform of the orbital modulations (black graph in the bottom frame of
Fig.~\ref{kic9406652-ep1}) is qualitatively explained by ellipsoidal
variations of the secondary star, considering the long period of this
system and the red TESS passband, where the secondary minimum is partially
filled in by reflection of light from the primary component. In contrast
to the easily explained orbital waveform, the shape of the superhump
waveform (red graph) is quite peculiar with two peaks of different extension
and decidedly pointed minima and maxima. 
 
In LC\#2 the strength of the fundamental superhump mode and its first
overtone is inverted. The superhump amplitude is enhanced, and the 
waveform -- still double humped but now superposed upon a single larger
maximum -- has a much more gentle (rounded) shape. The orbital waveform
has not changed much with respect to LC\#1. At
$P_{\rm pSH} = 0.28502(5)$~d
the superhump period is slightly but significantly shorter than in LC\#1.

\begin{figure}
	\includegraphics[width=\columnwidth]{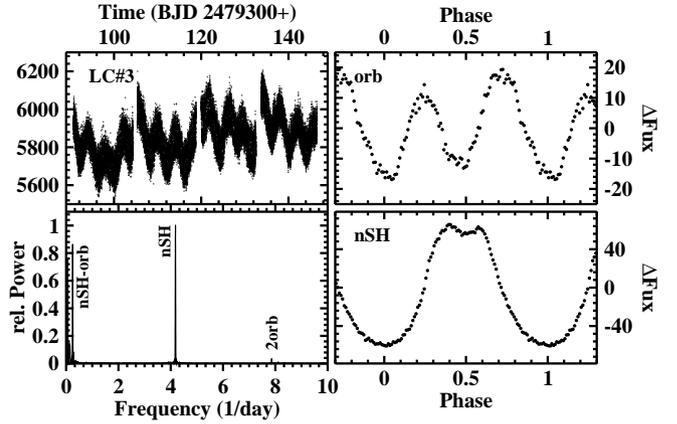}
      \caption[]{Properties of light curve LC\#3 of KIC~9406652: light
                 curve (upper left), low frequency part of the power
                 spectrum (lower left), orbital waveform (upper right) 
                 and superhump light curve (lower right).}
\label{kic9406652-ep2}
\end{figure}

In 2021 the light curve of KIC~9406652 was quite different (upper left frame
of Fig.~\ref{kic9406652-ep2}). It exhibits a clear periodicity on the time
scale of several days. The corresponding peak in the power spectrum (lower
left frame of the figure) indicates a period of
3.998(9)~d
which is close to the beat between the orbital period and the period 
$P_{\rm nSH} = 0.239197$~(8)~d 
corresponding to the main signal in the power spectrum. The latter is within 
the range of nSH periods seen in the Kepler data \citep{Kimura20}. 
Thus, the negative superhump is back and the positive one has gone. 
On the scale of the figure the orbital period is only manifest in
the power spectrum by its first overtone. This is easily explained by the
orbital waveform (upper right frame of the figure) which is somewhat different
than in LC\#1 and LC\#2, having maxima of different height and a deeper
secondary minimum.
The superhump waveform (lower right) is dominated by a strong maximum with
two small humps on top. 

\subsection{NSV 1907: Superhumps confirmed}

Discovered as a variable star a long time ago \citep{Hoffmeister63}, 
NSV~1907 (= CRTS~J051654.1+33252) remained largely unstudied until quite recently, when \citet{Hummerich17} identified it as a deeply eclipsing novalike 
cataclysmic variable, possibly of the RW~Sex subclass. The eclipses
permitted them to measured an accurate orbital period of 0.2761069(2)~d,
i.e., on the longer side of the CV period distribution. A secondary minimum
at eclipse phase 0.5 in their white light, $V$ and $B$ light curves attests 
to a non-negligible contribution of the secondary star at this long period.
\citet{Hummerich17} also observed a 4.2~d modulation which they interpret
as the nodal precession period of an accretion disk and which they use to
predict a period of 0.2591~d for a (negative) superhump.

\begin{figure}
	\includegraphics[width=\columnwidth]{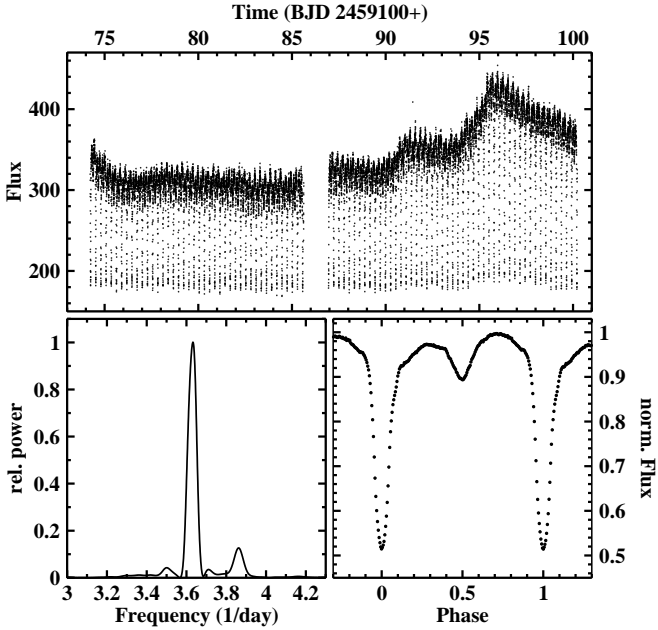}
      \caption[]{{\it Top:} Light curve of NSV~1907.
                {\it Bottom left:} Power spectrum of the light 
                curve in a small range around the orbital frequency.
                {\it Bottom right:} Average waveform of the orbital
                variations of NSV~1907 normalized such that the eclipse
                minimum has 50\% of the average flux of the two maxima.}
\label{nsv1907}
\end{figure}

The TESS light observations confirm the presence of the SH as well
as the nodal precession period. The light curve is shown in the upper frame
of Fig.~\ref{nsv1907}. Apart from the eclipses it is characterized by a strong
increase and subsequent decrease of the brightness of NSV~1907 in its second 
half. Nevertheless the flux at eclipse minimum remains almost the same, meaning 
that the light source responsible for these variations is largely eclipsed.
The out-of-eclipse flux level exhibits wiggles on the time scale of
a few days which grow much stronger during the phase of increased brightness
in the latter part of the light curve. A power spectrum of the data after
removal of the eclipses and the longer term variations contains a strong
peak at a frequency of
0.232(2)~d$^{-1}$. 
The corresponding period of 4.30(5)~d is very close to the period seen by
\citet{Hummerich17} and thus confirms their results.

At higher frequencies the power spectrum is dominated by signals at the 
orbital frequency and its overtones even after removal of the eclipses from the
light curves. The lower left frame of Fig.~\ref{nsv1907} shows a narrow
range around $F_{\rm orb}$. Apart from the orbital signal is
contains a weaker peak at
$F_{\rm SH} = 3.860(3)$~d$^{-1}$ ($P_{\rm SH} = 0.2591(1)$~d),
leaving no doubt that this is the superhump signal prediced by 
\citet{Hummerich17} but not seen directly in their data.

Finally, the orbital waveform, normalized in the same way as was done for 
AC~Cnc in Paper~I, is shown in the lower right frame of Fig.~\ref{nsv1907}.
It impressively confirms the presence of secondary eclipses which, not
surprisingly, are better defined in the redder TESS passband than in the
earlier observations. Before the secondary eclipse the hump in the waveform
has a significantly smaller amplitude that afterwards and also appears to
have some peculiar structure. I tested if the hump maxima are modulated
on the beat period between orbital and SH variations as has been seen
in AC~Cnc (Paper~I). This is indeed the case, but the effect is much smaller
than in AC~Cnc. 

\section{Discussion: A census of superhumps}
\label{Discussion}

An assessment of superhumps and their implications for the understanding of the 
structure, dynamics and evolution of CVs has been made many times in the past
\citep[e.g., Patterson 1998, 2001,][]{Patterson05}. A huge body of
observational information on pSHs in SU~UMa type dwarf nova has been
collected by \citet{Kato09} and in subsequent publications of this series.
Much of this work deals with dwarf novae and with pSHs, while specific
studies of nSHs and SH in non-outbursting CVs are considerably rarer. Among
observational papers with some emphasis on SHs in NLs and old novae I cite
\citet{Fuentes-Morales18}. For theoretical studies of nSHs I refer to
\citet{Thomas15} and citations therein. Concerning our theoretical 
understanding of pSHs, I mention the classical papers of \citet{Whitehurst88},
\citet{Whitehurst91} and \citet{Hirose91}. 

Table~\ref{Table: Superhump census} contains a census of the properties 
of positive and negative
superhumps in all novae and novalike variables for which I could find
reports on such variations in the literature. Apart from the orbital period
it lists the SH period (which is many cases is the average of several
slightly different values measured at different epochs) and the period excess
defined as $\epsilon = \left( P_{\rm SH} - P_{\rm orb} \right) / P_{\rm orb}$. The
table also
provides information about the SH waveform, the detection or not of the 
apsidal or nodal precession period in the light curves, and the frequency of
occurrence of SHs. I define three categories for the waveform: S 
indicates an approximately sinusoidal waveform which includes slight
deviations from a pure sine such as a sawtooth shape, DH stands for a double 
humped waveform, and C is used for more complex shapes. Of course, these
distinctions (in particular between DH and C) are sometimes subjective.
The presence or absence of variations
on the disk precession period is indicated by Y(es) or N(o) only in those cases 
where information in the literature permits a secure statement. The
occurrence of superhumps is categorized as permanent (P) whenever it is
seen in all available observations (but see below), and as transient
(T) when it is seen in some observations but not in others. 

\begin{table*}
\label{Table: Superhump census}	
\centering
	\caption{Summary of positive and negative superhump properties
                 observed in novalike variables and old novae. All periods
                 are expressed in days.}

\begin{tabular}{lllllllllllll}
\hline
Name & $P_{\rm orb}$ & \multicolumn{5}{l}{negative superhump} &
                     \multicolumn{5}{l}{positive superhump} & Ref.$^a$ \\
  &  & $P_{\rm nSH}$ & $\epsilon_{\rm nSH}^b$ & WF$^c$ & Prec$^d$ & Occ$^e$ & 
       $P_{\rm pSH}$ & $\epsilon_{\rm pSH}^b$ & WF$^c$ & Prec$^d$ & Occ$^e$ &  \\ 
\hline
CP Pup         &   
0.06139        &   
               &   
               &   
               &   
               &   
               &   
0.06250        &   
 0.0181        &   
S              &   
N              &   
T              &   
3              \\  
  
BK Lyn         &   
0.07494        &   
0.07280        &   
-0.0286        &   
S              &   
N              &   
T              &   
0.07848        &   
 0.0472        &   
S/C            &   
Y              &   
P              &   
1,4,5,6        \\  
  
V1974 Cyg      &   
0.08126        &   
               &   
               &   
               &   
               &   
               &   
0.08507        &   
 0.0469        &   
S              &   
Y              &   
P              &   
1,7,8,9        \\  
  
V348 Pup       &   
0.10184        &   
               &   
               &   
               &   
               &   
               &   
0.10740        &   
 0.0546        &   
S/C            &   
Y              &   
T              &   
1,10           \\  
  
V795 Her       &   
0.10825        &   
0.10474        &   
-0.0324        &   
C              &   
N              &   
T              &   
0.11619        &   
 0.0733        &   
S              &   
N              &   
T              &   
2,11-18        \\  
 [1ex]
  
KIC 8751494    &   
0.11438        &   
               &   
               &   
               &   
               &   
               &   
0.12249        &   
 0.0709        &   
S              &   
N              &   
P              &   
1,19,20        \\  
  
V592 Cas       &   
0.11506        &   
0.11193        &   
-0.0272        &   
S              &   
N              &   
T              &   
0.12239        &   
 0.0637        &   
S/DH           &   
N              &   
P              &   
1,21           \\  
  
DM Gem         &   
0.11570        &   
               &   
               &   
               &   
               &   
               &   
0.12423        &   
 0.0737        &   
S              &   
N              &   
P?             &   
2,22           \\  
  
V630 Sgr       &   
0.11793        &   
               &   
               &   
               &   
               &   
               &   
0.12417        &   
 0.0529        &   
?              &   
Y              &   
T              &   
23,24          \\  
  
LQ Peg         &   
0.11850        &   
               &   
               &   
               &   
               &   
               &   
0.12480        &   
 0.0532        &   
?              &   
Y              &   
P              &   
25,26          \\  
 [1ex]
  
V1084 Her      &   
0.12056        &   
0.11696        &   
-0.0299        &   
S              &   
Y              &   
?              &   
               &   
               &   
               &   
               &   
               &   
27             \\  
  
V442 Oph       &   
0.12443        &   
0.12090        &   
-0.0284        &   
S              &   
Y              &   
P?             &   
               &   
               &   
               &   
               &   
               &   
27             \\  
  
V4633 Sgr      &   
0.12557        &   
               &   
               &   
               &   
               &   
               &   
0.12823        &   
 0.0212        &   
S              &   
N              &   
T              &   
24,28          \\  
  
AH Men         &   
0.12721        &   
0.12355        &   
-0.0288        &   
S              &   
Y              &   
T              &   
0.13886        &   
 0.0916        &   
S              &   
N              &   
T              &   
1,29-31        \\  
  
MV Lyr         &   
0.13290        &   
0.12816        &   
-0.0357        &   
S              &   
N              &   
T              &   
0.13790        &   
 0.0376        &   
S              &   
Y              &   
T              &   
2,32,33        \\  
 [1ex]
  
DW UMa         &   
0.13661        &   
0.13264        &   
-0.0291        &   
S              &   
Y              &   
T              &   
0.14478        &   
 0.0598        &   
S              &   
Y              &   
P              &   
1,34           \\  
  
TT Ari         &   
0.13755        &   
0.13296        &   
-0.0334        &   
S              &   
Y              &   
P              &   
0.14927        &   
 0.0852        &   
S              &   
N              &   
T              &   
2,35-50        \\  
  
V603 Aql       &   
0.13822        &   
0.13390        &   
-0.0313        &   
S              &   
N              &   
T              &   
0.14548        &   
 0.0525        &   
S              &   
Y              &   
P              &   
51-58          \\  
  
V378 Peg       &   
0.13858        &   
0.13476        &   
-0.0276        &   
S              &   
               &   
P              &   
               &   
               &   
               &   
               &   
               &   
59,60          \\  
  
RR Cha         &   
0.14010        &   
0.13621        &   
-0.0278        &   
S              &   
               &   
P              &   
0.14442        &   
 0.0308        &   
               &   
N              &   
T              &   
1,61           \\  
 [1ex]
  
AQ Men         &   
0.14147        &   
0.13646        &   
-0.0354        &   
S              &   
Y              &   
P              &   
0.15047        &   
 0.0636        &   
C              &   
Y              &   
T              &   
2,62,63        \\  
  
LS Cam         &   
0.14238        &   
0.13753        &   
-0.0341        &   
               &   
Y              &   
P              &   
0.15485        &   
 0.0876        &   
               &   
N              &   
T              &   
64            \\  
  
V751 Cyg       &   
0.14458        &   
0.13936        &   
-0.0361        &   
S              &   
Y              &   
P              &   
               &   
               &   
               &   
               &   
               &   
1,65,66        \\  
  
RR Pic         &   
0.14503        &   
               &   
               &   
               &   
               &   
               &   
0.15770        &   
 0.0874        &   
S              &   
N              &   
T              &   
1,67,68        \\  
  
IM Eri         &   
0.14563        &   
0.13841        &   
-0.0496        &   
S              &   
Y              &   
P              &   
               &   
               &   
               &   
               &   
               &   
62             \\  
 [1ex]
  
PX And         &   
0.14634        &   
0.14150        &   
-0.0331        &   
S              &   
Y              &   
T              &   
               &   
               &   
               &   
               &   
               &   
1,69           \\  
  
V533 Her       &   
0.14737        &   
0.14289        &   
-0.0304        &   
               &   
N              &   
T              &   
0.15706        &   
 0.0658        &   
S              &   
Y              &   
T              &   
2,70           \\  
  
HS 1813+6122   &   
0.14752        &   
0.14095        &   
-0.0445        &   
S              &   
N              &   
T              &   
               &   
               &   
               &   
               &   
               &   
1,71           \\  
  
V2574 Oph      &   
0.14773        &   
0.14164        &   
-0.0412        &   
S              &   
               &   
T              &   
               &   
               &   
               &   
               &   
               &   
72             \\  
  
BB Dor         &   
0.14923        &   
0.14093        &   
-0.0556        &   
S              &   
Y              &   
P              &   
0.16330        &   
 0.0943        &   
S              &   
               &   
T              &   
1,73           \\  
 [1ex]
  
AO Psc         &   
0.14950        &   
               &   
               &   
               &   
               &   
               &   
0.16580        &   
 0.1090        &   
               &   
               &   
T              &   
1,74           \\  
  
BZ Cam         &   
0.15369        &   
               &   
               &   
               &   
               &   
               &   
0.15634        &   
 0.0172        &   
DH             &   
               &   
T              &   
1,75           \\  
  
V704 And       &   
0.15424        &   
0.14772        &   
-0.0423        &   
S              &   
Y              &   
T              &   
               &   
               &   
               &   
               &   
               &   
2              \\  
  
BH Lyn         &   
0.15588        &   
0.14700        &   
-0.0570        &   
               &   
               &   
T              &   
0.16772        &   
 0.0760        &   
S              &   
Y              &   
T              &   
2,76,77        \\  
  
BG Tri         &   
0.15844        &   
0.15150        &   
-0.0438        &   
S              &   
Y              &   
T              &   
0.17270        &   
0.09000        &   
S              &   
               &   
T              &   
78             \\  
 [1ex]
  
KR Aur         &   
0.16274        &   
0.15713        &   
-0.0345        &   
S              &   
               &   
T              &   
               &   
               &   
               &   
               &   
               &   
1,79           \\  
  
V1193 Ori      &   
0.16500        &   
0.15883        &   
-0.0374        &   
S              &   
N              &   
T              &   
0.17622        &   
 0.0680        &   
DH/C           &   
N              &   
T              &   
2              \\  
  
UU Aqr         &   
0.16580        &   
               &   
               &   
               &   
               &   
               &   
0.17510        &   
 0.0561        &   
               &   
               &   
T              &   
1,73           \\  
  
UX UMa         &   
0.19667        &   
0.18668        &   
-0.0508        &   
S              &   
               &   
T              &   
               &   
               &   
               &   
               &   
               &   
1,26,80        \\  
  
AY Psc         &   
0.21732        &   
0.20640        &   
-0.0502        &   
S              &   
               &   
T              &   
               &   
               &   
               &   
               &   
               &   
1,81           \\  
[1ex]
  
TV Col         &   
0.22860        &   
0.21611        &   
-0.0546        &   
S              &   
Y              &   
T              &   
               &   
               &   
               &   
               &   
               &   
2,82-85        \\  
  
RW Tri         &   
0.23188        &   
0.22190        &   
-0.0430        &   
S              &   
               &   
T              &   
               &   
               &   
               &   
               &   
               &   
2,26,86        \\  
  
KIC 9406652    &   
0.25451        &   
0.22945        &   
-0.0985        &   
S              &   
Y              &   
T              &   
0.29071        &   
 0.1422        &   
C              &   
N              &   
T              &   
1,87,88        \\  
  
NSV 1907       &   
0.27611        &   
0.25910        &   
-0.0616        &   
S              &   
Y              &   
P?             &   
               &   
               &   
               &   
               &   
               &   
1,89           \\  
  
RX 2133.7+5107 &   
0.29743        &   
0.28132        &   
-0.0542        &   
S              &   
N              &   
P              &   
               &   
               &   
               &   
               &   
               &   
1,90           \\  
[1ex]
  
RZ Gru         &   
0.41750        &   
               &   
               &   
               &   
               &   
               &   
0.52000        &   
 0.2455        &   
S              &   
               &   
               &   
2              \\  
  
\hline
\multicolumn{13}{l}{$^a$ References (see 
                         Table~5)} \\
\multicolumn{13}{l}{$^b$ Period excess defined as 
            $\epsilon = \left( P_{\rm SH}-P_{\rm orb} \right)/P_{\rm orb}$} \\
\multicolumn{13}{l}{$^c$ Waveform: S = sinusoidal; DH = double humped;
                         C = complex} \\
\multicolumn{13}{l}{$^d$ Precession period detected (Yes/No)} \\
\multicolumn{13}{l}{$^e$ Occurrence of superhump: P = permanent;
                         T = transient} \\
\end{tabular}
\end{table*}

\begin{table*}
\label{Table: Superhump references.}	
\centering
	\caption{References to Table~\ref{Table: Superhump census}	
}
\begin{tabular}{c}
\hline
(1)  This work;
(2)  Paper I;
(3) \citet{Patterson98b};
(4) \citet{Skillman93}
(5) \citet{Misselt95};\\
(6) \citet{Patterson13};
(7) \citet{Semeniuk94};
(8) \citet{Semeniuk95};
(9) \citet{Retter97};
(10) \citet{Rolfe00};\\
(11) \citet{Mironov83};
(12) \citet{Baidak85};
(13) \citet{Kaluzny89};
(14) \citet{Rosen89};
(15) \citet{Shafter90};\\
(16) \citet{Zhang91};
(17) \citet{Papadaki06};
(18) \citet{Simon12};
(19) \citet{Williams10};\\
(20) \citet{Kato13};
(21) \citet{Taylor98};
(22) \citet{Rodriguez-Gil05};
(23) \citet{Woudt01};\\
(24) \citet{Mroz15};
(25) \citet{Rude12};
(26) \citet{Bruch20};
(27) \citet{Patterson02};\\
(28) \citet{Lipkin08};
(29) \citet{Buckley93}
(30) \citet{Patterson95};
(31) \citet{Patterson98};
(32) \citet{Borisov92};\\
(33) \citet{Skillman95};
(34) \citet{Boyd17};
(35) \citet{Andronov92};
(36) \citet{Andronov99};\\
(37) \citet{Belova13};
(38) \citet{Bruch19b};
(39) \citet{Kim09};
(40) \citet{Kraicheva99};
(41) \citet{Roessiger88};\\
(42) \citet{Semeniuk87};
(43) \citet{Skillman98};
(44) \citet{Smak75};
(45) \citet{Sztanjo79};
(46) \citet{Tremko92};\\
(47) \citet{Udalski88};
(48) \citet{Volpi88};
(49) \citet{Weingrill09};
(50) \citet{Wu02};
(51) \citet{Bruch91};\\
(52) \citet{Bruch18};
(53) \citet{Haefner81};
(54) \citet{Haefner85};
(55) \citet{Hollander97};\\
(56) \citet{Patterson97};
(57) \citet{Patterson91};
(58) \citet{Patterson93};
(59) \citet{Kozhevnikov12};\\
(60) \citet{Ringwald12};
(61) \citet{Woudt02};
(62) \citet{Armstrong13};
(63) \citet{Ilkiewics21};\\
(64) \citet{Rawat22};
(65) \citet{Patterson01b};
(66) \citet{Papadaki09};
(67) \citet{Fuentes-Morales18};\\
(68) \citet{Schmidtobreick08};
(69) \citet{Stanishev02};
(70) \citet{McQuillin12};
(71) \citet{Rodriguez-Gil07};\\
(72) \citet{Kang06};
(73) \citet{Patterson05};
(74) \citet{Patterson01a}
(75) \citet{Kato01};
(76) \citet{Stanishev06};\\
(77) \citet{Patterson99};
(78) \citet{Stefanov22};
(79) \citet{Kozhevnikov07};
(80) \citet{deMiguel16};\\
(81) \citet{Guelsecen09};
(82) \citet{Augusteijn94};
(83) \citet{Barrett88};
(84) \citet{Hutchings81};
(85) \citet{Motch81a};\\
(86) \citet{Smak19};
(87) \citet{Gies13};
(88) \citet{Kimura20};
(89) \citet{Hummerich17};
(90) \citet{deMiguel17}.\\
\hline
\end{tabular}
\end{table*}

\begin{figure}
	\includegraphics[width=\columnwidth]{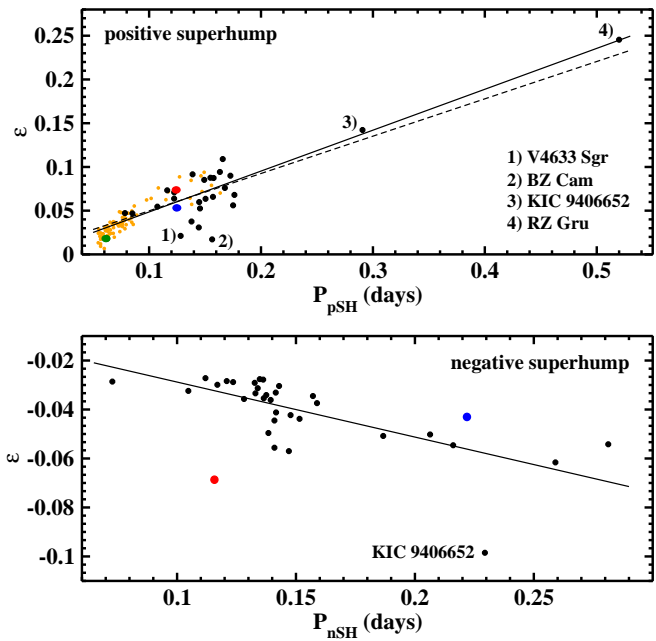}
      \caption[]{{\it Top:} Relation between the period excess and
                the superhump period for positive superhumps. The large
                dots represent data taken from 
                Table~\ref{Table: Superhump census}. For comparison,
                data taken from table~9 of \citet{Patterson05} are also 
                plotted as smaller orange dots. The solid line is a
                linear least squares fit to all data from 
                Table~\ref{Table: Superhump census}, excluding the low
                credential superhump systems V4633~Sgr and BZ~Cam.
                The dashed line is the same, additionally excluding the
                long period systems KIC~9406652 and RZ~Gru. The red,
                blue and green dots represent DM~Gem, LQ~Peg, and CP~Pup,
                respectively. For further details, see text.
                {\it Bottom:} The same for negative superhumps. The red
                and blue dots represent DM~Gem under the assumption that 
                it is a negative superhumper, and the low credential 
                system RW~Tri, respectively. The solid line is a linear
                least squares fit to all points except DM~Gem.}
\label{sh-epsilon-rel}
\end{figure}

Some of the stars
listed in the table may have low credentials as superhumpers because of
sparse of observations, weak indications for SHs, or alternative
explications for the observed variations. They are included in the
table anyway for the 
sake of completeness, but it should be kept in mind that they may not
exhibit genuine superhumps. In particular, I mention: (1) CP~Pup. 
\citet{Patterson98b} observed an unstable period which they interpreted as
being due to a SH. As I showed in Paper~I the TESS light curves of
CP~Pup exhibit a multitude of transient QPO-like variations in the respective 
period range which cannot be considered to be superhumps 
and which in more limited
terrestrial observations may mimick SHs. (2) BZ~Cam. This case is
similar to CP~Pup. \citet{Kato01} claim to have seen superhumps, but the
multitude of power spectrum peaks in the data of \citet{Patterson96} and
the time resolved power spectrum in Fig.~\ref{bzcam-ps} suggest only the
occurrence of QPO-like variations. (3) RZ~Gru. This star was identified in 
Paper~I as a system with pSHs, but the corresponding power spectrum 
peak may not be outstanding enough to provide convincing evidence. Moreover, 
at the long orbital period RZ~Gru is not expected to develop a precessing
accretion disk (however, I try to avoid a bias based on theoretical 
preconceptions). (4) RW~Tri. This system is claimed to have exhibited
nSHs restricted to the 1984 and possibly 1957 observing seasons 
\citep{Smak19}, but not
in other years \citep{Bruch20}. While this alone does not discredit it as 
a transient SH system, I consider the peaks in Smak's power
spectra, on which he bases his claim, not of suffient strength to convincingly
indicate a consistent periodicity. (5) V4633~Sgr. \citet{Fuentes-Morales18}
list this star as a superhumping old nova, but \citet{Lipkin08} interpret
the corresponding variations in an asynchronous polar scenario, where
the rotation of a magnetic white dwarf got out of synchronization with the 
orbital period as a
consequence of the recent nova outburst of V4633~Sgr in 1998. (6) LQ~Peg.
It is not clear if the consistent photometric variations in this star
are due to the orbital motion or a superhump. For a thorough discussion
of this issue, see \citet{Bruch20}. (7) AO~Psc.
As noticed in Sect.~\ref{AO Psc}, AO~Psc is only briefly mentioned as
a superhumper by \citet{Patterson01a}, while the announced publication of
details never occurred.

Including the systems with low credentials, 
Table~\ref{Table: Superhump census} contains 46 stars.
Of these, 30 exhibit positive and 33 negative superhumps. Thus, statistically 
among novae and novalike variable both species are about equally 
probable, unless there is an observational bias favouring the detection 
of one over the other. In nine stars both kinds of SHs have even been seen
simultaneously (V603~Aql, TT~Ari, LS~Cam, V592~Cas, RR~Cha, BB~Dor,
BK~Lyn, AQ~Men, DW~UMa).

\citet{Stolz84} were the first to note a linear relationship between
the period excess $\epsilon$ and the superhump period $P_{\rm SH}$ in 
SU~UMa type dwarf novae; a relationship which -- albeit less strictly -- 
is also valid for pSHs in other CVs. It has been refined and
discussed many times in subsequent years and sometimes replaced by a
relation between $P_{\rm orb}$ and $P_{\rm SH}$ 
\citep[e.g.,][]{Gaensicke09, Fuentes-Morales18}. Both relations are largely 
equivalent,
but if $\epsilon$ vs $P_{\rm SH}$ is linear, $P_{\rm orb}$ vs $P_{\rm SH}$
becomes notably nonlinear if a wider range of periods is regarded. Therefore, 
I use here the original notation of the Stolz-Schoembs relation. 

For all positive superhump systems listed in 
Table~\ref{Table: Superhump census}, $\epsilon$ is plotted as a function of
$P_{\rm pSH}$ as large dots in the upper frame of Fig.~\ref{sh-epsilon-rel}.
For comparison, the respective data taken from table~9 of \citet{Patterson05}
are drawn as smaller orange dots. The bulk of them refers to short period
SU~UMa type dwarf novae. Several lessons can be learned from this diagram.
As is already obvious from the data of \citet{Patterson05} and is also 
mentioned by \citet{Fuentes-Morales18}, the scatter of the points corresponding
to nova and novalike variables (i.e., almost all points with 
$P_{\rm pSH} > 0.1$~d) exhibit a much higher scatter than the dwarf novae
below the CV period gap. Thus, the Stolz-Schoembs relation becomes less
well defined for these systems. There are some outliers with low $\epsilon$
values, two of which (V4633~Sgr and BZ~Cam) can be identified with low
credential superhump systems. These two are ignored subsequently. In 
contrast, other systems with low 
credentials -- CP~Pup (green dot in the figure) and 
LQ~Peg (blue dot) -- follow the general trend well. For LQ~Peg this may be 
an indication
that the persistent light curve variability is indeed due to SHs and
not the orbital motion. DM~Gem (red dot) was found in Paper~I to exhibit two 
periods. But it could not be decided which of these is due to the orbital 
motion and which to SHs. In Fig.~\ref{sh-epsilon-rel} DM~Gem
follows the general trend; a good argument in favour of its interpretation 
of a positive (instead of negative) superhumper. 

Two systems, not previously discussed in the context
of the Stolz-Schoembs relation, extend this relation to much longer periods:
KIC~9406652 and RZ~Gru. While the credentials for superhumps in RZ~Gru may
not be totally convincing (see above) there is no doubt about the strong 
pSH in KIC~9406652 (see Sect.~\ref{KIC 9406652}). Excluding
these two stars, a linear least squares fit the to remaining points
yields the dashed line in Fig.~\ref{sh-epsilon-rel}. In spite of their
significantly longer periods the points corresponding to  KIC~9406652 and 
RZ~Gru lie only
$0.8 \sigma$ and $1.3 \sigma$, respectively above the extrapolated
$P_{\rm pSH} - \epsilon$ relation defined by the other systems, where 
$\sigma$ is the standard deviation of their scatter around the dashed line. 
Thus, both of the long period systems follow well the Stolz-Schoembs
relation for nova and novalike stars. The solid line is a least squares
fit to all data, yielding 
$\epsilon = 0.002(3) + 0.47(2)\, P_{\rm psh}$. 

KIC~9406652 and RZ~Gru present a challenge to theoretical explanations 
of pSHs. As mentioned in the Introduction, 
these are thought to be caused by the extra tidal
stresses in the outer regions of an asymmetric accretion disk when its
elongated part extends towards the secondary star. The disk can become
elliptical when the revolution period of particles in its outer part
reaches the 3:1 resonance with the orbital period \citep{Whitehurst88}.
This is expected to be possible only in systems with a small mass
ratio $q=M_{\rm WD}/M_{\rm sec}$. Just how small $q$ must be is a matter of
debate. Limits cited in the literature range from 0.22 to 0.39
\citep{Whitehurst91, Pearson06, Smak20}. The mass ratio of RZ~Gru is
unknown. Assuming the secondary star to have a mass according to the
semi-empirical mass-period relation of \citet{Knigge11} and the white
dwarf mass to be equal to the average mass of the compact object in CVs
\citep{Zorotovic11} the mass ratio is 0.41, higher than but still close to 
the upper limit of the theoretically permitted range.
Based on radial velocity measurements of
absorption and emission lines \citet{Gies13} derived
$q = 0.83 \pm 0.07$ for KIC~9406652, way beyond this range.

As also outlined in the Introduction, the phenomenological understanding
of nSHs as arising in a warped or tilted accretion disk is widely accepted. 
But their is no consensus on the mechanisms which cause the warp or 
inclination of the disk. In contrast to positive
superhumps it is therefore not possible to specify any limits imposed by
theory on the occurrence of negative superhumps and to confront such 
limits with observations. The $\epsilon - P_{\rm nSH}$ relation constructed from 
Table~\ref{Table: Superhump census} is shown in the lower frame of 
Fig.~\ref{sh-epsilon-rel}. I also insert as a red dot the location of
DM~Gem under the assumption that it is a negative superhumper. It falls
drastically below the general trend. This is thus a further argument in favor 
to a pSH nature of this system. The low credential system
RW~Tri (blue dot) follows the general trend reasonably well and may thus
in fact exhibit superhumps sporadically. This leaves KIC~9406652 as the
only true outlyer in the diagram with a much more negative period excess
than expected at its period. There is, however, no reason to suspect
anything to be wrong with its SH period. A linear least squares fit
to the data points (including KIC~9406652 but, of course, not DM~Gem) yields
$\epsilon = -0.006(2) - 0.22(1)\, P_{\rm nSH}$
and is shown as a solid line in the figure. I note that the ratio of the
inclination of this relation to the corresponding relation for pSHs
is $-0.48$, confirming the conventional wisdom that the
period deficit in negative superhumps systems is about half of that of
the period excess in their positive counterparts at a given period.

In most cases where the corresponding information is available the negative 
as well as the positive SHs have a roughly sinusoidal shape, consisting
of a single hump extending over all phases. Sometimes, it is distorted into
a saw tooth with the steeper side leading or trailing, or the maximum is broader
or narrower than the minimum. Often the waveform changes somewhat from one
epoch to another. More interesting are the rarer cases of double humped
or complex waveforms which occur more frequently in positive than in negative
superhump systems. Extreme examples are observed in AQ~Men \citep{Ilkiewics21},
V348~Pup (Fig.~\ref{v348pup-lc}) or KIC~9406652 (Fig.~\ref{kic9406652-ep1}).
Such waveforms constitute a valuable source of information about the
mechanisms leading to superhumps and the struture of superhumping accretion 
disks which has not yet been tapped adequately. 

Often superhumps are accompanied with variations on the beat period between
orbital and SH periods, i.e., at the precession period of the
accretion disk. Particularly impressive examples are V348~Pup 
(Fig.~\ref{v348pup-lc}) and DW~UMa (Fig.~\ref{dwuma}). But in about a
third of all cases variations on the beat period have not been detected.
Sometimes this may be explained by absent or only weak variations on the
orbital period, but there are counterexamples. The orbital and SH
signals are of comparable strength in the power spectrum of the first half 
of the TESS light curve of V795~Her (Paper I), yet no signal is seen at 
the beat frequency. In contrast, the orbital is much weaker than the
superhump signal in BK~Lyn, but the light curve
prominently exhibits variations on the beat period (Fig.~\ref{bklyn}).
Of especial interest are a few systems where supraorbital periods are not 
seen on the beat period but on multiples thereof. These are V603 Aql 
\citep{Bruch18}, RZ~Gru (Paper~I) and possibly V1974~Cyg 
(Sect.~\ref{V1974 Cyg}).

With a limited number of observations it is, of course, never possible to
be certain that superhumps in any CV are really permanent. Here, I classify 
a system tentatively as a permanent superhumper if observations at different
epochs are available and superhumps were always found when they have
been searched for in data of suitable quality and quantity, and unless they 
are substituted
by other long-term features in the light curve [such as the temporary
substitution of negative by positive superhumps or the temporary transition
into a low state in TT~Ari; \citet{Bruch20}]. In 
this sense, the most convincing permanent SH systems are V603~Aql
(pSH) and TT~Ari (nSH; see the numerous references cited in 
Table~\ref{Table: Superhump census}). Classifying a system as a transient
superhumper is much more straightforward: it is sufficient if superhumps are
seen at one or several epochs but not at others. Adopting this criterium,
transient superhumps are significantly more frequent than permanent ones. 


\section{Summary}
\label{Summary}

In this paper, I took advantage of the enormous richness of information
of the month long (or many times even longer) almost continuous high
cadence light curves provided by the TESS mission to explore the properties
and the temporal behaviour (periods, occurrence or absence, waveforms,
interplay with orbital periods) of superhumps, either negative of positive, 
in the majority of novalike variables and old novae where such phenomena 
were observed in the past and for which TESS data are available. The 
results of this study, in combination with Paper~I and information 
collected from the literature enabled a more complete census of the superhump
properties in these systems
than was possible hitherto. This compilation of old and new observational
attributes should serve to provide boundary conditions for physical models
for superhumps. In this context I draw special attention to the sometimes 
vexing morphological complexity and temporal variability of SH waveforms 
which merit more attention than they received so far.

As a corollary, for the eclipsing CVs among the targets of this study, 
eclipse epochs were derived from the TESS data. In some cases these were 
used together with additional eclipse epochs measured in archival terrestrial
light curves to update the sometimes decades old orbital ephemeries and to
discuss systematic or erratic period variations. 

It is common knowledge that pSHs are abundant -- even a defining characteristic
-- in superoutbursts of short period dwarf novae (SU~UMa stars). SHs -- both,
positive and negative ones -- are not observed as routinely in the longer
period NLs and old novae, but the fraction of such systems exhibiting SHs
is by no means small. The identification of several more such systems in
Paper~I suggests that many more superhumpers lurk amoung those systems
which have not yet been systematically investigated for SHs. TESS
light curves of a significant number of such NLs and old novae are available.
An effort to analyse these is currently underway and will hopefully lead to 
a third paper of this small series.

\section*{Acknowledgements}

This paper is based on data collected by the TESS and Kepler missions and 
obtained from the MAST data archive at the Space Telescope Science Institute 
(STScI). Funding for the missions is provided by the NASA Explorer Program
and the NASA Science Mission Directorate for TESS and Kepler, respectively. 
STScI is operated by the Association of Universities for Research in 
Astronomy, Inc., under NASA contract NAS 5-26555. Supportive data were
obtained from the data archives operated by the American Association of
Variable Star Observers and the Laborat\'orio Nacional de Astrof\'{\i}sica. 

\section*{Data availability}

All data used in the present study are publically available at the
Barbara A.\ Mikulski Archive for Space Telescopes 
(MAST:\\ https://mast.stsci.edu/portal/Mashub/clients/MAST/\\Portal.html),
the AAVSO web site (https://www.aavso.org) and the LNA Data Bank
(http://databank.lna.br).




\begin{thebibliography}{99}
\bibitem[\protect\citeauthoryear{Africano et al.}{1978}]{Africano78}
        Africano J.L., Nather R.E., Patterson J., Robinson E.L., Warner B.,
        1978, PASP, 90, 568 
\bibitem[\protect\citeauthoryear{Africano \& Wilson}{1976}]{Africano76}
        Africano J.L., Wilson J., 1976, PASP, 88, 8
\bibitem[\protect\citeauthoryear{Andronov et al.}{1999}]{Andronov99} 
        Andronov I.L., Arai K., Chinorova L.L., et al., 1999, 
        AJ, 117, 574
\bibitem[\protect\citeauthoryear{Andronov et al.}{1989}]{Andronov89}
        Andronov I.L., Kimeridze G.N., Richter G.A., Symkov V.P., 1989,
        IBVS, 3388
\bibitem[\protect\citeauthoryear{Andronov et al.}{1992}]{Andronov92} 
        Andronov I.L., Kolosov D.E., Movchan A.I., Rudenko A.N., 1992, 
        Soobshch. Spets. Astrofiz. Obs., 69, 79
\bibitem[\protect\citeauthoryear{Applegate}{1992}]{Applegate92}
        Applegate J.H., 1992, ApJ, 385, 621
\bibitem[\protect\citeauthoryear{Araujo-Betancor et al.}{2005}]
        {Araujo-Betancor05}
        Araujo-Betancor S. G\"ansicke B.T., Hagen, H.-J., et al., 2005,
        A\&A, 430, 629
\bibitem[\protect\citeauthoryear{Armstrong et al.}{2013}]{Armstrong13} 
        Armstrong E., Patterson J., Michelsen R., et al., 2013, MNRAS, 435, 707
\bibitem[\protect\citeauthoryear{Augusteijn et al.}{1994}]{Augusteijn94} 
        Augusteijn T., Heemskerk M.H.M., Zwarthoed G.A.A., van Paradijs J., 
        1994, A\&AS, 107, 219
\bibitem[\protect\citeauthoryear{Baidak et al.}{1985}]{Baidak85} 
        Baidak A.V., Lipunova N.A., Shugarov S.Yu., Moshkalev V.G., 
        Volkov, L.M., 1985, IBVS, 2676
\bibitem[\protect\citeauthoryear{Baptista et al.}{1995}]{Baptista95} 
        Baptista R., Horne K., Hilditch R.W., Mason K.O., Drew J.E.,
        1995, ApJ, 448, 393
\bibitem[\protect\citeauthoryear{Baptista et al.}{1994}]{Baptista94} 
        Baptista R., Steiner J.E., Cieslinski D., 1994, ApJ, 433, 332
\bibitem[\protect\citeauthoryear{Barrett et al.}{1988}]{Barrett88} 
        Barrett P., O'Donoghue D., Warner B., 1988, MNRAS, 233, 759
\bibitem[\protect\citeauthoryear{Beljawsky}{1926}]{Beljawsky26}
        Beljawsky P.R., 1926, Beobachtungs-Zirkular der Astron.\ Nachr., 6, 38
\bibitem[\protect\citeauthoryear{Belova et al.}{2013}]{Belova13} 
        Belova A.L., Suleimanov V.F., Bikmaev I.F., Khamitov I.M., Zhukov G.V.,
        Senio D.S., Belov I.Y., Sakhibullin N.A., 2013, Astron. Lett., 39, 111
\bibitem[\protect\citeauthoryear{Biryukov \& Borisov}{1990}]{Biryukov90}
        Biryukov W., Borisov G.V., 1990, Perem.\ Szvesdy, 1544, 1
\bibitem[\protect\citeauthoryear{Boeva et al.}{2021}]{Boeva21} 
        Boeva, S., Latev, G., Zamanov R., 2021, ATel, 14365
\bibitem[\protect\citeauthoryear{Bonnet-Bidaud et al.}{2006}]{Bonnet-Bidaud06} 
        Bonnet-Bidaud J.M., Mouchet M., de Martino D., Matt G., Motch C.,
        2006, A\&A, 445, 1037
\bibitem[\protect\citeauthoryear{Borisov}{1992}]{Borisov92} 
        Borisov G.V., 1992, A\&A, 261, 154
\bibitem[\protect\citeauthoryear{Boyd et al.}{2017}]{Boyd17} 
        Boyd D.R.S., de Miguel E. Patterson J., et al., 2017, MNRAS, 466, 3417
\bibitem[\protect\citeauthoryear{Bruch}{1991}]{Bruch91} 
        Bruch A., 1991, Acta Astron., 41,101
\bibitem[\protect\citeauthoryear{Bruch}{2019a}]{Bruch19a}
        Bruch A., 2019a, IBVS, 6257
\bibitem[\protect\citeauthoryear{Bruch}{2019b}]{Bruch19b}
        Bruch A., 2019b, MNRAS, 489, 2961
\bibitem[\protect\citeauthoryear{Bruch}{2020}]{Bruch20} 
        Bruch A., 2020, New Astr., 78, 101369
\bibitem[\protect\citeauthoryear{Bruch}{2021}]{Bruch21} 
        Bruch A., 2021, MNRAS, 503, 953
\bibitem[\protect\citeauthoryear{Bruch}{2022a}]{Bruch22a}
        Bruch A., 2022a, MNRAS, 509, 4669
\bibitem[\protect\citeauthoryear{Bruch}{2022b}]{Bruch22b}
        Bruch A., 2022b, MNRAS, 514, 4718
\bibitem[\protect\citeauthoryear{Bruch \& Cook}{2018}]{Bruch18} 
        Bruch A., Cook L.M., 2018, New Astr., 63, 1
\bibitem[\protect\citeauthoryear{Buckley et al.}{1993}]{Buckley93} 
        Buckley D.A.H., Remillard R.A., Tuohy I.R., Warner B., Sullivan D.J.,
        1993, MNRAS, 265, 926
\bibitem[\protect\citeauthoryear{Chen et al.}{2001}]{Chen01} 
        Chen A., O'Donogue D., Stobie R.S., Kilkenny D., Warner B., 2001
        MNRAS, 325, 89
\bibitem[\protect\citeauthoryear{Chote \& Sullivan}{2016}]{Chote16} 
        Chote P., Sullivan D.J., 2016, MNRAS, 458, 1393
\bibitem[\protect\citeauthoryear{Dai et al.}{2010}]{Dai10} 
        Dai Z.B., Qian S.B., Fern\'andez-Lajus E., Baume G.L., 2010,
        MNRAS, 409, 1195
\bibitem[\protect\citeauthoryear{Debosscher et al.}{2011}]{Debosscher11} 
        Debosscher J., Blomme J., Aerts C., De Ridder J., 2011, A\&A, 529, A89
\bibitem[\protect\citeauthoryear{Deeming}{1975}]{Deeming75} 
        Deeming T.J., 1975, Ap\&SS, 39, 137
\bibitem[\protect\citeauthoryear{de Miguel et al.}{2016}]{deMiguel16} 
        de Miguel E., Patterson J., Cejudo E., et al., 2016, MNRAS, 457, 1447
\bibitem[\protect\citeauthoryear{de Miguel et al.}{2017}]{deMiguel17} 
        de Miguel E., Patterson J., Jones J.L., et al., 2017, MNRAS, 467, 428
\bibitem[\protect\citeauthoryear{DeYoung \& Schmidt}{1993}]{DeYoung93} 
        DeYoung J.A., Schmidt R.E., 1993, IAU Circ., 5880
\bibitem[\protect\citeauthoryear{Diaz \& Steiner}{1990}]{Diaz90} 
        Diaz M.P., Steiner J.E., 1990, A\&A, 238, 170
\bibitem[\protect\citeauthoryear{Eastman et al.}{2010}]{Eastman10} 
        Eastman, J., Siverd, R., \& Gaudi, B.S., 2010, PASP, 122, 935
\bibitem[\protect\citeauthoryear{Froning et al.}{2003}]{Froning03} 
        Froning C.S., Long K.S., Baptista R., 2003, AJ, 126, 964
\bibitem[\protect\citeauthoryear{Fuentes-Morales et al.}{2018}]
        {Fuentes-Morales18} 
        Fuentes-Morales I., Vogt, N., Tappert C., 2018, MNRAS 474, 2493
\bibitem[\protect\citeauthoryear{G\"ansicke et al.}{2009}]{Gaensicke09}  
        G\"ansicke B.T., Dillon M., Southworth J., 2009, MNRAS, 397, 2170
\bibitem[\protect\citeauthoryear{Garnavich \& Szkody}{1988}]{Garnavich88}  
        Garnavich P., Szkody P., 1988, PASP, 100, 1522
\bibitem[\protect\citeauthoryear{Gies et al.}{2013}]{Gies13}  
        Gies D.R., Guo Z., Howell S.B., 2013, ApJ, 775, 64
\bibitem[\protect\citeauthoryear{Green et al.}{1986}]{Green86} 
        Green R.F., Schmidt M., Liebert J., 1986, ApJS, 61, 305 
\bibitem[\protect\citeauthoryear{Greenstein et al.}{1970}]{Greenstein70} 
        Greenstein J.L., Sargent A.I., Haug U., 1970, A\&A, 7, 1 
\bibitem[\protect\citeauthoryear{Greiner et al.}{2001}]{Greiner01} 
        Greiner J., Tovmassian G., Orio M., et al., 2001, A\&A, 376, 1031
\bibitem[\protect\citeauthoryear{G\"ulsecen et al.}{2009}]{Guelsecen09} 
        G\"ulsecen H., Retter A., Eseno\v{g}lu H., 2009, New Astr., 14, 330
\bibitem[\protect\citeauthoryear{Haefner}{1981}]{Haefner81} 
        Haefner R., 1981, IBVS, 2045
\bibitem[\protect\citeauthoryear{Haefner \& Metz}{1985}]{Haefner85} 
        Haefner R., Metz K., 1985, A\&A, 145, 311
\bibitem[\protect\citeauthoryear{Han et al.}{2017}]{Han17} 
        Han Z., Qian S., Voloshina I., Zhu L., 2017, Research A\&A, 17, 56
\bibitem[\protect\citeauthoryear{Heinze et al.}{2018}]{Heinze18} 
        Heinze A.N., Tonry J.L., Denneau, L., 2018, AJ, 156, 241
\bibitem[\protect\citeauthoryear{Hellier \& Robinson}{1994}]{Hellier94} 
        Hellier C., Robinson E.L., 1994, ApJ, 431, L107
\bibitem[\protect\citeauthoryear{Hirose et al.}{1991}]{Hirose91} 
        Hirose M., Osaki Y., Mineshige S., 1991, PASJ, 43, 809
\bibitem[\protect\citeauthoryear{Hoffmeister}{1963}]{Hoffmeister63} 
        Hoffmeister C., 1963, Astron.\ Nachr., 287, 169
\bibitem[\protect\citeauthoryear{Hollander et al.}{1997}]{Hollander97} 
        Hollander A., Kraakman H., van Paradijs, J., 1997, A\&AS, 101, 87
\bibitem[\protect\citeauthoryear{Honeycutt}{2001}]{Honeycutt01} 
        Honeycutt R.K., 2001, PASP, 113, 473
\bibitem[\protect\citeauthoryear{Honeycutt \& Kafka}{2004}]{Honeycutt04} 
        Honeycutt R.K., Kafka S., 2004, AJ, 128, 1279
\bibitem[\protect\citeauthoryear{Honeycutt et al.}{1994}]{Honeycutt94} 
        Honeycutt R.K., Robertson J.W., Turner G.W., Vesper D.N., 1994,  
        in Shafter A.W., ed., ASP Conf.\ Ser.\, 56, Interacting Binary Stars, 
        p.\ 277 
\bibitem[\protect\citeauthoryear{Howell et al.}{1991}]{Howell91} 
        Howell S.B., Szkody P., Kreidl T.J., Dobrzycka D., 1991, PASP, 103, 300
\bibitem[\protect\citeauthoryear{H\"ummerich et al.}{2017}]{Hummerich17} 
        H\"ummrich S., Gr\"obel R., Hambsch F.-J., et al., 2017
        New Astr., 50, 30
\bibitem[\protect\citeauthoryear{Hutchings et al.}{1981}]{Hutchings81} 
        Hutchings J.B., Crampton D., Cowley A.P., Thorstensen J.R., 
        Charles P.A., 1981, ApJ, 249, 680
\bibitem[\protect\citeauthoryear{Hutchings et al.}{1983}]{Hutchings83} 
        Hutchings J.B., Link R., Crampton D., 1983, PASP, 95, 265
\bibitem[\protect\citeauthoryear{I{\l}kiewicz et al.}{2021}]{Ilkiewics21} 
        I{\l}kiewics K., Scaringi S., Court J.M.C., et al., 2021, 
        MNRAS, 503, 4050
\bibitem[\protect\citeauthoryear{Kafka}{2021}]{Kafka21} 
        Kafka S., 2021, Observations from the AAVSO International Data Base
        available at https://www.aavso.org
\bibitem[\protect\citeauthoryear{Ka{\l}uzny}{1989}]{Kaluzny89} 
        Ka{\l}uzny J, 1989, Acta Astron., 39, 235
\bibitem[\protect\citeauthoryear{Kang et al.}{2006}]{Kang06} 
        Kang T.W., Retter A., Liu A., Richards M., 2006, AJ, 131, 1687
\bibitem[\protect\citeauthoryear{Kato et al.}{2009}]{Kato09}
        Kato T., Imada A, Uemura M., et al., 2009, PASJ, 61, S395
\bibitem[\protect\citeauthoryear{Kato et al.}{2002}]{Kato02a}
        Kato T., Ishioka R., Uemura M., 2002, PASJ, 54, 1003
\bibitem[\protect\citeauthoryear{Kato \& Maehara}{2013}]{Kato13}
        Kato T., Maehara H., 2013, PASJ, 65, 76
\bibitem[\protect\citeauthoryear{Kato \& Starkey}{2002}]{Kato02b} 
        Kato T., Starkey D.R., 2002, IBVS, 5358
\bibitem[\protect\citeauthoryear{Kato \& Uemura}{2001}]{Kato01} 
        Kato T., Uemura M., 2001, IBVS, 5077
\bibitem[\protect\citeauthoryear{Kim et al.}{2009}]{Kim09} 
        Kim Y., Andronov I.L., Cha S.M., Chinarova L.L., Yoon J.N., 2009, 
        A\&A, 496,765
\bibitem[\protect\citeauthoryear{Kimura et al.}{2020}]{Kimura20} 
        Kimura M., Osaki Y., Kato T., 2020, PASJ, 72, 94
\bibitem[\protect\citeauthoryear{Knigge et al.}{2011}]{Knigge11} 
        Knigge C., Baraffe I., Patterson J., 2011, ApJS, 194, 28
\bibitem[\protect\citeauthoryear{Kozhevnikov}{2007}]{Kozhevnikov07}
        Kozhevnikov V.P., 2007, MNRAS, 378, 957
\bibitem[\protect\citeauthoryear{Kozhevnikov}{2012}]{Kozhevnikov12} 
        Kozhevnikov V.P., 2012, New Astron., 17, 38
\bibitem[\protect\citeauthoryear{Kraichva et al.}{1999}]{Kraicheva99} 
        Kraicheva Z., Stanishev V., Genkov V., Iliev V., 1999, 
        A\&A, 351, 607
\bibitem[\protect\citeauthoryear{Kukarkin}{1977}]{Kukarkin77}
        Kukarkin B.V., 1977, MNRAS, 180, 5p
\bibitem[\protect\citeauthoryear{Lima et al.}{2021}]{Lima21}
        Lima I.J., Rodrigues C.V., Ferreira Lopes, C.E., et al.,
        2021, AJ, 161, 225
\bibitem[\protect\citeauthoryear{Lipkin \& Leibowitz}{2008}]{Lipkin08} 
        Lipkin Y.M, Leibowitz E.M., 2008, MNRAS, 387, 289
\bibitem[\protect\citeauthoryear{Lomb}{1976}]{Lomb76} 
        Lomb N.R., 1976, Ap\&SS, 39, 447
\bibitem[\protect\citeauthoryear{Mandel}{1965}]{Mandel65} 
        Mandel O.E., 1965, Perem.\ Szvesdy, 15, 475
\bibitem[\protect\citeauthoryear{McQuillin et al.}{2012}]{McQuillin12} 
        McQuillin R., Evans A, Wilson D., Maxtet P.F.L., Placco D., West R.G.,
        Hounsell R.A., Bode M.F., 2012, MNRAS, 419, 330
\bibitem[\protect\citeauthoryear{Mercado \& Honeycutt}{2002}]{Mercado02} 
        Mercado L., Honeycutt R.K., 2002, Bull.\ AAS, 34, 1162
\bibitem[\protect\citeauthoryear{Mironov et al.}{1983}]{Mironov83} 
        Mironov A.V., Moshkalev V.G., Shugarov S.Yu., 1983, IBVS, 2438
\bibitem[\protect\citeauthoryear{Misselt \& Shafter}{1995}]{Misselt95} 
        Misselt K.A., Shafter A.W., 1995, AJ, 109, 1757
\bibitem[\protect\citeauthoryear{Montgomery}{2009}]{Montgomery09} 
        Montgomery M.M., 2009, ApJ, 705, 603
\bibitem[\protect\citeauthoryear{Motch}{1981}]{Motch81a} 
        Motch C., 1981, A\&A, 200, 277
\bibitem[\protect\citeauthoryear{Motch et al.}{1998}]{Motch98} 
        Motch C., Guillout P., Haberl F., et al., 1998, A\&AS, 132,341
\bibitem[\protect\citeauthoryear{Motch \& Pakull}{1981}]{Motch81} 
        Motch C., Pakull M.W., 1981, A\&A, 101, L9
\bibitem[\protect\citeauthoryear{Mr\'oz et al.}{2015}]{Mroz15} 
        Mr\'oz P., Udalski A., Poleski R., et al., 2015, ApJS, 219, 26
\bibitem[\protect\citeauthoryear{Nather \& Robinson}{1974}]{Nather74}
        Nather R.E., Robinson E.L., 1974, ApJ, 190, 637
\bibitem[\protect\citeauthoryear{Neustroev et al.}{2011}]{Neustroev11} 
        Neustroev V.V., Suleimanov V.F., Borisov N.V., Belyakov K.V., 
        Shearer A., 2011, MNRAS, 410, 963
\bibitem[\protect\citeauthoryear{Norton et al.}{1996}]{Norton96} 
        Norton A.J., Beardmore A.P., Taylor P., 1996, MNRAS, 280, 973
\bibitem[\protect\citeauthoryear{Osaki \& Kato}{2013}]{Osaki13} 
        Osaki Y., Kato T., 2013, PASJ, 65, 95
\bibitem[\protect\citeauthoryear{Osaki \& Kato}{2014}]{Osaki14} 
        Osaki Y., Kato T., 2014, PASJ, 66, 15
\bibitem[\protect\citeauthoryear{Papadaki et al.}{2009}]{Papadaki09} 
        Papadaki C., Boffin H.M.J., Stanishev V., Boumis P., Akras S.,
        Sterken C., 2009, J.\ Astron.\ Data, 15, 1
\bibitem[\protect\citeauthoryear{Papadaki et al}{2006}]{Papadaki06} 
        Papadaki C., Boffin H.M.J., Sterken C., Stanishev V., Cuypers J., 
        Boumis P., Akras S., Alikakos J., 2006, A\&A, 456, 599
\bibitem[\protect\citeauthoryear{Patterson}{1995}]{Patterson95} 
        Patterson J., 1995, PASP, 107, 657
\bibitem[\protect\citeauthoryear{Patterson}{1998}]{Patterson98} 
        Patterson J., 1998, PASP, 110, 1132
\bibitem[\protect\citeauthoryear{Patterson}{1999}]{Patterson99} 
        Patterson J., 1999, in: S.\ Mineshige \& J.C.\ Wheeler (eds.)
        {\it Disk instabilities in close binary systems}, Tokyo Universal
        Academic Press, p.\ 61
\bibitem[\protect\citeauthoryear{Patterson}{2001}]{Patterson01a} 
        Patterson J., 2001, PASP, 113, 736
\bibitem[\protect\citeauthoryear{Patterson et al.}{2002}]{Patterson02} 
        Patterson J., Fenton W.H., Thorstensen J.R., et al., 2002, 
        PASP, 114, 1364 
\bibitem[\protect\citeauthoryear{Patterson et al.}{2005}]{Patterson05} 
        Patterson J., Kemp J., Harvey D.A., et al., 2005, PASP, 117, 1204
\bibitem[\protect\citeauthoryear{Patterson et al.}{1997}]{Patterson97} 
        Patterson J., Kemp, J., Saad J., Skillman D.R., Harvey D., Fried R., 
        Thorstensen J.R., Ashley R., 1997, PASP, 109, 468
\bibitem[\protect\citeauthoryear{Patterson et al.}{1996}]{Patterson96} 
        Patterson J., Patino R., Thorstensen J.R., 1996, AJ, 111, 2422
\bibitem[\protect\citeauthoryear{Patterson \& Price}{1981}]{Patterson81} 
        Patterson J., Price C.M., 1981, ApJ, 243, L83
\bibitem[\protect\citeauthoryear{Patterson \& Richman}{1991}]{Patterson91} 
        Patterson J., Richman H., 1991, PASP, 103, 735
\bibitem[\protect\citeauthoryear{Patterson et al.}{1993}]{Patterson93} 
        Patterson J., Thomas G., Skillman D.R., Diaz M., 1993, ApJS, 86, 235
\bibitem[\protect\citeauthoryear{Patterson et al.}{2001}]{Patterson01b} 
        Patterson J., Thorstensen J.R., Fried R., 2001, PASP, 113, 72
\bibitem[\protect\citeauthoryear{Patterson et al.}{2013}]{Patterson13} 
        Patterson J., Uthas H., Kemp J., et al., 2013, MNRAS, 434, 1902
\bibitem[\protect\citeauthoryear{Patterson \& Warner}{1998}]{Patterson98b} 
        Patterson J., Warner B., 1998, PASP, 110, 1026
\bibitem[\protect\citeauthoryear{Pearson}{2006}]{Pearson06} 
        Pearson K.J., 2006, MNRAS, 371, 235
\bibitem[\protect\citeauthoryear{Quigley \& Africano}{1978}]{Quigley78}
        Quigley R., Africano J., 1978, PASP, 90, 445
\bibitem[\protect\citeauthoryear{Ramsay et al.}{2016}]{Ramsay16} 
        Ramsay G., Hakala P., Wood M.A., 2016, MNRAS, 455, 2772
\bibitem[\protect\citeauthoryear{Rawat et al.}{2022}]{Rawat22}
        Rawat N., Pandey J.C., Joshi A., Yadava U., 2022, MNRAS, 512, 6054
\bibitem[\protect\citeauthoryear{Retter et al.}{1997}]{Retter97}
        Retter A., Leibowitz E.M., Ofek E.O., 1997, MNRAS, 286, 745
\bibitem[\protect\citeauthoryear{Ricker et al.}{2014}]{Ricker14}
        Ricker G.R., Winn, J.N., Vanderspek R., et al., 2014, 
        J.\, Astr.\, Tel.\, Instr.\, \& Systems, 1, 014003
\bibitem[\protect\citeauthoryear{Ringwald et al.}{1996}]{Ringwald96} 
        Ringwald F.A., Thorstensen J.R., Honeycutt R.K., Robertson J.W.,
        1997, MNRAS, 278, 125
\bibitem[\protect\citeauthoryear{Ringwald et al.}{2012}]{Ringwald12} 
        Ringwald F.A., Velasco K., Roveto J.J., Meyers M.E., 2012,
        New Astron., 17,433
\bibitem[\protect\citeauthoryear{Robinson et al.}{1991}]{Robinson91} 
        Robinson E.L., Shretone M.D., Africano J.L., 1991, AJ, 102, 1176
\bibitem[\protect\citeauthoryear{Rodr\'{\i}guez-Gil et al.}{2007}]
        {Rodriguez-Gil07}
        Rodr\'{\i}guez-Gil P., G\"ansicke B.T., Hagen, H.J., et al.,
        2007, MNRAS, 377, 1747
\bibitem[\protect\citeauthoryear{Rodr\'{\i}guez-Gil \& Potter}{2003}]
        {Rodriguez-Gil03}
        Rodr\'{\i}guez-Gil P., Potter S.B., 2003, MNRAS, 342, L1
\bibitem[\protect\citeauthoryear{Rodr\'{\i}guez-Gil et al.}{2012}]
        {Rodriguez-Gil12}
        Rodr\'{\i}guez-Gil P., Schmidtobreick, L., Long, K.S., G\"ansicke B.T.,
        Torres M.P.A., Rubio D\'{\i}ez M.M., Santander-Garc\'{\i}ia M., 2012
        MNRAS, 422, 2332
\bibitem[\protect\citeauthoryear{Rodr\'{\i}guez-Gil et al.}{2020}]
        {Rodriguez-Gil20}
        Rodr\'{\i}guez-Gil P., Shahbaz T., Torres M.P.A., G\"ansicke B.T.,
        Izquierdo P., Toloza O., \'Alvarez-Hern\'andez A., Steeghs D., 2020, 
        MNRAS, 494, 425
\bibitem[\protect\citeauthoryear{Rodr\'{\i}guez-Gil \& Torres}{2005}]
        {Rodriguez-Gil05} 
        Rodr\'{\i}guez-Gil P, Torres M.P.A., 2005, A\&A, 431, 289
\bibitem[\protect\citeauthoryear{R\"ossiger}{1988}]{Roessiger88} 
        R\"ossiger R., 1988, Sternwarte Sonneberg, Mitt. Veraenderliche 
        Sterne, 11, 112
\bibitem[\protect\citeauthoryear{Rolfe et al.}{2000}]{Rolfe00} 
        Rolfe D.J., Haswell C.A., Patterson J., 2000, MNRAS, 317, 759
\bibitem[\protect\citeauthoryear{Rosen et al.}{1989}]{Rosen89} 
        Rosen S.A., Branduardi-Raymont G., Mason K.O., Murdin P.G., 1989, 
        MNRAS, 237, 1037
\bibitem[\protect\citeauthoryear{Rosen et al.}{1994}]{Rosen94} 
        Rosen S.R., Clayton K.L., Osborne J.P., McGale P.A., 1994,
        MNRAS, 269, 913
\bibitem[\protect\citeauthoryear{Rubenstein et al.}{1991}]{Rubenstein91}
        Rubenstein E.P., Patterson J., Africano J.L., 1991, PASP, 103, 1258
\bibitem[\protect\citeauthoryear{Rude \& Ringwald}{2012}]{Rude12} 
        Rude G.D., Ringwald F.A., 2012, New Astr., 17, 453
\bibitem[\protect\citeauthoryear{Rutten et al.}{1992}]{Rutten92}
        Rutten R.G.M., van Paradijs J., Tinbergen J., 1992, A\&A, 260, 213
\bibitem[\protect\citeauthoryear{Saito \& Baptista}{2016}]{Saito16} 
        Saito R., Baptista R., 2016, MNRAS, 457, 198
\bibitem[\protect\citeauthoryear{Savitzky \& Golay}{1964}]{Savitzky64} 
        Savitzky A., Golay M.J.E., 1964, Analytical Chemistry, 36, 1627
\bibitem[\protect\citeauthoryear{Scargle}{1982}]{Scargle82} 
        Scargle J.D., 1982, ApJ, 263, 853
\bibitem[\protect\citeauthoryear{Schmidtobreick et al.}{2008}]
        {Schmidtobreick08} 
        Schmidtobreick L., Papadaki C., Tappert C., Ederoclite A., 2008,
        MNRAS, 389, 1345
\bibitem[\protect\citeauthoryear{Schwarzenberg-Czerny}{1991}]
        {Schwarzenberg-Czerny91} 
        Schwarzenberg-Czerny A., 1991, MNRAS, 253, 198
\bibitem[\protect\citeauthoryear{Semeniuk et al.}{1995}]{Semeniuk95} 
        Semeniuk I., de Young J.A., Pych W., Olech A., Ruszkowski M., 
        Schmidt R.E., 1995, Acta Astr., 45, 365
\bibitem[\protect\citeauthoryear{Semeniuk et al.}{1994}]{Semeniuk94} 
        Semeniuk I., Pych W., Olech A., Ruszkowski M., 1994, Acta Astr.,
        44, 277
\bibitem[\protect\citeauthoryear{Semeniuk}{1987}]{Semeniuk87} 
        Semeniuk I., Schwarzenberg-Czerny A., Duerbeck H., Hoffman M., Smak J., 
        Stepien K., Tremko J., 1987, Ap\&SS, 130, 167
\bibitem[\protect\citeauthoryear{Shafter}{1983}]{Shafter83} 
        Shafter A.W., 1983, ApJ, 267, 222
\bibitem[\protect\citeauthoryear{Shafter et al.}{1990}]{Shafter90} 
        Shafter A.W., Robinson E.L., Crampton D., Warner B., Prestage R.M., 
        1990, ApJ 354, 708
\bibitem[\protect\citeauthoryear{\v{S}imon et al.}{2012}]{Simon12} 
        \v{S}imon V., Polasek C., Strobl J., Hudec C., Blazek M., 2012, 
        A\&A, 540, A15
\bibitem[\protect\citeauthoryear{Singh et al.}{1993}]{Singh93}
        Singh J., Rao P.V., Agrawal P.C., Apparao K.M.V., Manchanda R.K.,
        Sanwal B.B., Sarma M.B.K., 1993, ApJ, 419, 337
\bibitem[\protect\citeauthoryear{Skillman et al.}{1998}]{Skillman98} 
        Skillman D.R., Harvey D.A., Patterson J., 1998, ApJ, 503, L67
\bibitem[\protect\citeauthoryear{Skillman \& Patterson}{1993}]{Skillman93} 
        Skillman D.R., Patterson J., 1993, ApJ, 417, 298
\bibitem[\protect\citeauthoryear{Skillman et al.}{1995}]{Skillman95} 
        Skillman D.R., Patterson J., Thorstensen J.R., 1995, PASP, 107, 545
\bibitem[\protect\citeauthoryear{Smak}{2019}]{Smak19}
        Smak J., 2019, Acta Astr., 69, 79
\bibitem[\protect\citeauthoryear{Smak}{2020}]{Smak20} 
        Smak J., 2020, Acta Astron., 70, 313
\bibitem[\protect\citeauthoryear{Smak \& St\c{e}pie\'n}{1975}]{Smak75} 
        Smak J., St\c{e}pie\'n K., 1975, Acta Astron., 25, 379
\bibitem[\protect\citeauthoryear{Stanishev et al.}{2002}]{Stanishev02} 
        Stanishev V.S., Kraicheva Z., Boffin H.M.J., Genkov V., 2002,
        A\&A, 394, 625
\bibitem[\protect\citeauthoryear{Stanishev et al.}{2006}]{Stanishev06} 
        Stanishev V., Kraicheva Z., Genkov V., 2006, A\&A, 455, 223
\bibitem[\protect\citeauthoryear{Stefanov et al.}{2022}]{Stefanov22} 
        Stefanov S.Y., Latev G., Boeva S., Moyseev M., 2022, MNRAS, 516, 2775
\bibitem[\protect\citeauthoryear{Stolz \& Schoembs}{1984}]{Stolz84} 
        Stolz V., Schoembs R., 1984, A\&A, 132, 187
\bibitem[\protect\citeauthoryear{Szkody et al.}{1989}]{Szkody89} 
        Szkody P., Howell S.B., Mateo M., Kreidl T.J., 1989, PASP, 101, 899
\bibitem[\protect\citeauthoryear{Szkody et al.}{2013}]{Szkody13} 
        Szkody P., Mukadam A.S., G\"ansicke B.T., et al., 2013, ApJ, 775, 66
\bibitem[\protect\citeauthoryear{Sztanjo}{1979}]{Sztanjo79} 
        Sztanjo M., 1979, IBVS, 1710
\bibitem[\protect\citeauthoryear{Taylor et al.}{1998}]{Taylor98} 
        Taylor C.J., Thorstenson J.R., Patterson J., et al., 1998, 
        PASP, 110, 1148
\bibitem[\protect\citeauthoryear{Thomas \& Wood}{2015}]{Thomas15} 
        Thomas D.M., Wood M.A., 2015, ApJ, 803, 55
\bibitem[\protect\citeauthoryear{Thorstensen et al.}{2010}]{Thorstensen10} 
        Thorstensen J.R., Peters C.S., Skinner J.N., 2010, PASP, 122, 1285
        ApJ, 359, 204
\bibitem[\protect\citeauthoryear{Tremko et al.}{1992}]{Tremko92} 
        Tremko J., Andronov I.L., Luthard R., Pajdosz G., Patkos, L. 
        R\"ossiger S., Zola S., 1992, IBVS 3763
\bibitem[\protect\citeauthoryear{Tuohy et al.}{1990}]{Tuohy90} 
        Tuohy I.R., Remillard R.A., Brissenden R.J.V., Bradt H.V., 1990, 
        ApJ, 359, 204
\bibitem[\protect\citeauthoryear{Udalski}{1988}]{Udalski88} 
        Udalski A., 1988, Acta Astron., 38, 315
\bibitem[\protect\citeauthoryear{van Houten}{1966}]{vanHouten66} 
        van Houten C.J., 1966, Bull.\ Astron.\ Inst.\ Neth., 18, 439 
\bibitem[\protect\citeauthoryear{van Zyl et al.}{2004}]{vanZyl04} 
        van Zyl L., Warner B., O'Donogue D., et al., 2004, MNRAS, 350, 307
\bibitem[\protect\citeauthoryear{Vogt}{1974}]{Vogt74}
        Vogt N., 1974, A\&A, 36, 369
\bibitem[\protect\citeauthoryear{Volkov et al.}{1986}]{Volkov86}
        Volkov I.M., Shugarov S.Yu., Seregina T.M., 1986, 
        Astr.\ Tsirk., 1418, 3
\bibitem[\protect\citeauthoryear{Volpi et al.}{1998}]{Volpi88} 
        Volpi A., Natali G., D'Antona F., 1988, A\&A 193, 87
\bibitem[\protect\citeauthoryear{Warner}{1986}]{Warner86} 
        Warner B., 1986, MNRAS, 219, 347
\bibitem[\protect\citeauthoryear{Weingrill et al.}{2009}]{Weingrill09} 
        Weingrill J., Kleinschuster G., Kuschnik R., Matthews J.M., Moffat A., 
        Rucinski S., Sasselov D., Weiss W.W., 2009, Comm. Astroseism. 159, 114
\bibitem[\protect\citeauthoryear{Whitehurst}{1988}]{Whitehurst88} 
        Whitehurst R., 1988, MNRAS, 232, 35
\bibitem[\protect\citeauthoryear{Whitehurst \& King}{1991}]{Whitehurst91} 
        Whitehurst R., King, A., 1991, MNRAS, 249, 25
\bibitem[\protect\citeauthoryear{Williams et al.}{2010}]{Williams10} 
        Williams, K.A., de Martino D., Silvotti R., et al., 2010, AJ, 139, 2587
\bibitem[\protect\citeauthoryear{Wood et al.}{2011}]{Wood11} 
        Wood M.A., Still M.D., Howell S.B., Cannizzo J.K., Smale A.P.,
        2011, ApJ, 741, 105
\bibitem[\protect\citeauthoryear{Woudt \& Warner}{2001}]{Woudt01} 
        Woudt P., Warner B., 2001, MNRAS, 328, 159
\bibitem[\protect\citeauthoryear{Woudt \& Warner}{2002}]{Woudt02} 
        Woudt P., Warner B., 2002, MNRAS, 335, 44
\bibitem[\protect\citeauthoryear{Wu et al.}{2002}]{Wu02} 
        Wu X., Li Z., Ding Y., Zhang Z., Li Z., 2002, ApJ 569, 418
\bibitem[\protect\citeauthoryear{Yang et al.}{2017}]{Yang17} 
        Yang M.T., Chou Y., Ngeow C.-C., 2017, PASP, 129, 4202
\bibitem[\protect\citeauthoryear{Zhang et al.}{1991}]{Zhang91} 
        Zhang E., Robinson E.L., Ramseyer T.F., Shretone M.D., Stiening R.F., 
        1991, ApJ, 381, 534
\bibitem[\protect\citeauthoryear{Zorotovic et al.}{2011}]{Zorotovic11} 
        Zorotovic M., Schreiber M.R., G\"ansicke B.T. 2011, A\&A, 536, A42
\end{thebibliography}


\appendix

\section{Additional eclipse timings}
\label{Additional eclipse timings}

\begin{table*}
\label{Table: UU Aqr eclipse epochs}
	\centering
	\caption{UU~Aqr eclipse epochs (zero point for cycle counts
                 as defined by Eq.~\ref{UU Aqr: linear ephemeris}).}

\begin{tabular}{rrrrrrrrrr}
\hline
Epoch (BJD)  & Cycle & Epoch (BJD) & Cycle &
Epoch (BJD)  & Cycle & Epoch (BJD) & Cycle & Epoch (BJD) & Cycle \\
(2400000+)   & No.   & (2400000+)  & No.   &
(2400000+)   & No.   & (2400000+)  & No.   & (2400000+)  & No.   \\
\hline
51755.7281 &      0 & 55415.8404 &  22375 & 
56215.4214 &  27263 & 57249.7401 &  33586 & 58368.6292 &  40426 \\
51756.5464 &      5 & 55416.6578 &  22380 & 
56507.5749 &  29049 & 57262.4991 &  33664 & 58369.4481 &  40431 \\
51756.7100 &      6 & 55416.8217 &  22381 & 
56507.7392 &  29050 & 57609.4530 &  35785 & 58369.6107 &  40432 \\
53234.6594 &   9041 & 55417.6403 &  22386 & 
56508.5574 &  29055 & 57617.4687 &  35834 & 58370.4295 &  40437 \\
54322.4696 &  15691 & 55417.8032 &  22387 & 
56508.7214 &  29056 & 57627.6106 &  35896 & 58370.5924 &  40438 \\ [1ex]
54323.4510 &  15697 & 55418.6214 &  22392 & 
56509.5399 &  29061 & 57627.7741 &  35897 & 58370.7575 &  40439 \\
54325.4128 &  15709 & 55418.7846 &  22393 & 
56509.7027 &  29062 & 57628.5922 &  35902 & 58371.5731 &  40444 \\
54357.4749 &  15905 & 55469.4949 &  22703 & 
56510.6839 &  29068 & 57628.7564 &  35903 & 58371.7378 &  40445 \\
54365.4905 &  15954 & 55778.4979 &  24592 & 
56510.6842 &  29068 & 57629.5739 &  35908 & 58372.5563 &  40450 \\
54728.4749 &  18173 & 55795.5097 &  24696 & 
56523.4439 &  29146 & 57629.7363 &  35909 & 58372.7188 &  40451 \\ [1ex]
54731.4211 &  18191 & 55799.5995 &  24721 & 
56563.5208 &  29391 & 57630.5556 &  35914 & 58373.5371 &  40456 \\
54734.5267 &  18210 & 55799.7637 &  24722 & 
56563.5209 &  29391 & 57630.7185 &  35915 & 58373.7004 &  40457 \\
54734.6915 &  18211 & 55800.5812 &  24727 & 
56563.6837 &  29392 & 57642.4966 &  35987 & 58374.5178 &  40462 \\
54735.3454 &  18215 & 55800.7451 &  24728 & 
56563.6837 &  29392 & 57991.7407 &  38122 & 58377.6263 &  40481 \\
54736.3271 &  18221 & 55801.5633 &  24733 & 
56872.6888 &  31281 & 57995.6656 &  38146 & 58378.4439 &  40486 \\ [1ex]
54810.2672 &  18673 & 55801.7268 &  24734 & 
56872.8515 &  31282 & 58349.6548 &  40310 & 58378.6062 &  40487 \\
54830.2252 &  18795 & 55893.3309 &  25294 & 
56874.6512 &  31293 & 58349.8171 &  40311 & 58726.5426 &  42614 \\
55059.3979 &  20196 & 56157.6775 &  26910 & 
56874.8140 &  31294 & 58350.4714 &  40315 & 58726.7062 &  42615 \\
55106.3460 &  20483 & 56159.4761 &  26921 & 
56893.4627 &  31408 & 58351.6177 &  40322 & 58733.5766 &  42657 \\
55415.6770 &  22374 & 56160.4586 &  26927 & 
57249.5761 &  33585 & 58351.7801 &  40323 &    &    \\
\hline
\end{tabular}
\end{table*}

\begin{table*}
\label{Table: V348 Pup eclipse epochs}	\
\centering
	\caption{V348~Pup eclipse epochs (zero point for cycle counts
                 as defined by \citet{Dai10}).}

\begin{tabular}{rrrrrrrrrr}
\hline
Epoch (BJD)  & Cycle & Epoch (BJD) & Cycle &
Epoch (BJD)  & Cycle & Epoch (BJD) & Cycle & Epoch (BJD) & Cycle \\
(2400000+)   & No.   & (2400000+)  & No.   &
(2400000+)   & No.   & (2400000+)  & No.   & (2400000+)  & No.   \\
\hline
56631.6506 &  78948 & 56633.6873 &  78968 & 
57008.7600 &  82651 & 57728.7619 &  89721 & 57729.7803 &  89731 \\
56633.5858 &  78967 & 56633.7889 &  78969 & 
57728.6603 &  89720 & 57729.6783 &  89730 &            &        \\
\hline
\end{tabular}
\end{table*}

\begin{table*}
\label{Table: RW Tri eclipse epochs}
	\centering
	\caption{RW Tri eclipse epochs (zero point for cycle counts
                 as defined by Eq.~\ref{RW Tri: linear ephemeris}).}

\begin{tabular}{rrrrrrrrrr}
\hline
Epoch (BJD)  & Cycle & Epoch (BJD) & Cycle &
Epoch (BJD)  & Cycle & Epoch (BJD) & Cycle & Epoch (BJD) & Cycle \\
(2400000+)   & No.   & (2400000+)  & No.   &
(2400000+)   & No.   & (2400000+)  & No.   & (2400000+)  & No.   \\
\hline
53672.6223 &      0 & 56228.4405 &  11022 & 
57328.2638 &  15765 & 57337.5394 &  15805 & 57409.6550 &  16116 \\
54392.3882 &   3104 & 56609.4252 &  12665 & 
57329.4233 &  15770 & 57338.6984 &  15810 & 57410.5828 &  16120 \\
54419.5180 &   3221 & 56619.3963 &  12708 & 
57329.6556 &  15771 & 57339.3944 &  15813 & 57415.6840 &  16142 \\
54447.3442 &   3341 & 56636.3240 &  12781 & 
57329.8875 &  15772 & 57340.3218 &  15817 & 57416.6113 &  16146 \\
54835.2862 &   5014 & 56922.4676 &  14015 & 
57330.5834 &  15775 & 57342.4087 &  15826 & 57419.3941 &  16158 \\ [1ex]
55063.4584 &   5998 & 56933.3661 &  14062 & 
57330.8145 &  15776 & 57343.3362 &  15830 & 57420.3215 &  16162 \\
55106.3577 &   6183 & 56935.4533 &  14071 & 
57331.2785 &  15778 & 57344.2639 &  15834 & 57421.4813 &  16167 \\
55172.4441 &   6468 & 57314.3506 &  15705 & 
57331.5104 &  15779 & 57345.4232 &  15839 & 57422.4087 &  16171 \\
55487.3422 &   7826 & 57315.7423 &  15711 & 
57332.6698 &  15784 & 57391.7998 &  16039 & 57423.3364 &  16175 \\
55490.3565 &   7839 & 57317.5974 &  15719 & 
57332.9016 &  15785 & 57397.8289 &  16065 & 57424.4957 &  16180 \\ [1ex]
55533.4868 &   8025 & 57319.6843 &  15728 & 
57333.3658 &  15787 & 57399.6844 &  16073 & 57424.7273 &  16181 \\
55822.4132 &   9271 & 57320.3800 &  15731 & 
57334.2930 &  15791 & 57400.3797 &  16076 & 57623.4522 &  17038 \\
55828.4424 &   9297 & 57321.3071 &  15735 & 
57334.5248 &  15792 & 57400.6118 &  16077 & 57642.4662 &  17120 \\
55867.3985 &   9465 & 57321.5389 &  15736 & 
57334.7571 &  15793 & 57401.7710 &  16082 & 57645.4807 &  17133 \\
55881.3116 &   9525 & 57324.5538 &  15749 & 
57335.4523 &  15796 & 57402.6986 &  16086 & 57684.9010 &  17303 \\ [1ex]
55950.4122 &   9823 & 57326.4088 &  15757 & 
57335.9160 &  15798 & 57403.3943 &  16089 & 57685.5961 &  17306 \\
55953.4270 &   9836 & 57326.6406 &  15758 & 
57336.3799 &  15800 & 57403.6260 &  16090 & 57685.8285 &  17307 \\
55957.3700 &   9853 & 57327.3361 &  15761 & 
57336.6115 &  15801 & 57407.5681 &  16107 & 57686.7559 &  17311 \\
56200.3826 &  10901 & 57327.5684 &  15762 & 
57336.8436 &  15802 & 57408.7277 &  16112 & 58431.3325 &  20522 \\
\hline
\end{tabular}
\end{table*}

\begin{table*}
\label{Table: UX UMa eclipse epochs}
	\centering
	\caption{UX UMa eclipse epochs (zero point for cycle counts
                 as defined by Eq.~\ref{UX UMa: linear ephemeris}).}

\begin{tabular}{rrrrrrrrrr}
\hline
Epoch (BJD)  & Cycle & Epoch (BJD) & Cycle &
Epoch (BJD)  & Cycle & Epoch (BJD) & Cycle & Epoch (BJD) & Cycle \\
(2400000+)   & No.   & (2400000+)  & No.   &
(2400000+)   & No.   & (2400000+)  & No.   & (2400000+)  & No.   \\
\hline
43311.7185 & -40718 & 56688.7090 &  27299 & 
57116.6640 &  29475 & 57143.8064 &  29613 & 57525.7421 &  31555 \\
43660.8097 & -38943 & 56725.2894 &  27485 & 
57117.6483 &  29480 & 57144.0029 &  29614 & 57526.7250 &  31560 \\
43998.6925 & -37225 & 56725.4859 &  27486 & 
57118.4352 &  29484 & 57144.3959 &  29616 & 57529.6749 &  31575 \\
48779.7710 & -12915 & 56731.3853 &  27516 & 
57119.4182 &  29489 & 57144.7893 &  29618 & 57530.6580 &  31580 \\
51319.7821 &      0 & 56733.3520 &  27526 & 
57119.6150 &  29490 & 57145.7726 &  29623 & 57785.9373 &  32878 \\ [1ex]
52363.5161 &   5307 & 56733.5494 &  27527 & 
57119.8115 &  29491 & 57145.9689 &  29624 & 57804.8179 &  32974 \\
52551.3369 &   6262 & 56734.3363 &  27531 & 
57120.4025 &  29494 & 57146.7558 &  29628 & 57809.5381 &  32998 \\
53466.6455 &  10916 & 56734.5324 &  27532 & 
57121.3857 &  29499 & 57147.7394 &  29633 & 57811.8981 &  33010 \\
53473.9221 &  10953 & 56738.2695 &  27551 & 
57121.5827 &  29500 & 57148.5256 &  29637 & 57883.6832 &  33375 \\
53476.8723 &  10968 & 56738.4663 &  27552 & 
57121.7789 &  29501 & 57150.6887 &  29648 & 57888.7966 &  33401 \\ [1ex]
53477.8555 &  10973 & 56739.2526 &  27556 & 
57122.3686 &  29504 & 57150.8855 &  29649 & 57890.7632 &  33411 \\
53481.7887 &  10993 & 56739.4495 &  27557 & 
57122.5651 &  29505 & 57152.4585 &  29657 & 57907.6769 &  33497 \\
53495.7527 &  11064 & 56741.4160 &  27567 & 
57122.7608 &  29506 & 57152.8521 &  29659 & 57939.3410 &  33658 \\
53760.8655 &  12412 & 56741.6130 &  27568 & 
57122.9581 &  29507 & 57153.4429 &  29662 & 58154.8926 &  34754 \\
54154.4047 &  14413 & 56742.3995 &  27572 & 
57123.3518 &  29509 & 57153.8352 &  29664 & 58173.3798 &  34848 \\ [1ex]
54187.4456 &  14581 & 56742.5968 &  27573 & 
57123.7447 &  29511 & 57154.4258 &  29667 & 58173.5762 &  34849 \\
54512.3469 &  16233 & 56743.3830 &  27577 & 
57124.5319 &  29515 & 57154.6222 &  29668 & 58175.3468 &  34858 \\
54912.7701 &  18269 & 56749.4800 &  27608 & 
57124.7287 &  29516 & 57157.5729 &  29683 & 58175.5431 &  34859 \\
54933.6164 &  18375 & 56751.4463 &  27618 & 
57125.3192 &  29519 & 57158.5561 &  29688 & 58179.8698 &  34881 \\
55279.3648 &  20133 & 56752.4297 &  27623 & 
57125.5155 &  29520 & 57160.5220 &  29698 & 58181.8367 &  34891 \\ [1ex]
55279.7582 &  20135 & 56754.3964 &  27633 & 
57126.4985 &  29525 & 57162.4906 &  29708 & 58192.6534 &  34946 \\
55298.6385 &  20231 & 56754.5929 &  27634 & 
57126.6951 &  29526 & 57163.4728 &  29713 & 58212.5173 &  35047 \\
55309.6520 &  20287 & 56756.3637 &  27643 & 
57127.4820 &  29530 & 57164.4564 &  29718 & 58246.7381 &  35221 \\
55616.4587 &  21847 & 56757.3465 &  27648 & 
57127.6771 &  29531 & 57165.4392 &  29723 & 58527.3874 &  36648 \\
55617.4423 &  21852 & 56757.5433 &  27649 & 
57128.6623 &  29536 & 57166.4216 &  29728 & 58527.5843 &  36649 \\ [1ex]
55621.3754 &  21872 & 56763.4428 &  27679 & 
57128.8586 &  29537 & 57166.6203 &  29729 & 58554.7254 &  36787 \\
55993.2806 &  23763 & 56785.6671 &  27792 & 
57129.6454 &  29541 & 57168.3892 &  29738 & 58554.9221 &  36788 \\
55993.4773 &  23764 & 56789.6006 &  27812 & 
57129.8413 &  29542 & 57168.5861 &  29739 & 58559.4450 &  36811 \\
55994.2636 &  23768 & 56789.7974 &  27813 & 
57130.6284 &  29546 & 57169.5706 &  29744 & 58559.6411 &  36812 \\
55994.4605 &  23769 & 56792.3540 &  27826 & 
57131.4158 &  29550 & 57169.7664 &  29745 & 58620.6092 &  37122 \\ [1ex]
55995.2469 &  23773 & 56794.3202 &  27836 & 
57131.8085 &  29552 & 57170.5534 &  29749 & 58627.6898 &  37158 \\
55995.4438 &  23774 & 56802.7776 &  27879 & 
57132.0061 &  29553 & 57170.7493 &  29750 & 58898.8991 &  38537 \\
56012.7505 &  23862 & 56807.6943 &  27904 & 
57132.3987 &  29555 & 57171.5348 &  29754 & 58915.2223 &  38620 \\
56346.5016 &  25559 & 56811.6282 &  27924 & 
57132.5952 &  29556 & 57173.5031 &  29764 & 58915.4194 &  38621 \\
56351.6146 &  25585 & 57078.7074 &  29282 & 
57132.7923 &  29557 & 57177.4367 &  29784 & 58922.4999 &  38657 \\ [1ex]
56353.3848 &  25594 & 57081.6568 &  29297 & 
57132.9895 &  29558 & 57190.4161 &  29850 & 58923.2869 &  38661 \\
56353.5812 &  25595 & 57081.8539 &  29298 & 
57133.3822 &  29560 & 57191.4004 &  29855 & 58923.4825 &  38662 \\
56372.4621 &  25691 & 57083.8209 &  29308 & 
57133.7748 &  29562 & 57192.5799 &  29861 & 58941.5776 &  38754 \\
56376.3951 &  25711 & 57092.8677 &  29354 & 
57133.9721 &  29563 & 57193.5634 &  29866 & 58941.7736 &  38755 \\
56378.3616 &  25721 & 57098.5713 &  29383 & 
57134.3648 &  29565 & 57194.5476 &  29871 & 59283.7852 &  40494 \\ [1ex]
56381.3120 &  25736 & 57099.3580 &  29387 & 
57134.7581 &  29567 & 57198.4803 &  29891 & 59293.8150 &  40545 \\
56381.5088 &  25737 & 57099.5551 &  29388 & 
57135.5457 &  29571 & 57199.4629 &  29896 & 59310.7292 &  40631 \\
56382.2954 &  25741 & 57099.7514 &  29389 & 
57135.7420 &  29572 & 57202.4135 &  29911 & 59317.6125 &  40666 \\
56383.2791 &  25746 & 57100.3415 &  29392 & 
57135.9383 &  29573 & 57207.5266 &  29937 & 59317.8093 &  40667 \\
56383.4753 &  25747 & 57100.5382 &  29393 & 
57136.5285 &  29576 & 57208.5098 &  29942 & 59332.7564 &  40743 \\ [1ex]
56385.4419 &  25757 & 57102.7016 &  29404 & 
57136.7258 &  29577 & 57209.4956 &  29947 & 59352.6205 &  40844 \\
56386.4253 &  25762 & 57103.2916 &  29407 & 
57138.4943 &  29586 & 57222.4738 &  30013 & 59364.6172 &  40905 \\
56388.3921 &  25772 & 57103.4888 &  29408 & 
57139.6757 &  29592 & 57223.4576 &  30018 & 59605.3423 &  42129 \\
56391.3420 &  25787 & 57104.4708 &  29413 & 
57140.4630 &  29596 & 57224.4399 &  30023 & 59605.5396 &  42130 \\
56391.5389 &  25788 & 57107.4216 &  29428 & 
57140.6594 &  29597 & 57226.4081 &  30033 & 59622.4531 &  42216 \\ [1ex]
56393.3089 &  25797 & 57107.6197 &  29429 & 
57140.8548 &  29598 & 57227.3907 &  30038 & 59639.3667 &  42302 \\
56395.2752 &  25807 & 57108.7979 &  29435 & 
57141.4453 &  29601 & 57227.7848 &  30040 & 59639.5644 &  42303 \\
56454.6696 &  26109 & 57108.9958 &  29436 & 
57141.6419 &  29602 & 57467.3300 &  31258 & 59675.7512 &  42487 \\
56686.3487 &  27287 & 57109.7807 &  29440 & 
57141.8388 &  29603 & 57467.5267 &  31259 & 59703.6790 &  42629 \\
56686.5451 &  27288 & 57110.7647 &  29445 & 
57142.4293 &  29606 & 57470.6735 &  31275 & 59703.8751 &  42630 \\ [1ex]
56687.3310 &  27292 & 57111.7481 &  29450 & 
57142.6257 &  29607 & 57471.6567 &  31280 &    &    \\
56687.5284 &  27293 & 57111.9446 &  29451 & 
57142.8231 &  29608 & 57494.6675 &  31397 &    &    \\
56688.3147 &  27297 & 57113.7153 &  29460 & 
57143.4127 &  29611 & 57520.6280 &  31529 &    &    \\
56688.5115 &  27298 & 57114.8952 &  29466 & 
57143.6087 &  29612 & 57520.8249 &  31530 &    &    \\
\hline
\end{tabular}
\end{table*}


\bsp	
\label{lastpage}
\end{document}